\documentclass[sn-mathphys,Numbered]{sn-jnl}

\usepackage{graphicx}%
\usepackage{multirow}%
\usepackage{amsmath,amssymb,amsfonts}%
\usepackage{amsthm}%
\usepackage{mathrsfs}%
\usepackage[title]{appendix}%
\usepackage{xcolor}%
\usepackage{textcomp}%
\usepackage{manyfoot}%
\usepackage{booktabs}%
\usepackage{algorithm}%
\usepackage{algorithmicx}%
\usepackage{algpseudocode}%
\usepackage{listings}%
\usepackage{hyperref}

\graphicspath{{Images/}}

\newcommand{\eg}{\emph{e.g.}\ }
\newcommand{\ie}{\emph{i.e.}\ }

\newcommand{\calK}{\ensuremath{\mathcal{K}}}
\newcommand{\kmax}{\ensuremath{k_{\max}}}

\newcommand{\xmin}{\ensuremath{x_{\min}}}
\newcommand{\xmax}{\ensuremath{x_{\max}}}

\newcommand{\Frag}{\mytext{Frag.}}

\newcommand{\floor}[1]{\lfloor #1 \rfloor}
\newcommand{\ceil}[1]{\lceil #1 \rceil}
\newcommand{\mytext}[1]{\ensuremath{\mathrm{#1}}}
\newcommand{\Prob}[1]{\ensuremath{\mathrm{Pr}}\left( #1 \right)}

\begin{document}

\title[Article Title]{Robustness of steady state and stochastic cyclicity in generalized coalescence-fragmentation models}


\author*[1,2]{\fnm{Brennen T.} \sur{Fagan}}\email{brennen.fagan@york.ac.uk}

\author[1]{\fnm{Niall J.} \sur{MacKay}}\email{niall.mackay@york.ac.uk}

\author[1,3]{\fnm{A. Jamie} \sur{Wood}}\email{jamie.wood@york.ac.uk}

\affil[1]{\orgdiv{Department of Mathematics}, \orgname{University of York}, \orgaddress{\street{Heslington Lane}, \city{York}, \postcode{YO10 5DD},  \country{UK}}}

\affil[2]{\orgdiv{Leverhulme Centre for Anthropocene Diversity}, \orgname{University of York}, \orgaddress{\street{Heslington Lane}, \city{York}, \postcode{YO10 5DD},  \country{UK}}}

\affil[3]{\orgdiv{Department of Biology}, \orgname{University of York}, \orgaddress{\street{Heslington Lane}, \city{York}, \postcode{YO10 5DD},  \country{UK}}}

\abstract{Processes of coalescence and fragmentation are used to understand the time-evolution of the mass distribution of various systems and may result in a steady state or in stable deterministic or stochastic cycles.
Motivated by applications in insurgency warfare we investigate coalescence-fragmentation systems. 
We begin with a simple model of size-biased coalescence accompanied by shattering into monomers. 
Depending on the parameters this model has an approximately power-law-distributed steady state or stochastic cycles of alternating gelation and shattering. 
We conduct stochastic simulations of this model and its generalizations to include different kernel types, accretion and erosion, and various distributions of non-shattering fragmentation. 
Our central aim is to explore the robustness of the steady state and gel-shatter cycles to these variations. 
We show that an approximate power-law steady state persists with the addition of accretion and erosion, and with partial rather than total shattering. 
However, broader distributions of fragment sizes typically vitiate both the power law steady state and gel-shatter cyclicity.
This work clarifies features shown in coalescence/fragmentation model simulations and elucidates the relationship between the microscopic dynamics and observed phenomena in this widely applicable interdisciplinary model type.}



\maketitle
\pagebreak
\section{Introduction}

One of the most prominent quantitative results in political science is due to the physicist Lewis Fry Richardson \cite{Richardson60_SDQ}, who observed that deaths caused by other humans are well-approximated by power laws across a wide range of sizes, with an exponent of around 2.5 for murders (1 death) up to small wars (1000 deaths), and a smaller exponent around 1.5 for larger wars. A body of recent work, exemplified by \cite{Bohorquez09_Common} for the FARC insurgency in Colombia, and summarized in \cite{Spagat18_WarEventSizes}, has demonstrated that similar distributions occur for event-size distributions for a wide range of modern wars and insurgencies. Although it has been challenged -- there are cases where the drop-off is faster than expected \cite{Zwetsloot18_TRL} -- Richardson's law remains a broadly applicable truth.

Why is this so? 
The belief is that event sizes approximately follow the distributions of the sizes of the human (typically insurgent) groupings which cause them. 
Distributions which are approximately power-law across most of their range are common in wide range of phenomena and can result from a range of generative processes including, most prominently, preferential attachment {\cite{Simon55_Skew}.
More broadly, processes of coalescence, or (in finite populations) of coalescence balanced by fragmentation (CF), produce distributions that have approximately power-law steady states. 
Most simply, a multiplicative kernel, in which the probability of two groups coalescing is proportional to the product of their sizes, yields a power-law exponent of 2.5 (see below).
Yet it would seem rather too strong to believe that the complexity of human grouping for deadly conflict is captured by so simple a model. 
More reasonable, \emph{a priori}, is some form of universality class -- that a wide range of such models, with varied rules for coalescence and fragmentation, all produce steady-state distributions with exponents around the observed values.
Furthermore we may conjecture that the common structure of observed distributions results from a broad sense in which human conflict, especially at the small scales below those of nation-states, is self-organizing. 
There is a wide literature exploring this using deterministic mean-field (DMF) methods, summarized in \cite{Ruszczycki2009_CF} which presented analysis of various generalizations of (\ref{eq:GenCoFr}) below, and to which we refer the reader for additional background.

From the perspective of applications to complex systems, the crucial requirement is to understand what features are necessary for genuinely emergent phenomena. 
Beyond the steady state, the most salient observed features are {\em gelation}, the absorption of most of the population into a single large group, and the {\em shattering} of large groups into individuals. 
Again this can be studied mathematically in the DMF approach \cite{Banasiak19_GelShatter}. 
However, we have already reported that in stochastic simulations of a fixed population a new phenomenon occurs, that of {\em stochastic gel-shatter cycles} \cite{Fagan21_gel-shatter}, in which steady gelation is followed by stochastic shattering, in certain suitable parameter regimes of models with multiplicative size-biased kernels \cite{Fagan21_gel-shatter}. 

The primary purpose of the present paper is to investigate the robustness of an approximately power-law steady state and the occurrence and extent of stochastic cyclicity through stochastic simulation studies of a variety of CF processes. 
We begin with size-biased coalescence accompanied by shattering fragmentation, and then extend to include: different types of kernel; accretion and erosion (`Becker-D\"oring dynamics'); and a range of non-shattering fragmentation processes.
Whilst these processes are motivated by microscopic rules with a clear justification in our primary application area of operations research, our work is much more generally applicable either in principle or via direct analogy to alternate microscopic behaviour.

Alongside reporting our own simulations we draw together a range of contributory results from the literature. 
CF models have been used in a wide variety of systems, ranging from the physical interactions between asteroids and dust \cite{Tanaka96_SelfSimilar,Birnstiel11_Astrophysics} to probabilistic \cite{Aldous99_CoalProbabilists}, economic \cite{Pushkin04_BankMergers}, biological \cite{Datta11_MarineCoFr} and social structures such as the insurgent warfare with which we began \cite{Bohorquez09_Common,DHulst00_EconoCoFr}. 
Previous expectations have typically been for fragmentation to balance coalescence in such models, yielding a steady state that follows a truncated power-law distribution (\ie a power-law distribution with an exponential cutoff).
Recent literature has questioned this assumption. 
With coalescence, resupply, and sedimentation without fragmentation, oscillations in the DMF kinetics have been observed \cite{Ball12_CoalOscillations}. 
Hopf bifurcations have been observed both in \cite{Ball12_CoalOscillations} and when employing Becker-D\"{o}ring mechanics \cite{Pego20_BeckDoerCycles}. 
Deterministic oscillations also occur in systems with fragmentation and when using kernels of the form $K(i,j)=(i/j)^{a} + (j/i)^{a}$ \cite{Matveev17_OscillationsCoFr}.

However, and crucially for the present paper, a DMF treatment is not enough.
In fully stochastic simulations we see steady states becoming fragile, with some time-asymmetric stochasticity (and thus moving away from detailed balance), and in some regimes the new phenomenon of gel-shatter cyclicity, which must now be included alongside deterministic cycles and power-law steady states as a known stable outcome of a CF process. 
Thus it is only with simulations of the full stochastic process, with statistical identification both of the steady-state distribution and of stochastic cyclicity, that we can describe the full phenomenology of these models.

The classical approach begins with a DMF treatment of coalescence-fragmentation systems of the form
\begin{align}\label{eq:GenCoFr}
    \frac{dn_k}{dt} = & \frac{1}{2} \sum_{i = 1} ^ {k-1} K(i, (k - i)) n_i n_{k-i} 
        - n_k\sum_{i = 1} ^ \infty K(i, k) n_i + \nonumber \\
        & + \frac{1}{k} \sum_{i = k}^{\infty} F(i, k) i n_i 
        - n_k \sum_{i = 1}^{k} F(k, i)\mbox{ },
\end{align}
where $n_k$ is the density of clusters of size $k \geq 1$, $K(i, j)$ is the (symmetric) coalescence rate kernel for clusters of sizes $i$ and $j$ to form clusters of size $i + j$, and $F(i, k)$ is the fragmentation rate kernel from clusters of size $i$ to clusters of size $k$.
These systems are known to yield (truncated) power-law distributions when in steady state,  $ \frac{dn_k}{dt} = 0 \;\forall k$ \cite{DHulst00_EconoCoFr,Clauset10_Wiegel,Kyprianou18_Universality}.
We adopt as our basic model the multiplicative coalescence and shattering fragmentation kernels
\begin{equation}\label{eq:kern_Kij}
    K_{\mathrm{mult}}(i, j) = \hat{K} \frac{ij}{M^2},
\end{equation}
and
\begin{equation}\label{eq:kern_Fij}
    F_{\mathrm{mult}}(i, k) = \hat{F} \frac{i}{M} \delta_{k, 1},
\end{equation}
where $\hat{K}$ and $\hat{F}$ are the constant reaction rates, normalized for system size $M$,  the maximum possible number of monomers in the closed system.
We will also use $N = \sum_i n_i$ to measure the number of groups (aggregated clusters) in the system.
When this system has a steady state, its power law distribution (probability density function) can be found analytically, with an exponent of $\alpha = 2.5$.

Central to our approach, in the light of the results on oscillatory behaviour noted above, is empirically to identify cyclic departures from steady state, so we introduce a cyclicity order parameter $\calK$, defined by
\begin{equation}
    \calK = \frac{\sum_{t} \text{sgn}(\kmax(t) - \kmax(t-1))}{t},
\end{equation}
where $\text{sgn}$ is the sign function and $\kmax$ is the size of the largest cluster in the system.
This lies between $-1$ and $+1$, and is the proportion of computational time steps for which the largest cluster becomes larger, minus that for which it becomes smaller. 
In the steady state, or in regimes of symmetric stochastic variation, $\calK$ is approximately zero. 
In regimes of slow build-up followed by sudden fragmentation, by contrast, $\calK$ becomes significantly different from zero, and in extreme cases approaches $+1$. 
Negative $\calK$ are seldom observed and never persistent. 
Thus $\calK$ measures the amount of asymmetric stochastic cyclicity in the system, and thereby diagnoses slow-fast cycles, analogous to some relaxation oscillations such as the `Oregonator' \cite{field1974}. 
The empirical $\calK$ is naturally accompanied by the theoretical dimensionless number $r$, the ratio of the characteristic time-to-gelation to the characteristic time-to-fragmentation \cite{Fagan21_gel-shatter}.

This paper is structured as follows. 
We begin with a discussion of our summary statistics in Section \ref{sec:SumStat}, needed in order to frame and analyse the problem. 
With these well understood, in Section \ref{sec:Base} we study variations in the construction of the model including different kernel types and a wide range of parameters, enabling us to establish our base model.
In Section \ref{sec:StSt} we study what happens when we add various secondary processes to a particular case of the multiplicative-kernel base model that has a clear steady state with low cyclicity, testing the robustness of the system to changes in the nature of the underlying dynamics. 
In Section \ref{sec:GlSh} we add the same secondary processes to a particular case of the multiplicative-kernel base model that has distinctive gel-shatter cyclicity, again in order to explore its robustness. 
We conclude with a discussion of the implications in Section \ref{sec:Conc}, with a particular emphasis on problems in operations research.

\section{Statistical Methodology} \label{sec:SumStat}
In order to study the behaviour of the system as it evolves over time, we need to first establish a standard method of how we will statistically analyse data emerging from our microscopic simulations. 
The method we adopt is to fit a power-law distribution at each time step in order to not be averaging raw data.
We then need to understand how our power-law distribution summary statistics behave.

To fit the power-law distribution we use maximum likelihood estimation for the estimation of the exponent and the maximum. For the minimum, we either use the smallest possible value $\xmin=1$, the monomer of the system, (method `MLE') or Kolmogorov-Smirnov maximum likelihood estimation (`KS-MLE') \cite{Clauset09_PowerLaw}, which selects the minimum that corresponds to the power-law distribution with the least Kolmogorov-Smirnov distance (the maximum difference between the cumulative distribution functions) to the empirical data.
In effect, KS-MLE fits only the power-law tail of the data, while MLE fits a power-law to the entirety of the data.
(We also examined the performance of power-law distributions with exponential cutoffs and observed very little difference in fitted summary statistics.) 

To characterise the behaviour of the estimators, we examine how they work in a model system with $M = 10^4$ and $\hat F = 0.2 = 1 - \hat K$. (Note that these give $r = 2.5 \times 10^3$; steady states are typically observed if $r > 10^3$ \cite{Fagan21_gel-shatter}.)
Four simulations of this system can be seen in Figure \ref{fig:examplesteadystate}, whose four panels show: a heat map of the locations the system visits (after burn-in) in $(\kmax/{M}, {N}/{M})$ space; $\kmax$ over time; the {KS-MLE} $\alpha$ values over time; and the {MLE} $\alpha$ values over time.
In practice, this system has a small amount of cyclicity $\calK = 0.123$ on average across the simulations, which can be seen upon close examination of the time series.
The region explored is quite narrow, with $\kmax/{M}$ within $[0.0023, 0.0554]$ and ${N}/{M}$ within $[0.6318, 0.7261]$.
Of these simulations, $99\%$ of time steps sampled remained within $[0.0035, 0.0276]$ and $[0.6427, 0.7106]$ respectively, consistent with steady-state and noise behaviour (in comparison to cyclic behaviour). 
The average cyclicity and $95\%$ intervals for {KS-MLE} and {MLE} $\alpha$ are collected in the first row of Table \ref{tab:cofr:robustnessSS}, which collects analogous values for each model we later consider in Section \ref{sec:StSt}. (We take large ($>5$) and small ($<1$) estimates of the {KS-MLE} $\alpha$ to indicate failures to fit the data.)

\begin{figure} 
    \begin{minipage}{\linewidth}
        \begin{minipage}{0.49\linewidth}
            \includegraphics[width=\linewidth]{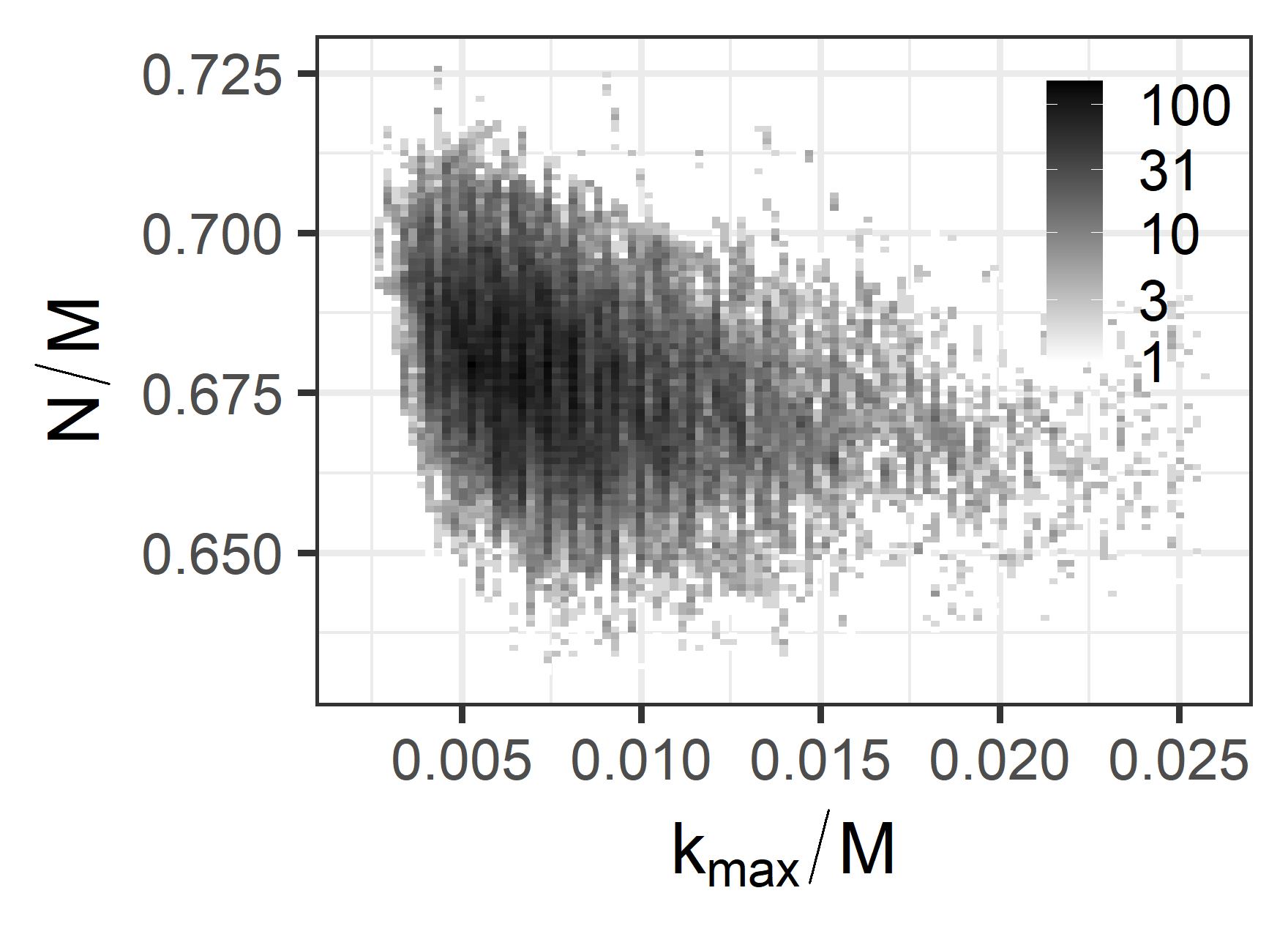}
        \end{minipage}
        \hspace{0.01\linewidth}
        \begin{minipage}{0.49\linewidth}
            \includegraphics[width=\linewidth]{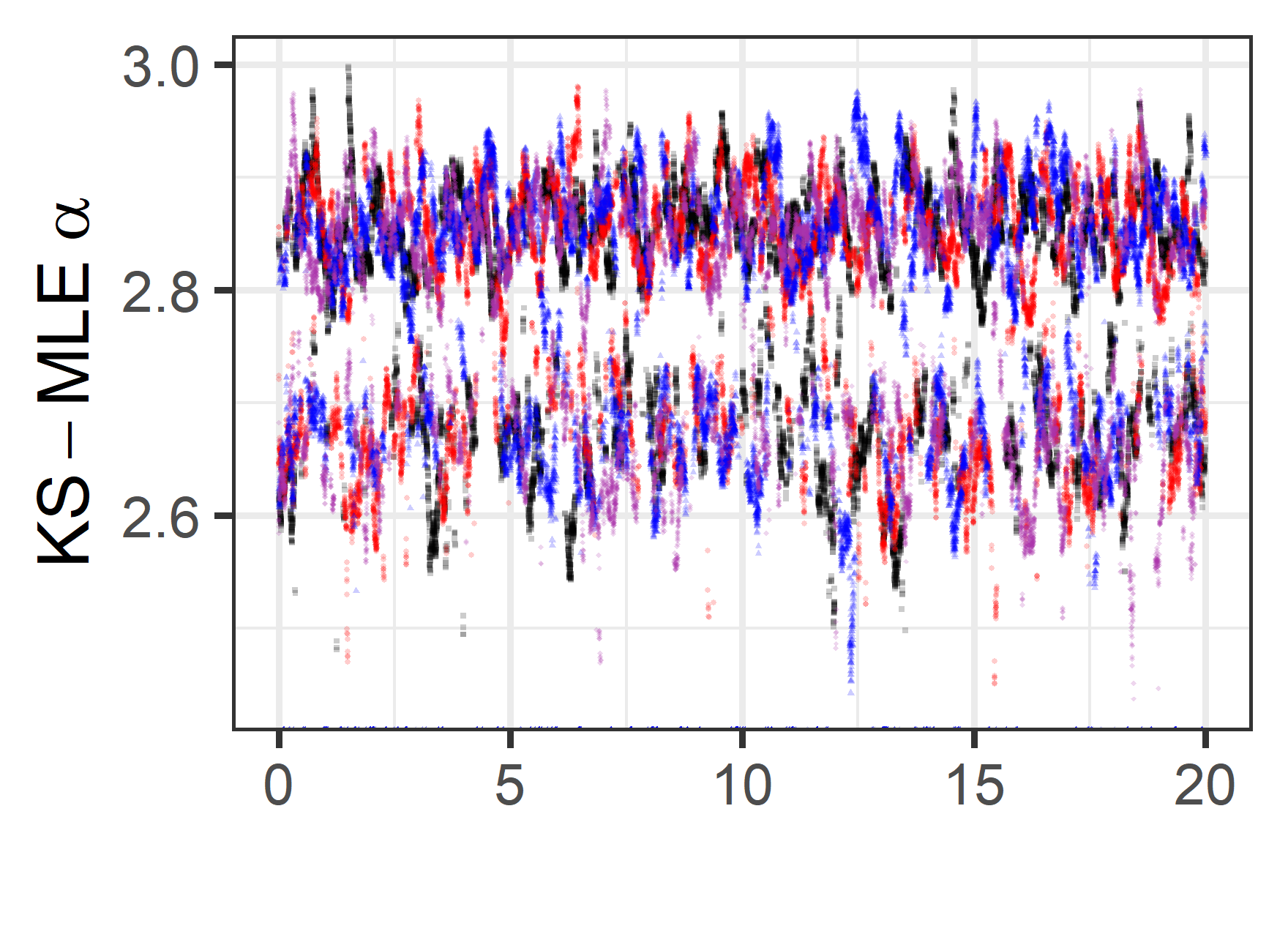}
        \end{minipage}
        \hspace{0.01\linewidth}
        \begin{minipage}{0.49\linewidth}
            \includegraphics[width=\linewidth]{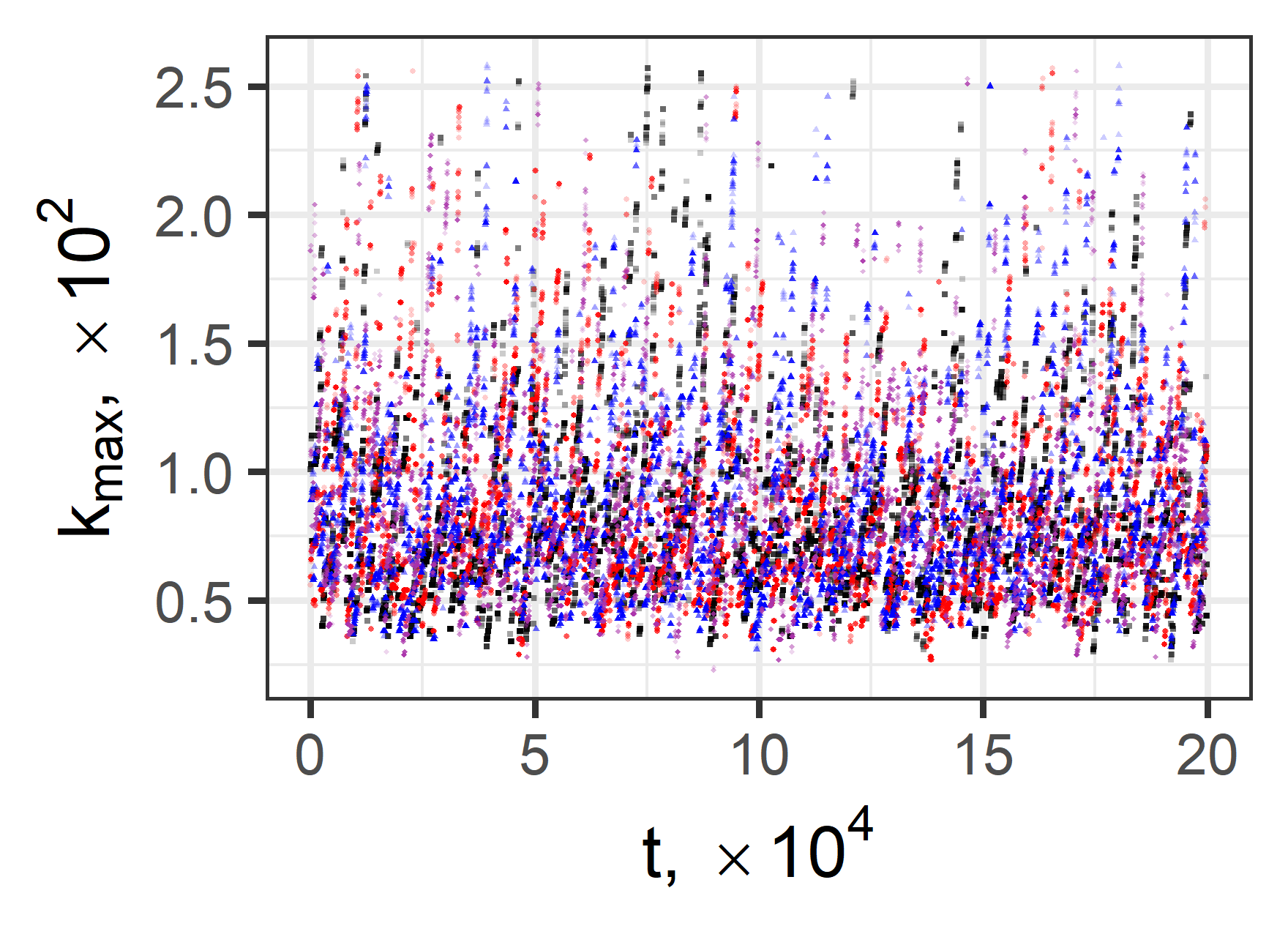}
        \end{minipage}
        \hspace{0.01\linewidth}
        \begin{minipage}{0.49\linewidth}
            \includegraphics[width=\linewidth]{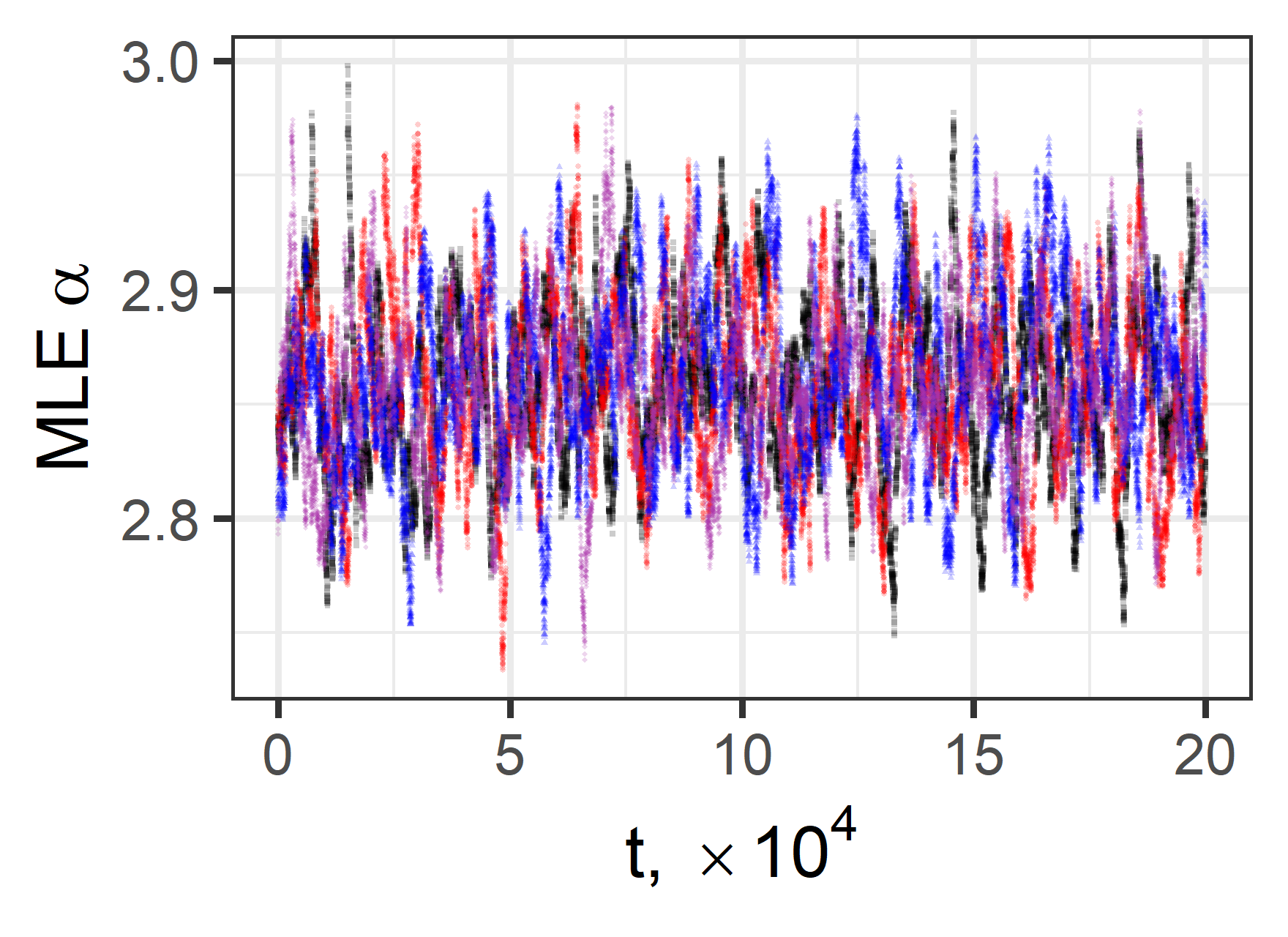}
        \end{minipage}
        \hspace{0.01\linewidth}
    \end{minipage}
    \caption[Coalescence and fragmentation, summary statistics, ${M} = 10^4$, $\Prob{\Frag} = 0.20$.
    ]{
    Coalescence and fragmentation summary statistics, ${M} = 10^4$, $\hat{F} = 0.20$. 
    Four simulations were conducted (black, blue, red, and purple).
    From top-left and proceeding counter-clockwise, we present a heatmap of the locations of the simulations in the $(\frac{\kmax}{M}, \frac{N}{M})$ plane, the time series of $\kmax$, the time series of the {MLE} $\alpha$ estimate, and the time series of the {KS-MLE} $\alpha$ estimate.
    We have trimmed large ($>5$) and small ($<1$) estimates of the {KS-MLE} $\alpha$, corresponding to failures to fit the data.
    }
    \label{fig:examplesteadystate}
\end{figure}

\begin{figure} 
    \begin{minipage}{\linewidth}
        \centering
        \begin{minipage}{0.55\linewidth}
            \includegraphics[width = \textwidth]{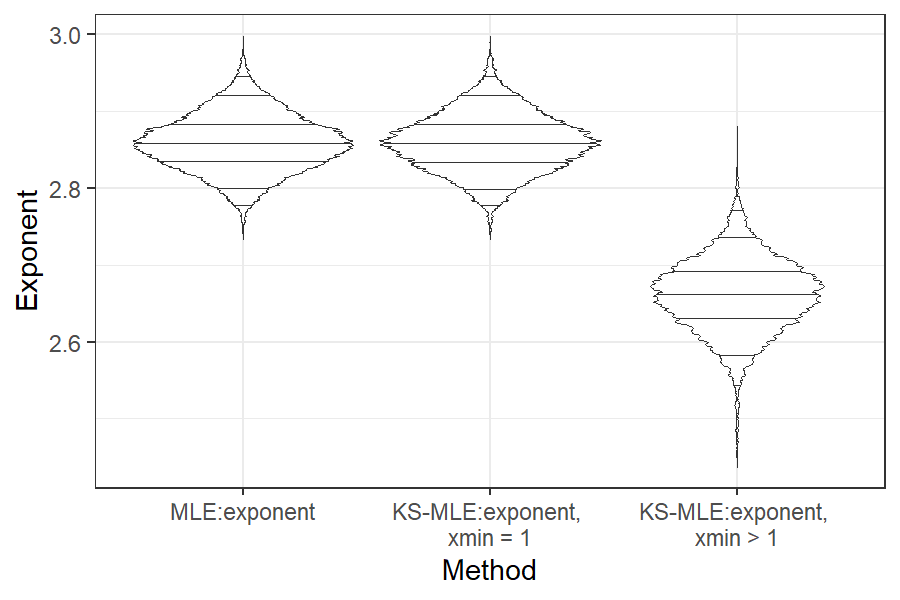}
        \end{minipage}      
        \hspace{0.01\linewidth}
        \begin{minipage}{0.4\linewidth}
            \begin{tabular}{|l|ccc|}\hline
            Stat. & {MLE} & {KS-MLE} & {KS-MLE} \\
                  &     & $\xmin  = 1$ & $\xmin  > 1$ \\\hline
            0.99  & 2.95 & 2.95 & 2.77 \\
            0.95  & 2.92 & 2.92 & 2.73 \\
            0.75  & 2.88 & 2.88 & 2.69 \\
            Mean  & 2.86 & 2.86 & 2.66 \\
            0.50  & 2.86 & 2.86 & 2.66 \\
            0.25  & 2.83 & 2.83 & 2.63 \\
            0.05  & 2.80 & 2.80 & 2.58 \\
            0.01  & 2.78 & 2.78 & 2.54 \\\hline
            \end{tabular}
        \end{minipage}
    \end{minipage}
    \caption[Coalescence and fragmentation, exponent summary statistics.]{Coalescence and fragmentation exponent summary statistics.
    We present the distribution of estimated exponents for Figure \ref{fig:examplesteadystate}.
    Together, these simulations constitute $80,000$ samples.
    (Left) Violin plots compare how the fitted exponents are distributed.
    (Right) Quantiles and the means for the fitted exponents.
    The quantiles correspond to the horizontal lines within the violin plots.
    {MLE} $\alpha$ and {KS-MLE} $\alpha$ agree when the {KS-MLE} $\xmin  = 1$.
    The theoretical result indicates $\alpha = 2.5$. 
    }
    \label{fig:steadystate:methodviolins}
\end{figure}

Despite the proximity to the steady state, the system has a distinctive two-layer pattern in the fitted {KS-MLE} $\alpha$ which is simply explained.
Recall that the {MLE} $\xmin$ is always set to $1$, but the {KS-MLE} $\xmin$ is allowed to vary to optimise the fit, and
the {KS-MLE} $\alpha$ is highly sensitive to it.
The higher values arise when $\xmin  = 1$ is preferred by the {KS-MLE} algorithm, while lower values arise when $\xmin  \geq 2$. Looking at the `violin' distribution plots in Figure \ref{fig:steadystate:methodviolins}, we first note that where {MLE} and {KS-MLE} both use $\xmin  = 1$, the {MLE} $\alpha$ is equal to the {KS-MLE} $\alpha$; this occurs in around $62\%$ of samples. 
Where the {MLE} and {KS-MLE} $\xmin$ differ, there is no obvious relationship between the {MLE} $\alpha$ and {KS-MLE} $\alpha$, although there is a correlation (of $0.454$). 
Henceforth, when we refer to {KS-MLE} $\alpha$, we mean cases with $\xmin > 1$.

The bias of the fitted exponents away from the expected $2.5$ level is systematic; varying $M$ shows that it persists for large values of $M$ and is not due to finite size.
To check that it is not an artefact we performed a simple experiment to compare our simulation with the theoretical steady state: we used the (truncated, normalised) theoretical steady state of the corresponding CF system and randomly created partitions of ${M}$.
To create each partition, we drew cluster sizes repeatedly from the normalised solution, treating it as the probability distribution, until the sum of the cluster sizes drawn exceeds ${M}$.
We then discarded the last cluster and instead added the missing mass as an additional cluster.
We performed this procedure $1,000$ times in Figure \ref{fig:steadystate:methodviolinsSimulated} with the same values of the parameters as before.
Agreement with Figure \ref{fig:steadystate:methodviolins} is good, and
we conclude that the bias is inherent to the fitting procedures used.

This point is of real importance: there is a substantial difference between the power-law exponent obtained in the calculation of the theoretical solution and the power-law exponent obtained by the simulation, due to the finite nature of the system that the solution is expected to describe.
The exponents obtained through the MLE and KS-MLE power-law fits indicate that the latter has a positive bias away from the theoretical solution's exponent, and studies that rely on these methods of fitting should expect this positive bias to be present.

\begin{figure}[t] 
    \begin{minipage}{\linewidth}
        \centering
        \begin{minipage}{0.55\linewidth}
            \includegraphics[width = \textwidth]{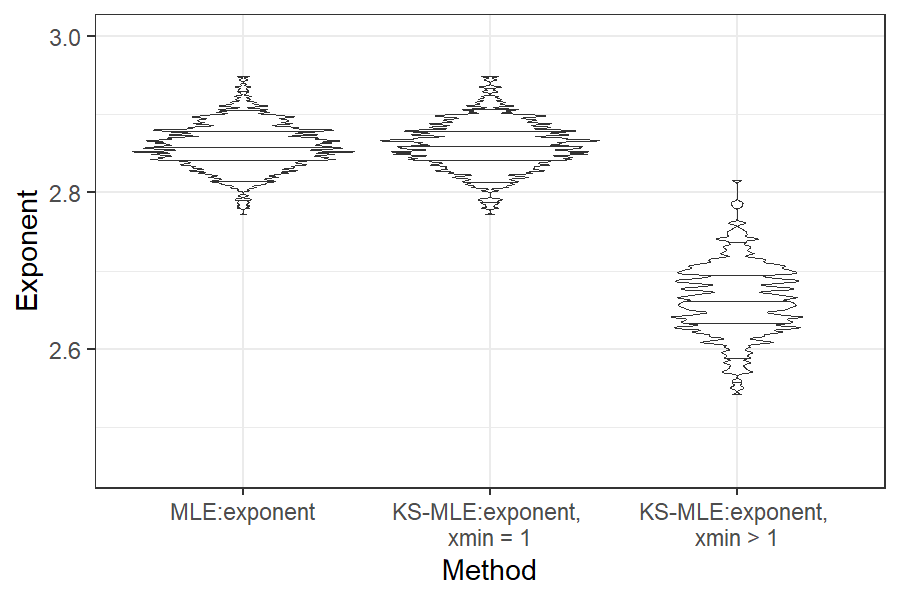}
        \end{minipage}      
        \hspace{0.01\linewidth}
        \begin{minipage}{0.4\linewidth}
            \begin{tabular}{|l|ccc|}\hline
            Stat. & {MLE} & {KS-MLE} & {KS-MLE} \\
                  &     & $\xmin  = 1$ & $\xmin  > 1$ \\\hline
            0.99  & 2.93 & 2.93 & 2.77 \\
            0.95  & 2.91 & 2.91 & 2.74 \\
            0.75  & 2.88 & 2.88 & 2.70 \\
            Mean  & 2.86 & 2.86 & 2.66 \\
            0.50  & 2.86 & 2.86 & 2.66 \\
            0.25  & 2.84 & 2.84 & 2.63 \\
            0.05  & 2.81 & 2.81 & 2.59 \\
            0.01  & 2.79 & 2.79 & 2.55 \\\hline
            \end{tabular}
        \end{minipage}
    \end{minipage}
    \caption[Coalescence and fragmentation, exponent simulated summary statistics.]{Coalescence and fragmentation exponent simulated summary statistics.
    We present the distribution of summary statistics of $1,000$ samples from the theoretical coalescence and fragmentation solution with ${M} = 10^4$ and $\hat{F} = 0.20$. 
    Compare with Figure \ref{fig:steadystate:methodviolins} for which we observe good agreement, indicating deviations from the expected exponent $\alpha = 2.5$ are systematic.
    }
    \label{fig:steadystate:methodviolinsSimulated}
\end{figure}

\section{Analysis of the base model} \label{sec:Base}

Now that we have described our summary statistics and their expected behaviours, we turn our attention to how varying the base model's underlying rules affects the overall behaviour of the system.
We begin by discussing aspects of the basic construction of the model, including different types of kernel, and for the full range of the controlling parameters.

We then discuss in Section \ref{sec:StSt:BD} what happens to a particular case of the base model with multiplicative kernel, which has a clear steady state with no gel-shatter cyclicity, when we include various secondary processes. 
These processes act so as to redistribute small amounts of mass through the system; the effects we observe establish that the system is sensitive to behaviour in the bulk.
In Sections \ref{sec:StSt:Frag} and \ref{sec:StSt:PL} we consider the behaviour of the fragmentation function.
With these two cases, we describe how the movement of mass from the (right) tail into the bulk influences the behaviour of the system.
We summarise these findings in Section \ref{sec:StSt:Summary} in anticipation of analogous analysis for the cyclic regime.

Size-biased coalescence and shattering fragmentation are clearly not the only possible kernels for the base model, but they are the most intuitive to motivate for most applications.
For example, Smoluchowski's original work used a constant coalescence kernel $K(i,j) = 1$ (without fragmentation) \cite{Smoluchowski16_Coagulation}, and results for constant, additive, and multiplicative (size-biased) kernels have been known since at least the 1960s for both discrete and continuous mass-distributions. 
An accessible introduction is \cite{Aldous99_CoalProbabilists}.
Given the variety of kernels and underlying rules they represent, we first establish some of the sensitivity of the model to the form of the rules, and follow with a brief review of the sensitivity of the model to variation of the parameters, following our previous work \cite{Fagan21_gel-shatter}.
Together, these set the stage for us to robustly interpret further modifications to the system.

First, we consider the significance of the kernels in the system.
It is well-known that systems with multiplicative coalescence kernels and, more generally, homogeneous kernels exhibit gelation (defined in infinite systems as the divergence of the second moment, \cite{Aldous99_CoalProbabilists}, p10).
Recall that equations (\ref{eq:GenCoFr}) track the density $n_k$ and size $k$ of clusters within the system. 
One could then calculate the mass, $M = \sum_k k n_k$, the total number of units, among which $kn_k$ units have coalesced into the $n_k$ units of size $k$. 
For purely coalescent systems (\ie $F(i, k) = 0$) with kernels that exhibit gelation, it is possible for the system to coalesce so rapidly that mass escapes to a cluster of infinite size in finite time, in which case the infinite-sized cluster is referred to as a gel.
Typically the remainder of the system is referred to as the sol.
In this case, analytically the gel is not tracked by the size-biased sum of the system cluster sizes, resulting in the mass $M$ appearing to decrease in the system even if the equations otherwise appear to conserve mass \cite{Aldous99_CoalProbabilists}.
In coalescence of random graphs, the gel appears spontaneously as a giant connected component of the system, prior to which the system is made up of isolated graphs. Its formation marks a shift in the size of the largest component; for multiplicative coalescence kernels $K(i, j) \propto ij$ this shift is from order $\log M$ to order $M$ \cite{Erdos1960-RandGraphs}. 
In our systems (where time is measured by the number of events that have occurred thus far, similar to the number of edges that have appeared thus far in a random graph \cite{Erdos1960-RandGraphs}), `gelation' appears as creation of a large cluster that rapidly increases in size. 
This simulated gel (of size $k_{\max}$) cannot become truly infinite-sized in our simulation, but still separates from the sol, in the sense that the gap between the size of the gel and the next largest cluster is much larger than the next such gap.

It is possible for fragmentation to prevent gelation, but the effectiveness must depend on the relative strength of the fragmentation and coalescence kernels.
For example, employing a fragmentation barrier, above which clusters must fragment and restock the smaller cluster sizes, leads to varying steady-state power-law distributions for the sizes of clusters dependent on how fragments and coalescence occur \cite{Birnstiel11_Astrophysics}. 
For certain choices of kernel functions, exact steady-state solutions are possible and well-known, including when the reactions are reversible and satisfy detailed balance \cite{Wattis06}, such as growth and decay by monomers only (in the presence of sub-linear rate kernels and finite system size) \cite{Ball86_BeckerDoering}. 
Alternatively, the steady state can also be reached if a constant source of small clusters is provided to the system to counter-balance the removal by the gel \cite{Pushkin02_Self}.
For coalescence kernels of the form $K(i, j) = \hat{K} (i j) ^{\alpha}$ for $\alpha \geq 0$ and fragmentation of clusters to monomers, a steady-state solution is known to be approximately $k^\alpha n_k \approx \frac{1}{\sqrt{\pi}} n_1  (k + 1) ^ {-3/2}$ and is known exactly for $F(i, k) \propto i$ with $\alpha = 1$ \cite{DHulst00_EconoCoFr,Clauset10_Wiegel,Brilliantov15_SizeDistributionAndPL,Johnson16_NewOnline}. Many extensions and variants exist, such as non-binary coalescence \cite{Kyprianou18_Universality}.
Despite the existence of steady-state solutions, however, 
it is not given that fragmentation actually prevents gelation, although conditions do exist for some forms of fragmentation, such as binary fragmentation \cite{Costa95_StrongFragmentation}.

So what happens when systems with shattering fragmentation evolve through time? In Figure \ref{fig:varyingkernels}, we plot (normalized by the total mass $M$) the number of clusters against the size of the largest cluster, explored over 20,000 steps in four different simulations, each choosing coalescence or fragmentation to be either multiplicative, as above in equations (\ref{eq:kern_Kij}, \ref{eq:kern_Fij}), or constant,  
\begin{equation}\label{eq:Kkern_const}
    K_{\mathrm{const}}(i, j) =\frac{\hat{K}}{N^2},
\end{equation}
and
\begin{equation}\label{eq:Fkern_const}
    F_{\mathrm{const}}(i, k) = \frac{\hat{F}}{N} \delta_{k, 1}.
\end{equation}

\begin{figure} 
    \begin{minipage}{\linewidth}
        \begin{minipage}{0.49\linewidth}
            \includegraphics[width=\linewidth]{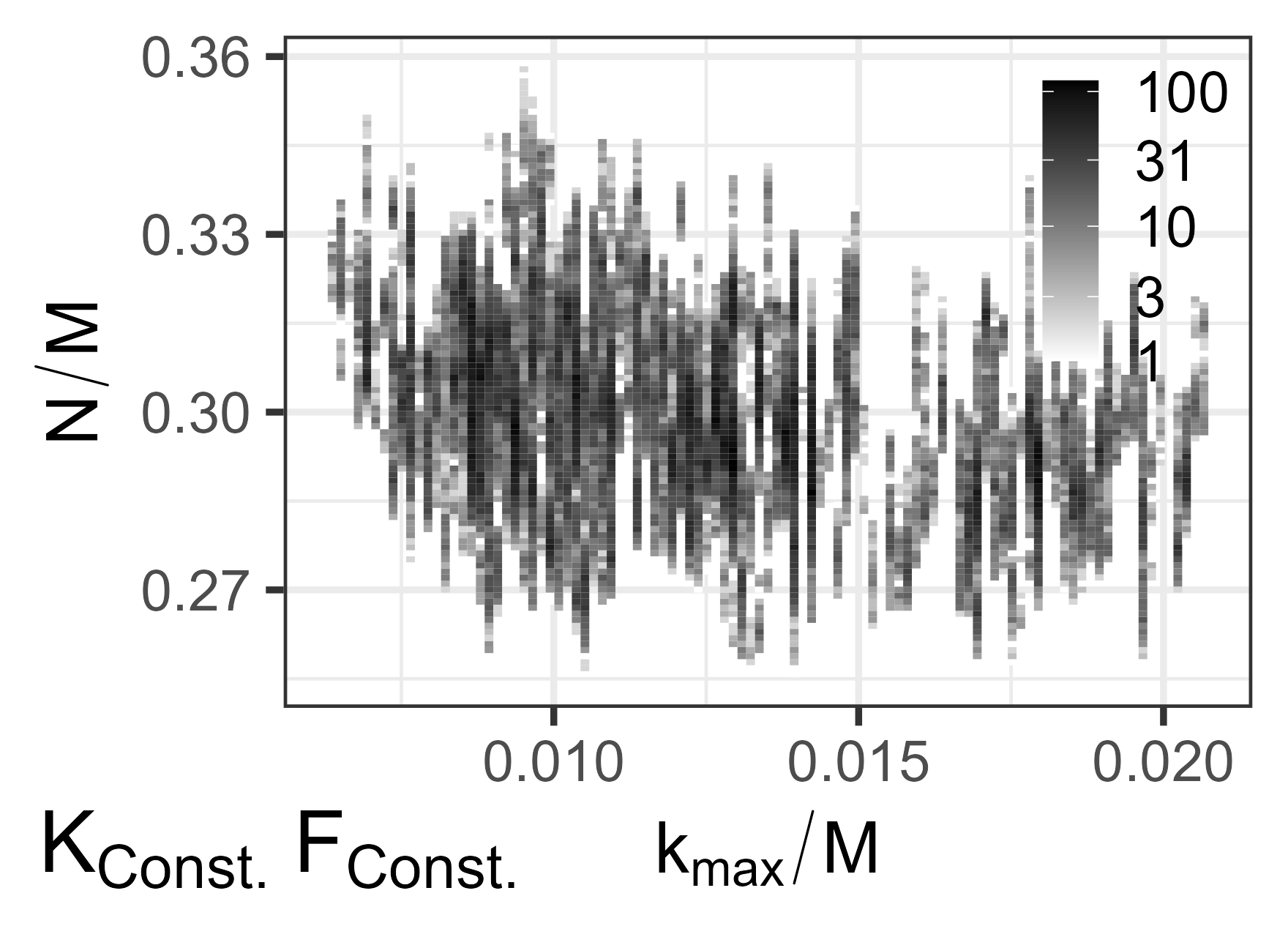}
        \end{minipage}
        \hspace{0.01\linewidth}
        \begin{minipage}{0.49\linewidth}
            \includegraphics[width=\linewidth]{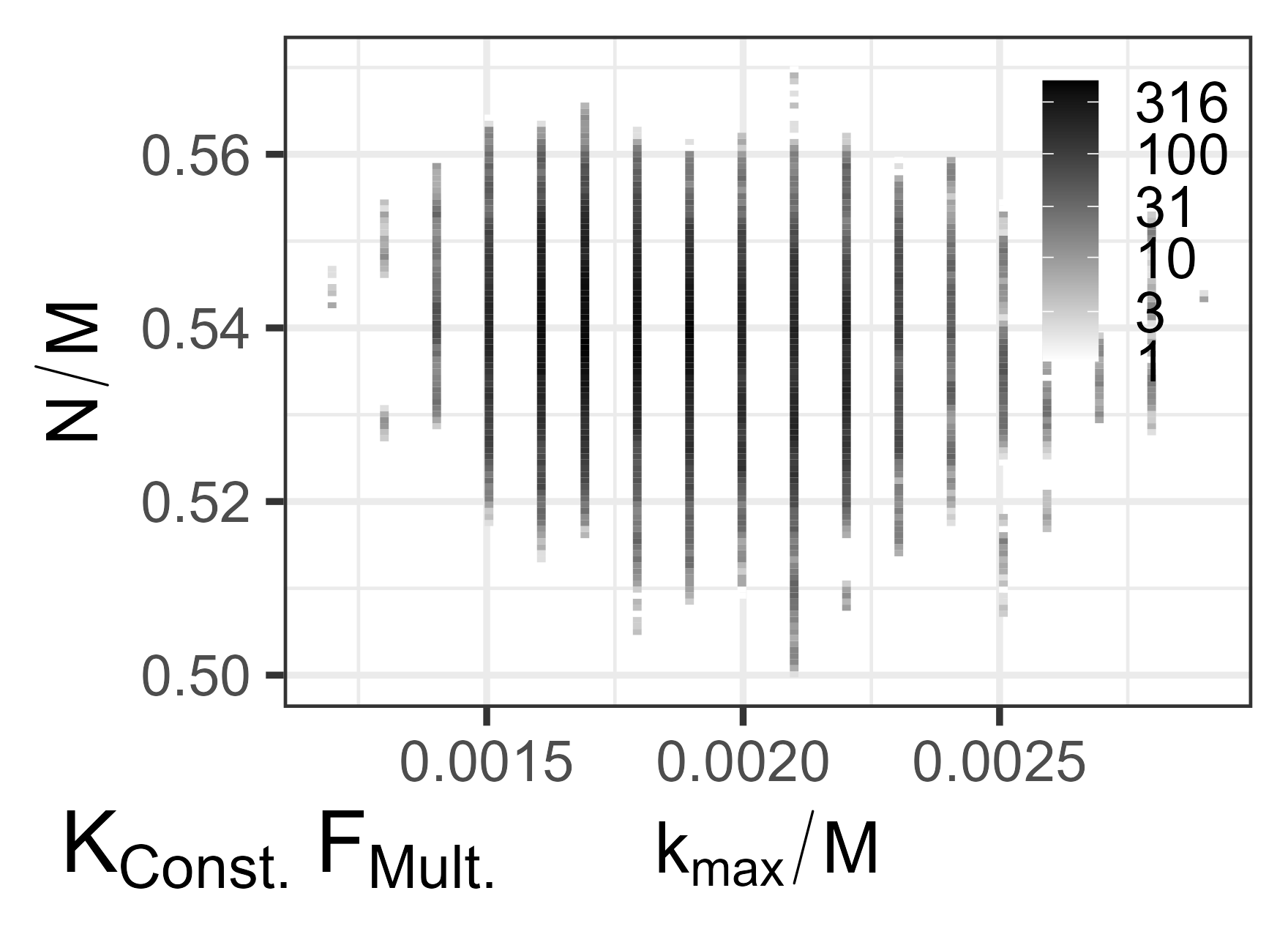}
        \end{minipage}
        \hspace{0.01\linewidth}
        \begin{minipage}{0.49\linewidth}
            \includegraphics[width=\linewidth]{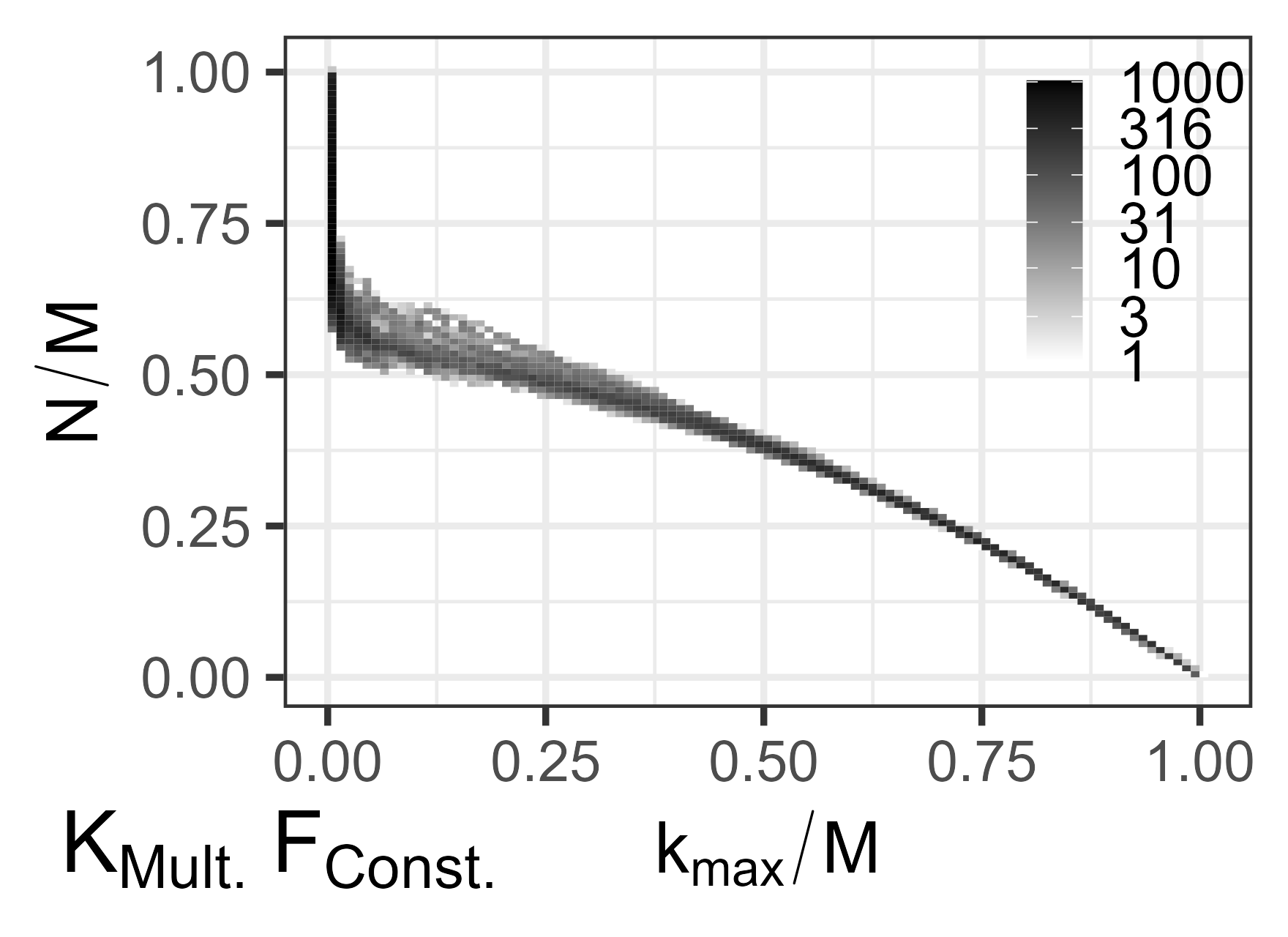}
        \end{minipage}
        \hspace{0.01\linewidth}
        \begin{minipage}{0.49\linewidth}
            \includegraphics[width=\linewidth]{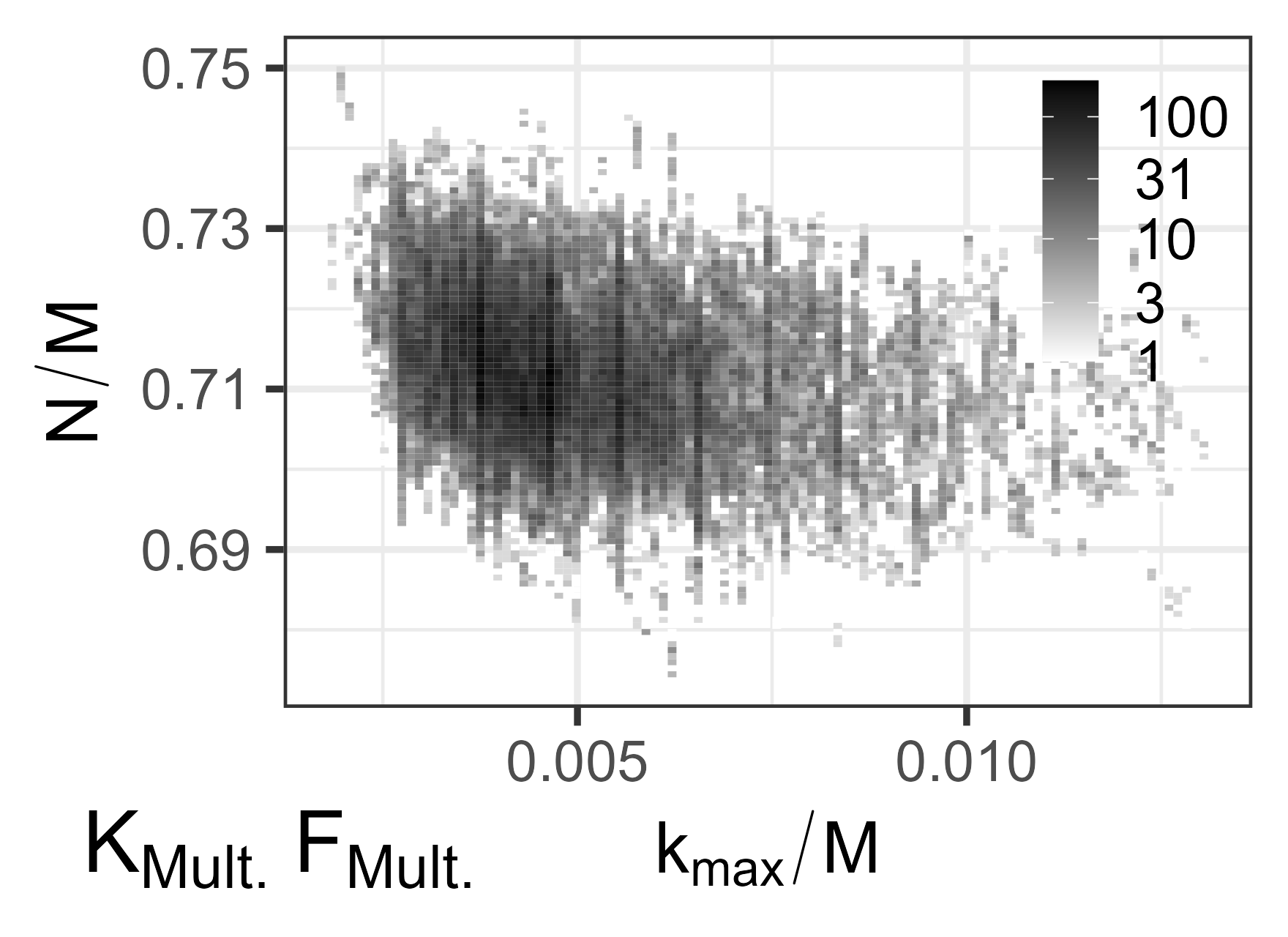}
        \end{minipage}
        \hspace{0.01\linewidth}
    \end{minipage}
    \caption[Coalescence and fragmentation, varying kernels.
    ]{
    Space explored by coalescence and fragmentation simulations with varying kernels, ${M} = 10^4$, $\hat{F} = 0.30$.
    Top-left: constant coalescence and fragmentation kernels, Eq. \ref{eq:Kkern_const} and \ref{eq:Fkern_const}. Top-right: we increase the strength of fragmentation by using a multiplicative fragmentation kernel, Eq. \ref{eq:kern_Fij}. Bottom-left: multiplicative coalescence, Eq. \ref{eq:kern_Kij}, constant fragmentation. Bottom-right: multiplicative kernels for both coalescence and fragmentation.
    When the type of kernel is the same, the system approaches a non-trivial steady state. When it is instead unbalanced the system either remains mostly disaggregated (top-right) or is forced into cyclicity by rapid coalescence followed by fragmentation (bottom-left).
    }
    \label{fig:varyingkernels}
\end{figure}

The top two plots use constant coalescence and the bottom two plots multiplicative coalescence, while the left-hand plots use constant fragmentation and the right-hand plots multiplicative fragmentation. 
We begin with the top-left panel, in which both coalescence and fragmentation are constant, $K_{\mathrm{const}}$ and $F_{\mathrm{const}}$. 
The system aggregates around $2\%$ of its total mass into a single cluster, but the variation is large. Constant kernels act more slowly than, but similarly to, to their multiplicative equivalents. The bottom-right panel, in which both coalescence and fragmentation are multiplicative, yields a mostly disaggregated system with some occasional large clusters. The bottom-left panel has multiplicative coalescence $K_{\mathrm{mult}}$ but constant fragmentation $F_{\mathrm{const}}$. The largest cluster grows rapidly until the only possible event is fragmentation, at which point the entire system is shattered to clusters of unit size.
This behaviour is similar to, but more extreme than, the behaviour we will explore in Section \ref{sec:GlSh}. On the other hand, when the fragmentation kernel is multiplicative but the coalescence kernel is constant, in the top-right panel, the system never begins to aggregate beyond very small values.

Figure \ref{fig:varyingkernels} establishes that the form and balance of the kernels are important determinants of the system dynamics.
We now restrict to multiplicative kernels, determining how the parameters influence the dynamics of the system, and extending the brief treatment in \cite{Fagan21_gel-shatter}.
We have already noted the existence of gel-shatter cycles as well as a steady state for coalescence and fragmentation systems.
Fixing the scaling of the coalescence and fragmentation kernels, the transition between regimes is determined by the interplay between $M$, $\hat K$, and $\hat F$.
In particular, the dimensionless parameter 
\begin{equation} r = \frac{\hat F M}{\hat K} \end{equation}
is the ratio of characteristic times of gelation to shattering, and thus captures the balance between the two processes.

When $r$ is small, $r < 0.1$, the system is dominated by rapid gelation and must wait until fragmentation resets the system, similar to when the coalescence kernel is multiplicative and the fragmentation kernel is constant.
We expect the cyclicity order parameter $\calK$ to be low, $\calK < 0.1$, in this regime, as the system rapidly gels and then stalls until a fragmentation event finally occurs.
When $r$ is large, $r > 10^3$, the system is instead dominated by fragmentation.
When a gel does emerge, its size relative to that of the other clusters and the frequency of fragmentation rapidly removes the gel from the system, forcing the system to dwell primarily in the pre-gel phase.
As $r$ increases further, the system begins to resemble the case with $K_{\mathrm{const}}$ and $F_{\mathrm{mult}}$.
Again, $\calK$ is expected to be low in this region, $\calK < 0.1$, as the coalescence routinely creates small clusters that are nearly immediately shattered and, otherwise, shattering fragmentation events do nothing to clusters of size 1.

In between the two regimes we find gel-shatter cycles, the subject of Section \ref{sec:GlSh} and previous work \cite{Fagan21_gel-shatter}.
Here, the system visits a broad region of the $(k_{\max},N)$ plot due to the presence of gelation and infrequent but not rare fragmentation.
Fragmentation occurs often enough to prevent the system from stalling with all of the mass in a single cluster, but not so often as to prevent the formation of the gel itself.
This means that there are many small step increases in the size of the largest cluster alongside a few large step decreases, resulting in a comparatively large $\calK$. These three regimes are contrasted in Figure \ref{fig:StSt_Space_3Regimes}.

\begin{figure} 
    \centering
    \includegraphics[width=0.32\textwidth]{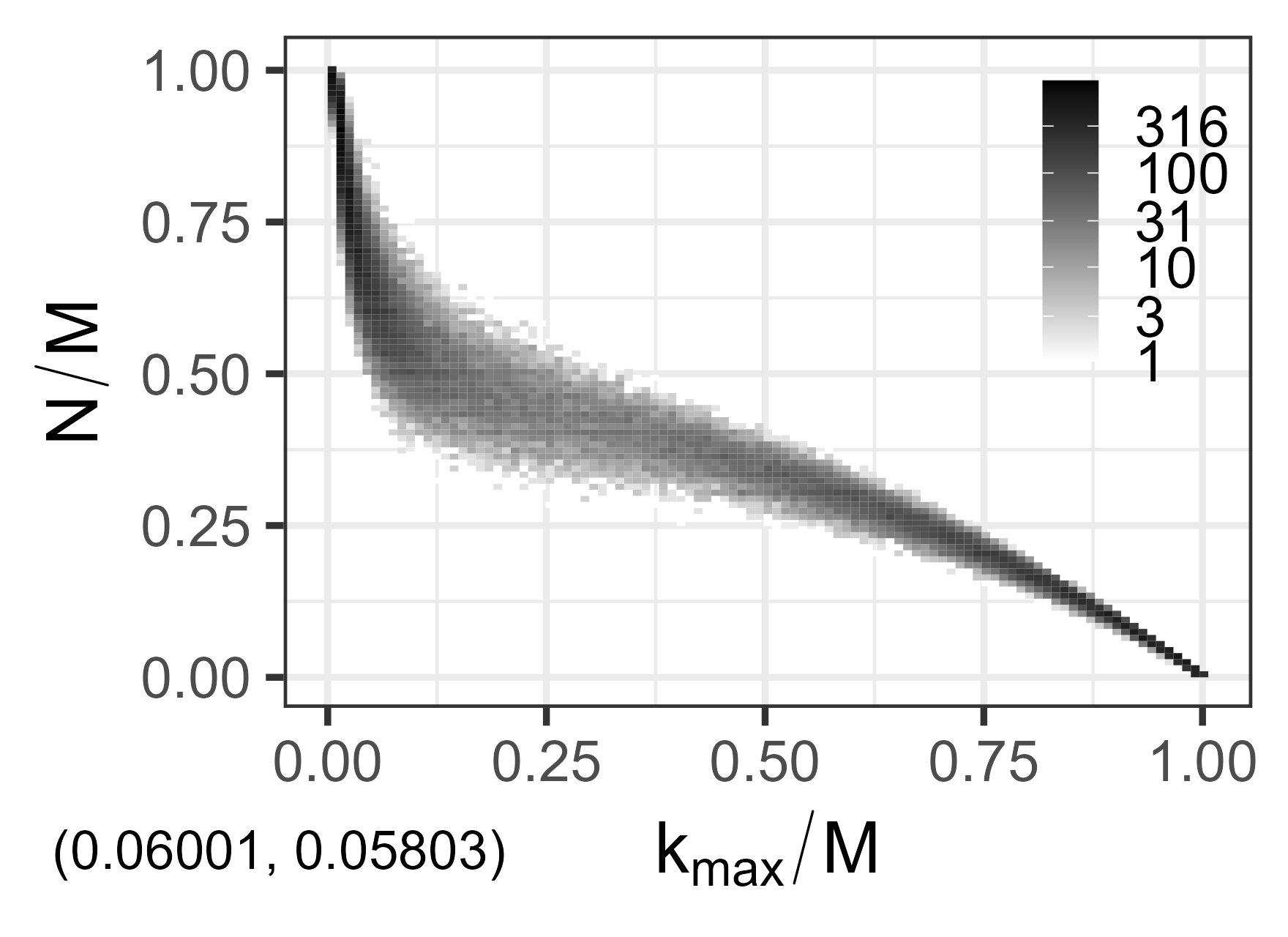}
    \includegraphics[width=0.32\textwidth]{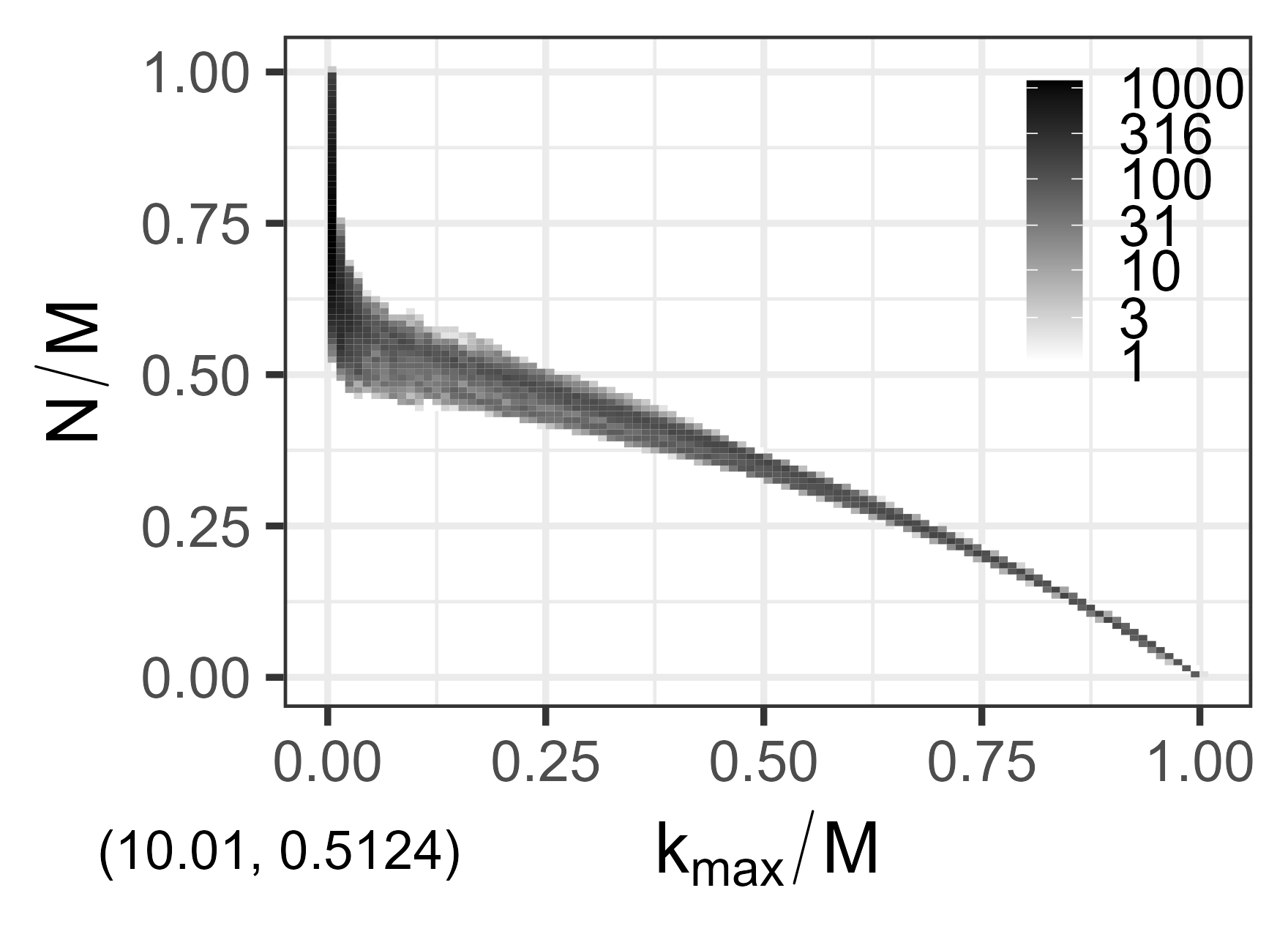}
    \includegraphics[width=0.32\textwidth]{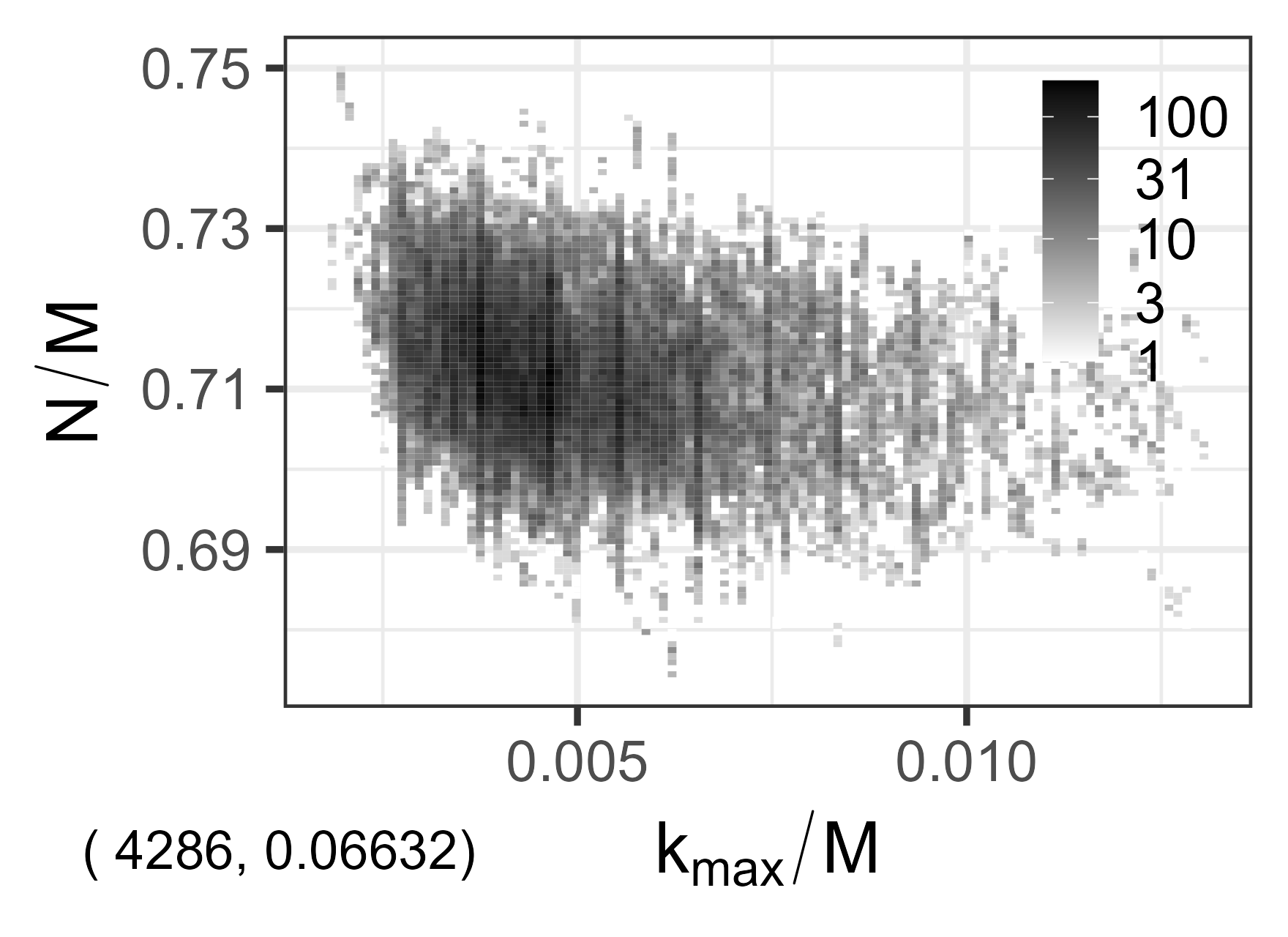}
    \caption{Shapes of spaces explored in various regimes, kernels $K_{\mathrm{mult}}$ and $F_{\mathrm{mult}}$. We vary simulation parameters $M$ and $\hat{F}$ to showcase the different regimes, obtaining different $r$ and $\calK$, shown in parentheses at bottom left of each panel. In the first plot with $M = 300$ and $\hat{F} = 0.0002$, we obtain forced fragmentation cycles, in which the system quickly coalesces into a single cluster (bottom right of plot) and then must wait for a rare fragmentation event to reset it. In the second plot with $M = 10000$ and $\hat{F} = 0.001$, the system continuously grows and fragmentation is stochastic and common, preventing stalling and creating unforced gel-shatter cycles. In the third plot, $M = 10000$ and $\hat{F} = 0.3$, fragmentation dominates, preventing the emergence of a gel. See \cite{Fagan21_gel-shatter} for further details.}
    \label{fig:StSt_Space_3Regimes}
\end{figure}

As we have shown in this section, coalescence and fragmentation systems have many built-in modelling assumptions that determine their behaviour.
The underlying rules and parameter values naturally have broad impacts, and determine whether and where the system's steady-state behaviour occurs.
Of prime importance is the type of kernels used and whether they admit the formation of large clusters or not.
If they do, then the balance of the competition between coalescence and fragmentation needs to be correct; too much of either, due to parameters or kernels, and the system stalls.

\section{Effect of rule variations on base model with steady state}
\label{sec:StSt}

Having established the influence of kernel type on the system, and as a means to gauge the importance of unmodelled small systematic perturbations in a system of interest, we next consider the effects of process variation by inclusion of a simpler set of {\em additional} rules in the particular multiplicative-kernel base model case of Section \ref{sec:SumStat}}, which has $M = 10^4$ and $\hat{F} = 0.2$. This case has an approximate power-law steady state with $\alpha$ confidence intervals KS-MLE $[2.568, 2.754]$ and MLE $[2.788, 2.932]$. The model is near the edge of cyclicity without fully engaging with gel-shatter cycles, $r = 2.5 \times 10^3$ and $\calK = 0.123$. (See Figure \ref{fig:examplesteadystate} for the behaviour of this system.)

First in Section \ref{sec:StSt:BD} we redistribute small amounts of mass through the system, establishing that it is sensitive to behaviour in the bulk.
Then in Sections \ref{sec:StSt:Frag} and \ref{sec:StSt:PL} we consider the behaviour of the fragmentation function, describing how the movement of mass from the (right) tail into the bulk influences the behaviour of the system.
We summarise in Section \ref{sec:StSt:Summary} to prepare ourselves for the cyclic regime.

\subsection{Becker-D\"{o}ring dynamics}\label{sec:StSt:BD}

The first variation we consider, which we refer to as Becker-D\"{o}ring dynamics \cite{Pego20_BeckDoerCycles,Becker35_Original,Ball86_BeckerDoering}, involves only the movement of monomers between clusters in the population.
We refer to the addition of monomers to randomly-chosen clusters as `accretion' and the removal of monomers from randomly-chosen clusters as `erosion'.
These dynamics are motivated by the importance of individuals in social applications such as warfare modelling, although the terminology is that of the physical sciences -- accretion might be simply termed `recruitment', while erosion is an individual leaving, through death or disillusionment for example.
Our implementation performs fixed numbers $n_{\mathrm{ac}}$ of accretion and $n_{\mathrm{er}}$ of erosion events each time step (either $0$, $1$, $3$ or $9$), and we choose which clusters to accrete or erode at random with probability proportional to their size.
Mass remains conserved in our implementation: accretion does not occur if there are no uncoalesced monomers in the system, and erosion does not occur if there are no non-monomers in the system.
We note, however, that {MLE} methods are expected to be more susceptible to perturbations that affect $n_1$
while {KS-MLE} should be more robust. We write $\alpha_{(n_{\mathrm{ac}},n_{\mathrm{er}})}$ and $\calK_{(n_{\mathrm{ac}},n_{\mathrm{er}})}$. 

\begin{figure} 
    \begin{minipage}{\linewidth}
        \begin{minipage}{0.49\linewidth}
            \includegraphics[width=\linewidth]{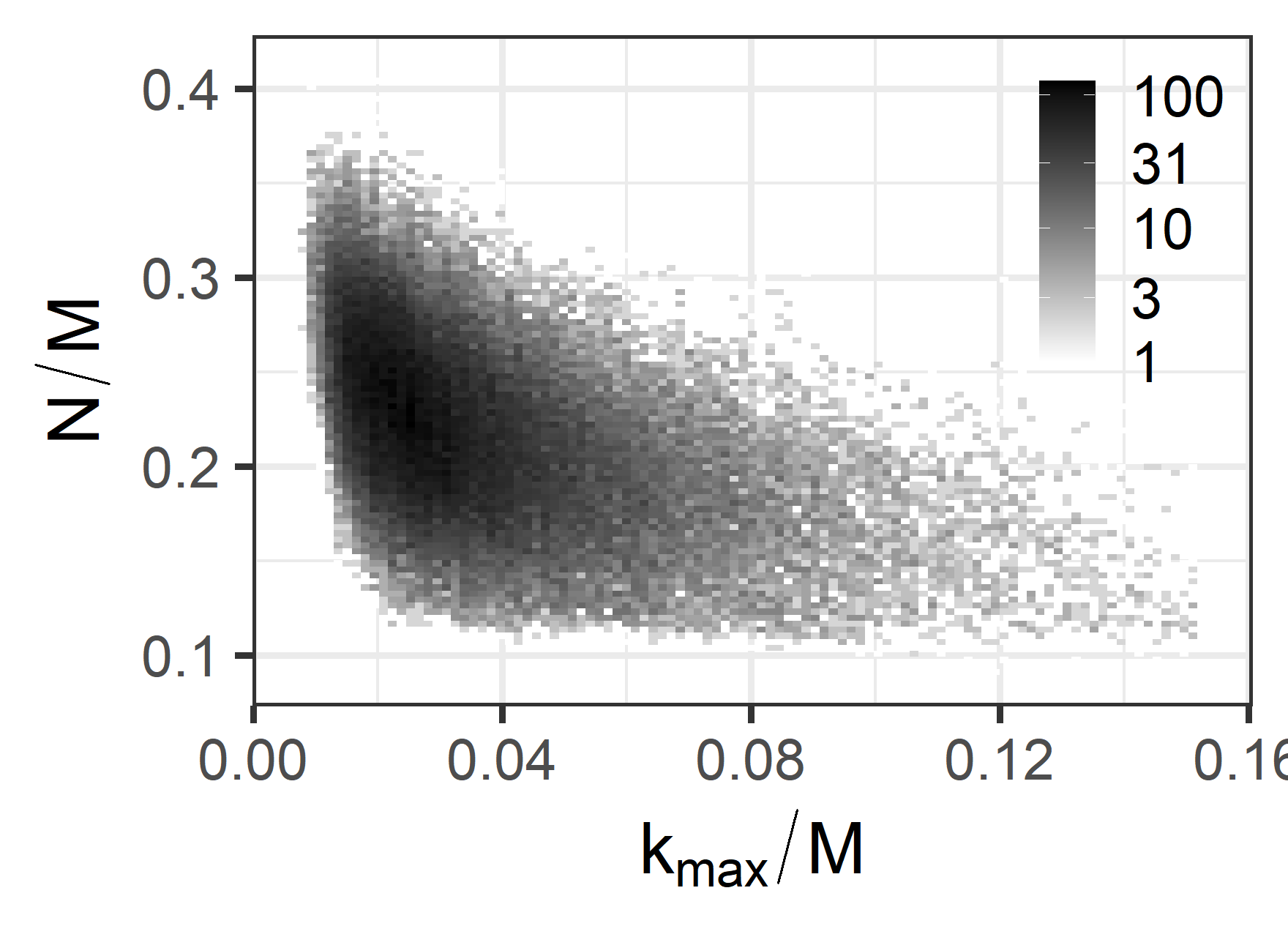}
        \end{minipage}
        \hspace{0.01\linewidth}
        \begin{minipage}{0.49\linewidth}
            \includegraphics[width=\linewidth]{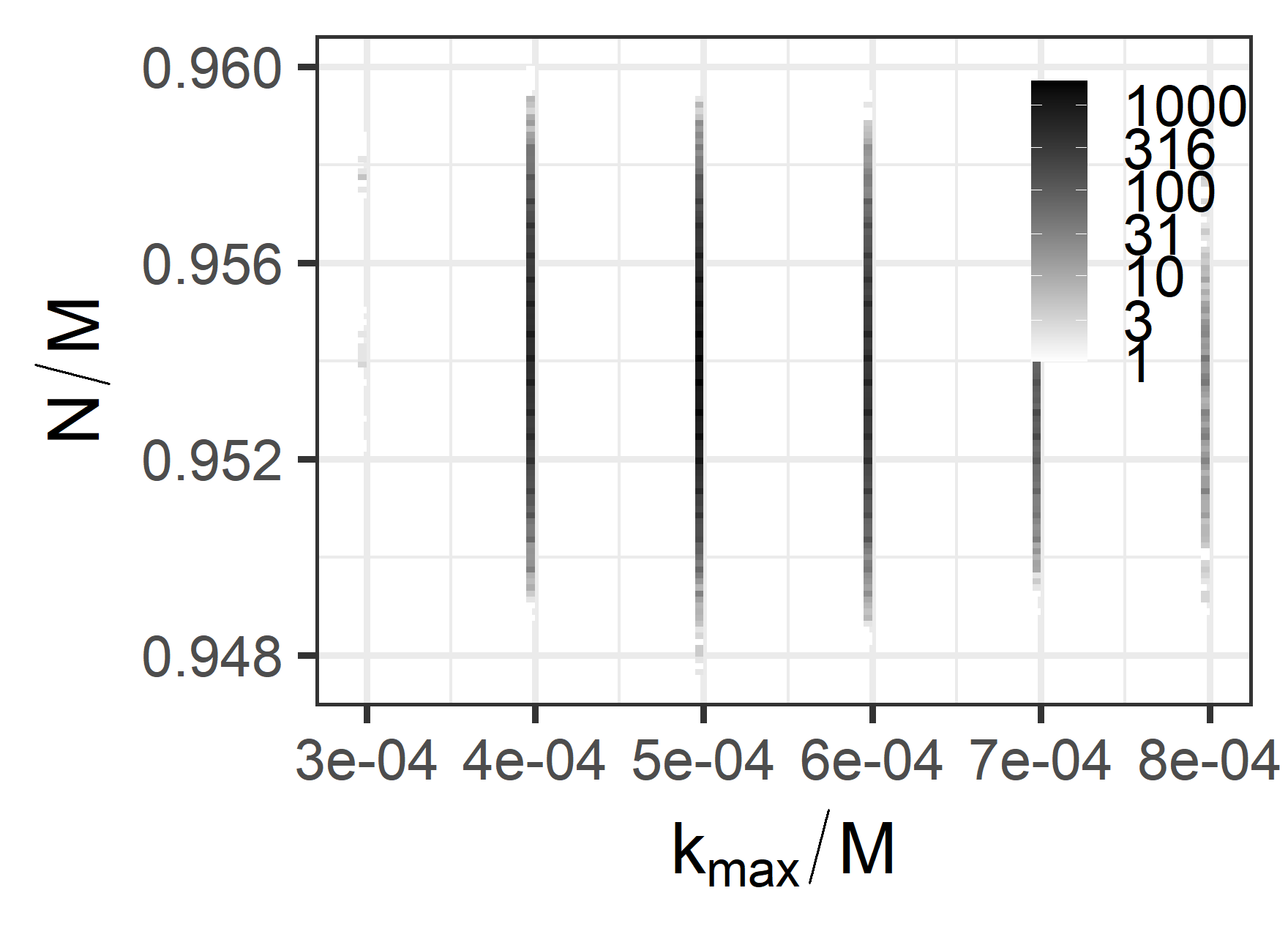}
        \end{minipage}
        \hspace{0.01\linewidth}
        \begin{minipage}{0.49\linewidth}
            \includegraphics[width=\linewidth]{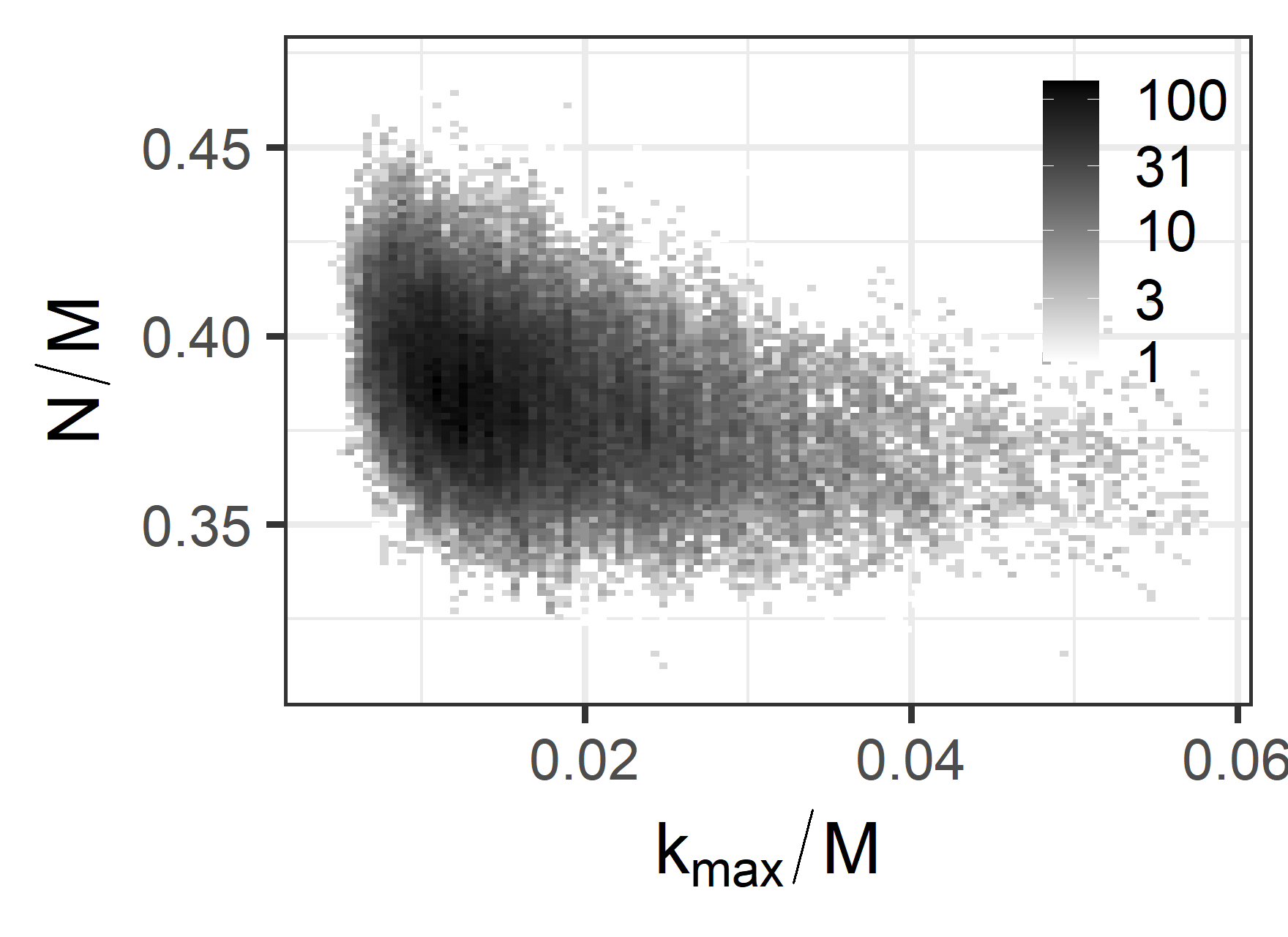}
        \end{minipage}
        \hspace{0.01\linewidth}
        \begin{minipage}{0.49\linewidth}
            \includegraphics[width=\linewidth]{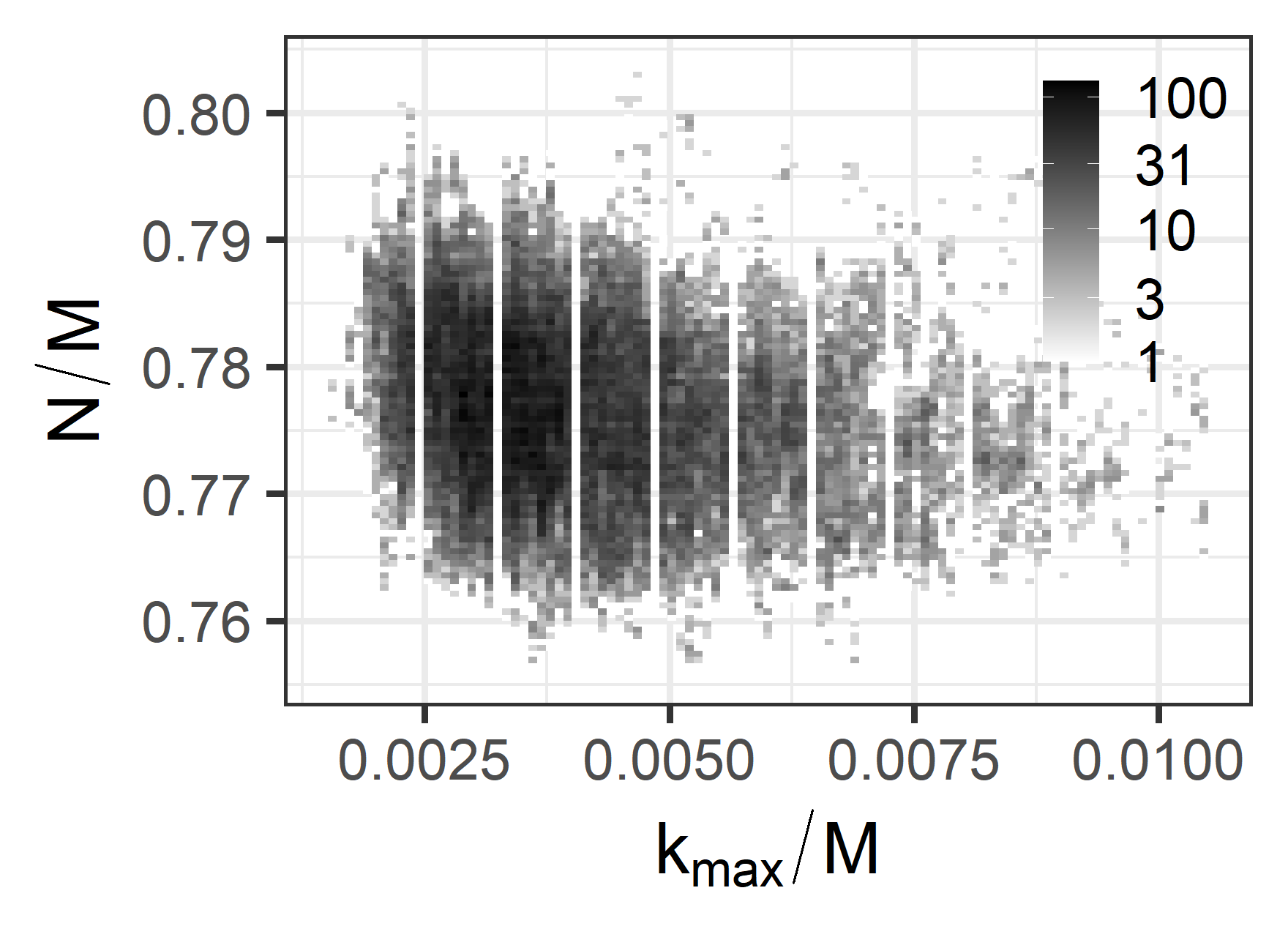}
        \end{minipage}
        \hspace{0.01\linewidth}
    \end{minipage}
    \caption[Coalescence and fragmentation, state space, ${M} = 10^4$, $\Prob{\Frag} = 0.2$, accretion and erosion.]{
    Clockwise from top-left: accretion, erosion amongst any clusters, erosion amongst unique clusters, and combined accretion and erosion.
    In each case, we set $n_{\mathrm{ac}} = 9$ and $n_{\mathrm{er}} = 9$ to best show the effects.
    }
    \label{fig:ststacat}
\end{figure}

Accretion results in an increase in $\calK$ and a decrease in $\alpha$: $\calK_{(9, 0)} = 0.831$, mean {KS-MLE} $\alpha_{(9, 0)}=2.33$, and mean {MLE} $\alpha_{(9, 0)}=1.72$.
More comparable to our earlier statistics, we have $\calK_{(3, 0)} = 0.564$, KS-MLE $\alpha_{(3, 0)} \in [2.248,2.682]$, and MLE $\alpha_{(3, 0)} \in [1.906,2.136]$ in $95\%$ of samples.
These reflect the increase in number of larger clusters and the tendency to increase the largest cluster's size even when the coalescence events do not target this cluster.
The increase in $\calK$ is somewhat deceptive, as can be seen in Figure \ref{fig:ststacat} (top-left), especially in comparison to Figure \ref{fig:examplesteadystate} and the system studied in Section \ref{sec:GlSh}, Figure \ref{fig:examplegelshatter}.
The system is certainly more cycle-like, given the shape of the state space visited has elongated, but it has not fully escaped steady state.
Accretion shows that the system's supposed power-law can be heavily weighted towards larger clusters by increasing the recruitment amongst monomers only.

Erosion generally shows the opposite behaviour to accretion, but its effectiveness relies on an implementation detail: do we remove from distinct clusters within a time step, or can the same cluster have multiple removals? 
To do the former greatly restricts the effects of erosion, as can be seen in Figure \ref{fig:ststacat} (top-right and bottom-right) where $n_{\mathrm{er}} = 9$, and still results in a steady-state-like system. The summary statistics are
$\calK_{(0, 3)} = 0.007$, KS-MLE $\alpha_{(0, 3)} \in [2.849,3.047]$, and MLE $\alpha_{(0, 3)} \in [3.122,3.229]$ in $95\%$ of samples.
To do the latter prevents the system from growing beyond $\kmax = 8$, and a power-law is a poor description of this case with $n_{\mathrm{er}} = 9$.
When $n_{\mathrm{er}} = 3$, the summary statistics are
$\calK_{(0, 3)} =-0.011$, KS-MLE $\alpha_{(0, 3)} \in [2.118,4.848]$, and MLE $\alpha_{(0, 3)} \in [3.528,3.635]$ in $95\%$ of samples.

Overall, the system is less robust to erosion, the frequent loss of monomers from clusters, than to accretion, frequent gain of monomers by clusters. The exact implementation can have a drastic effect, with more frequent erosion reducing cyclicity and making the system more steady-state-like. 

For completion, combining accretion and erosion with $n_{\mathrm{ac}} = n_{\mathrm{er}}$ results in a hybrid system that is similar to but more aggregated than the original system, Figure \ref{fig:ststacat} (bottom-left).
There is a modest increase in $\calK$: $\calK_{(1, 1)} = 0.148$, $\calK_{(3, 3)} = 0.179$, and $\calK_{(9, 9)} = 0.215$ when using erosion without the uniqueness restriction.
This reflects a slightly more cyclic system that is still not far from steady state.
The system also suffers from a depletion of monomers, so while KS-MLE $\alpha_{(3, 3)} \in [2.468,2.806]$, MLE $\alpha_{(3, 3)} \in [2.178,2.315]$.

\subsection{Weakening Fragmentation}\label{sec:StSt:Frag}

We have learned that accretion and erosion can be quite influential, but can also balance each other (resulting in a generally more aggregated system without greatly affecting the gelation-related properties) and generally make coalescence and fragmentation work more efficiently.
Against this, the models' assumptions on coalescence and fragmentation already take forms that are probably more efficient than in real systems. (For example, collisions might join two clusters imperfectly, producing some fragments; or shattering may fail to produce only monomers.) 
Overall this prompts the question, how efficient must the system be to be well-approximated by shattering fragmentation?
(Coalescence has been treated elsewhere in the literature, \eg \cite{Tanaka96_SelfSimilar}.)
For example, in an insurgency war situation complete shattering could be interpreted as a perfect response of a cell to being compromised -- idealised but implausible.

To attempt to answer this question, we consider a suite of variations of fragmentation. 
We begin with halving a cluster (\ie a size-$n$ cluster of fragments into one of $\floor{\frac{n}{2}}$ and one of $\ceil{\frac{n}{2}}$ where $\floor{\cdot}$ and $\ceil{\cdot}$ are the floor and ceiling functions).  
We then consider the case where one of these two clusters is shattered, before altering the proportion of the original cluster shattered from one-half to one-tenth or nine-tenths. 
We will see that even a small amount of shattering can significantly reset a system while keeping it near steady state.
Next we create the fragments using a single partition drawn uniformly at random from the set of possible partitions of the original cluster.
This motivates us finally to consider a specific intermediate possibility, repeated fragmentation of a cluster or \emph{stick-breaking}, in Section \ref{sec:StSt:PL}.

\begin{figure} 
    \begin{minipage}{\linewidth}
        \begin{minipage}{0.49\linewidth}
            \includegraphics[width=\linewidth]{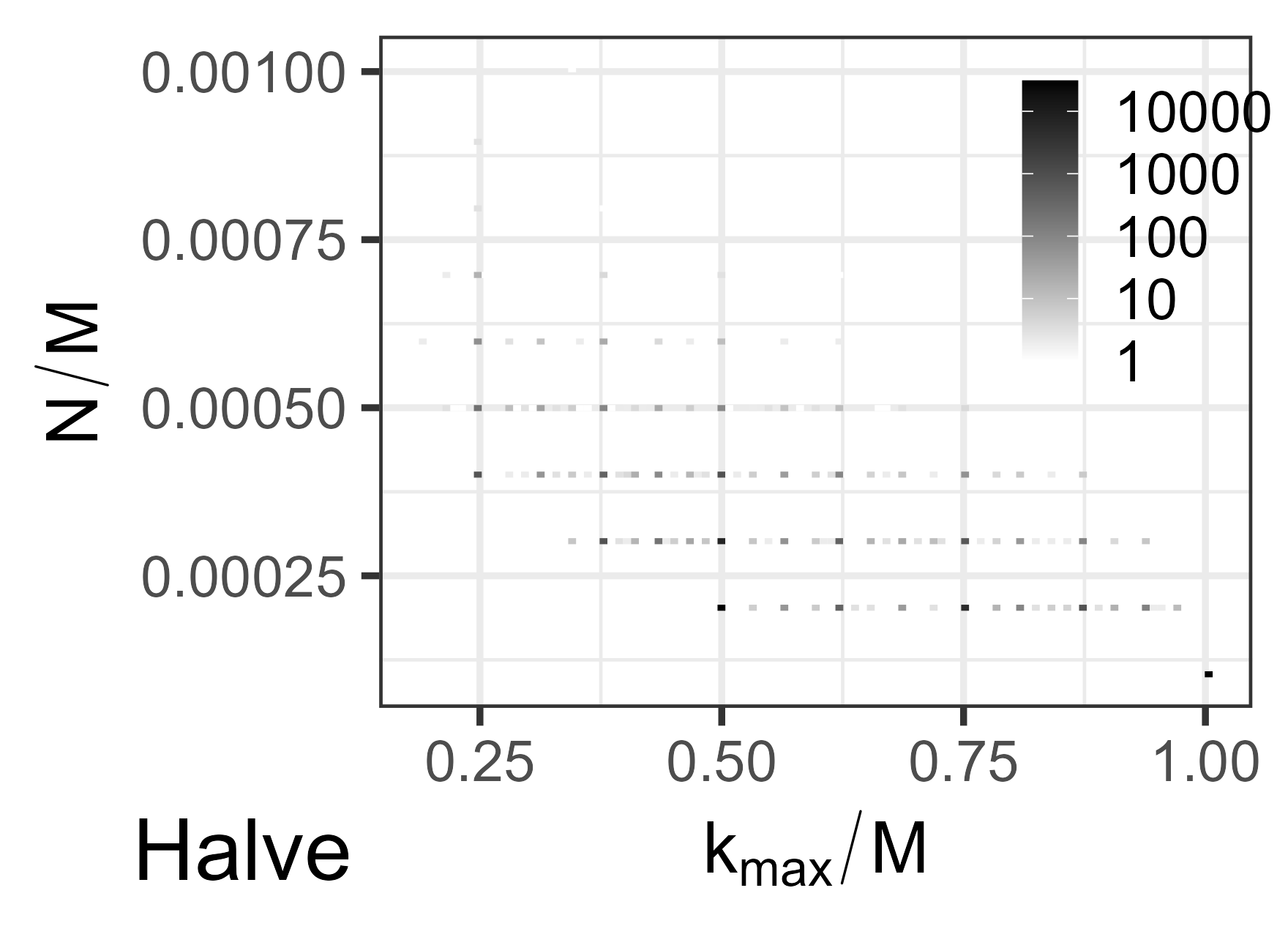}
        \end{minipage}
        \hspace{0.01\linewidth}
        \begin{minipage}{0.49\linewidth}
            \includegraphics[width=\linewidth]{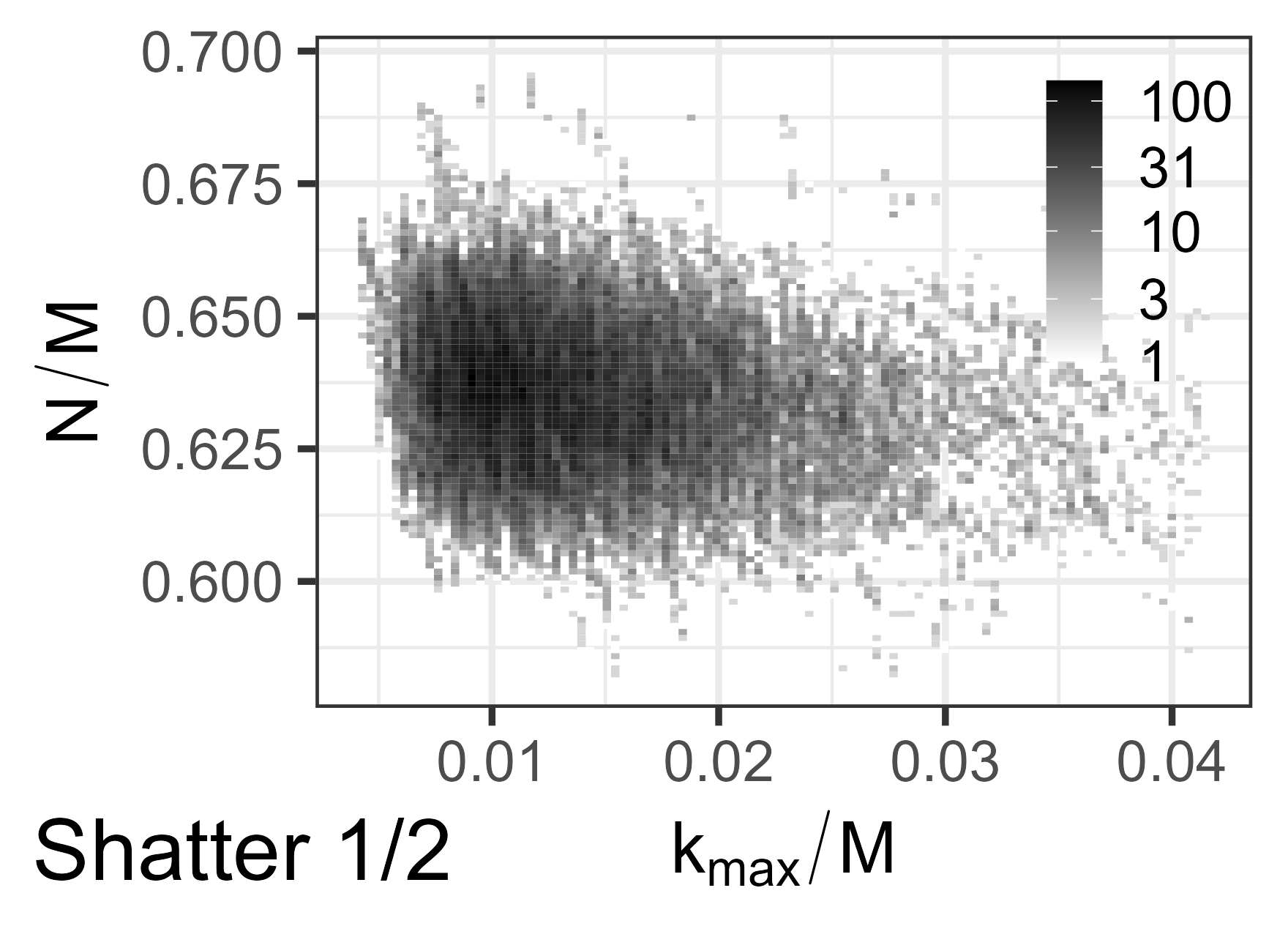}
        \end{minipage}
        \hspace{0.01\linewidth}
        \begin{minipage}{0.49\linewidth}
            \includegraphics[width=\linewidth]{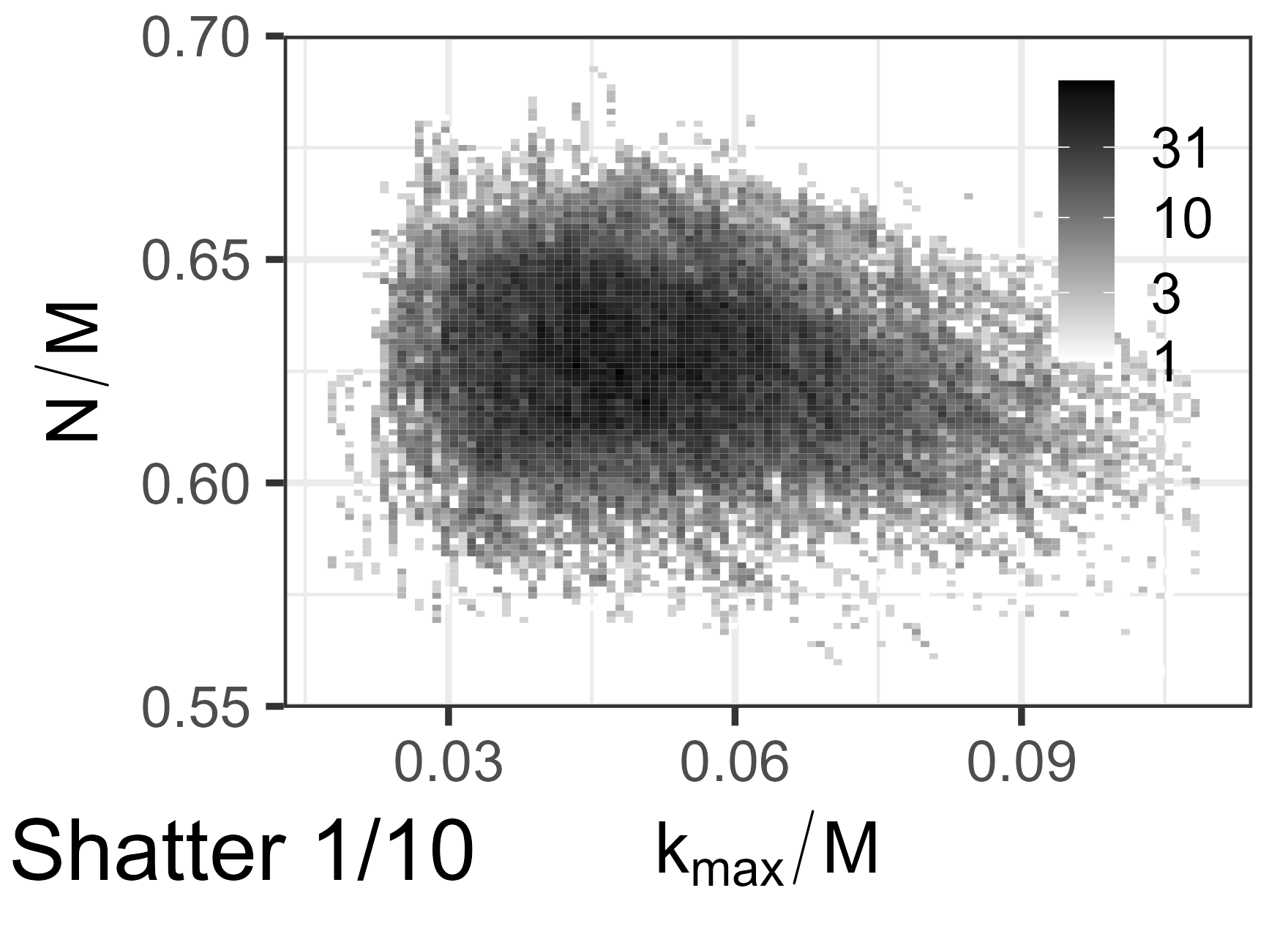}
        \end{minipage}
        \hspace{0.01\linewidth}
        \begin{minipage}{0.49\linewidth}
            \includegraphics[width=\linewidth]{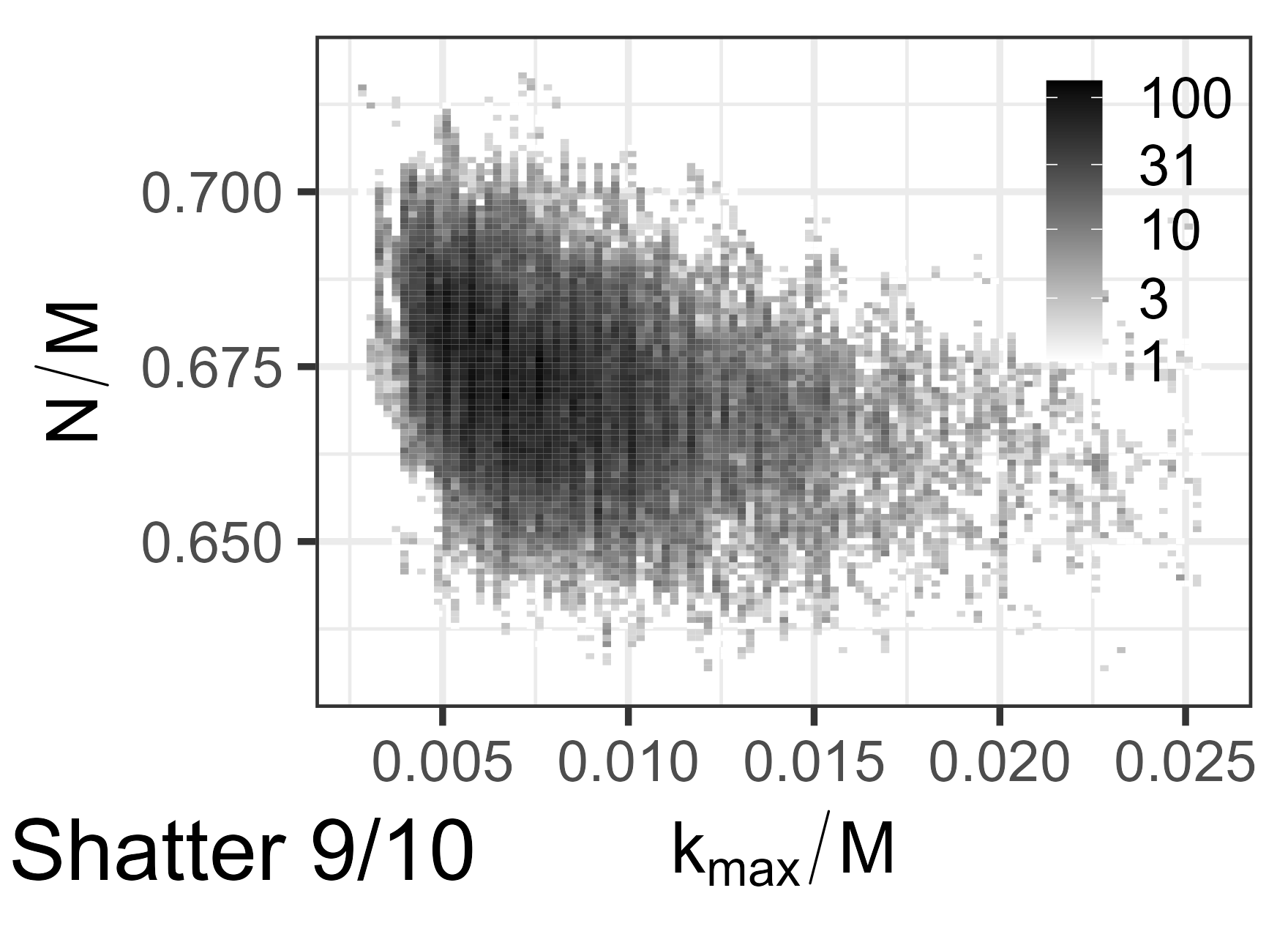}
        \end{minipage}
        \hspace{0.01\linewidth}
    \end{minipage}
    \caption[Coalescence and fragmentation, state space, ${M} = 10^4$, $\Prob{\Frag} = 0.2$, Split and Shatter Fragmentations.]{
    Clockwise from top-left: halving fragmentation, shattering half fragmentation, shattering nine-tenths fragmentation, and shattering one-tenth fragmentation.
    }
    \label{fig:ststSplit}
\end{figure}

Halving a cluster instead of shattering it is perhaps the furthest removed from shattering fragmentation: not only is it reversible in a single (binary coalescence) step, but it leaves both newly formed clusters as aggregated entities. 
It is perhaps no surprise then that coalescence and such a weak form of fragmentation leave the system in a very small number of clearly discrete states, as seen in Figure \ref{fig:ststSplit} (top-left). 
We find $\calK = -0.005$; the system spends slightly more time halving the largest group then it does growing it.
To apply power-law distribution estimators would be clearly incorrect here, even as proxies for other processes. If we attempt to do so, approximately $79\%$ of KS-MLE fits ($\xmin > 1$) report $\alpha = 0$.

Instead, consider the case where we shatter some portion of the original cluster to be fragmented. This still leaves the remainder as an aggregated cluster, but requires more -- potentially many more -- coalescent steps to reverse the damage caused by the fragmentation. 
In Figure \ref{fig:ststSplit} we consider shattering 0.5 of the cluster (top-right), 0.1 of the cluster (weakening fragmentation, bottom-left), and 0.9 of the cluster (strengthening fragmentation, bottom-right). 
Each result looks consistently steady-state-like (compare Figure \ref{fig:examplesteadystate}, top-left), but we are also seeing a smooth but modest deformation in the shapes of the plots (note that the scales on the horizontal axes differ), with the largest clusters tending to be smaller as the extent of shattering increases.
This is reflected in the summary statistics. Shattering $10\%$ of a cluster results in $95\%$ of KS-MLE $\alpha$ estimates in $[2.491, 2.721]$ and MLE $\alpha$ estimates in $[2.759, 2.953]$. 
Shattering $50\%$ yields estimates of $[2.431, 2.659]$ and $[2.727, 2.875]$ respectively, while shattering $90\%$ yields $[2.548, 2.744]$ and $[2.783, 2.918]$. 
Compared to the original coalescence and fragmentation estimates of $\left[2.568, 2.754\right]$ and $\left[2.788, 2.932\right]$, the differences are quite small. 
While the occupied region of the $(\frac{\kmax}{M}, \frac{N}{M})$ plane is somewhat different, it seems that even relatively small amounts of shattering can fairly well approximate full shattering for some purposes. 
That said, increasing shattering has a pronounced effect on the tendency of the system to produce cyclic or cycle-like behaviour.
Shattering $10\%$, $50\%$ and $90\%$ of the cluster yield $\calK = 0.431$, $0.186$, and $0.126$ respectively.
The summary statistic $\calK$ of cyclicity is thus more affected by the proportion shattered than is the $\alpha$ of the steady state, with highly inefficient shattering affecting the power law very little but creating much greater stochastic cyclicity hidden behind it.


So far, coalescence and fragmentation appear to be satisfactorily robust to perturbations to the exact form of the fragmentation process, as long as some shattering is present. But once again this may be an idealization motivated by the physical sciences -- in the social sciences it could be that fragmenting is more typically into a range of group sizes, and complete return to individual autonomy is implausible.
So, does the system remain robust when we move away from shattering? And how do the sizes of the clusters formed complicate the situation?

There are many ways to fragment a cluster, a problem which is equivalent to partitioning a natural number.
We begin first with `probabilistic divide and conquer', which selects a partition uniformly at random from the set of partitions \cite{Arratia16_PartitionsUAR}. 
Our implementation results on average in $25$ monomers ($95\%$: $[0, 88]$) and $94.6$ clusters ($95\%$: $[53, 165]$) when partitioning 1,000 in 1,000 trials, so only slightly fewer clusters, but the clusters are larger in the median than for shattering one-tenth.
(Note that shattering one-tenth would result in $101$ clusters, with $100$ monomers, shattering one-half $501$ or $500$, and shattering nine-tenths $901$ or $900$.)
Despite the similar number of clusters, the exact size of clusters is important.
Partitioning fragmentation results in $95\%$ of KS-MLE $\alpha$ estimates in $[2.22, 2.62]$ and  $95\%$ of MLE $\alpha$ estimates of $[2.09, 2.19]$.
Furthermore, the space itself is far more aggregated on average than that of one-tenth shattering fragmentation. 
While one-tenth shattering fragmentation had around $6254$ clusters on average, partitioning fragmentation has around $4032$, although this is counterbalanced by a higher average $\xmax$ for one-tenth shattering fragmentation ($532$ vs $204$).

So we now know that the exact form of the fragmentation does matter, and that even shattering just one-tenth of a cluster makes the system fairly close to that in which the entire cluster is shattered.
On the other hand, this seems to be serving as a proxy for the ease of re-assembly of a fragmented cluster. Due to the size-biased nature of the kernels, obtaining monomers makes it harder to re-assemble than if one obtains clusters of varying sizes.
Unfortunately, partitioning uniformly at random does not give us a clear control parameter to explore how the results are changing. 
In the next section, we explore a method of partitioning that does have such a parameter, allowing us to explore its effect on our summary statistics.

\subsection{Stick-Breaking and Chinese Restaurants}\label{sec:StSt:PL}

One simple form of fragmentation, analogous to stick-breaking processes \cite{Pitman06_CombinatorialStochasticProcesses}, is to repeatedly fragment a portion of a cluster. 
We take a stick (respectively, cluster), snap it in two at a random point (resp.\ divide it into two parts), discard one, and then repeat the process on the remainder. 
The question then becomes how to pick our random point. We begin here with a uniform distribution, before considering the natural extension to the beta-binomial distribution. 
We then consider a more general parametrized form of fragmentation, the Chinese Restaurant Process (CRP).

We can also use stick-breaking for coalescence, although a little more specification is needed. When two clusters meet, our implementation gives the larger cluster the first fragment broken off the smaller cluster, which must be of size at least 1, after which the remainder of the smaller cluster experiences stick-breaking fragmentation as described above.

\begin{figure} 
    \begin{minipage}{\linewidth}
        \begin{minipage}{0.49\linewidth}
            \includegraphics[width=\linewidth]{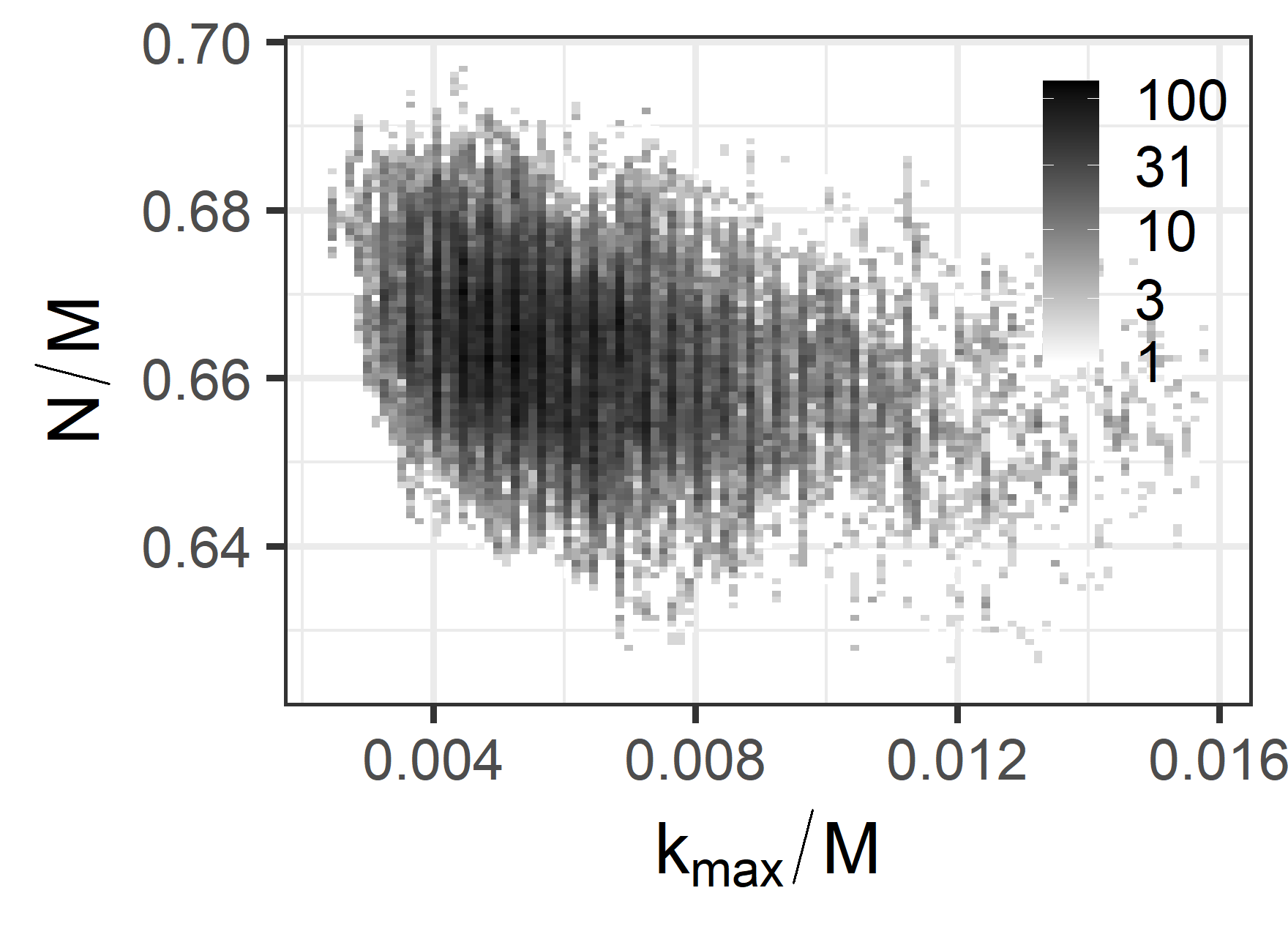}
        \end{minipage}
        \hspace{0.01\linewidth}
        \begin{minipage}{0.49\linewidth}
            \includegraphics[width=\linewidth]{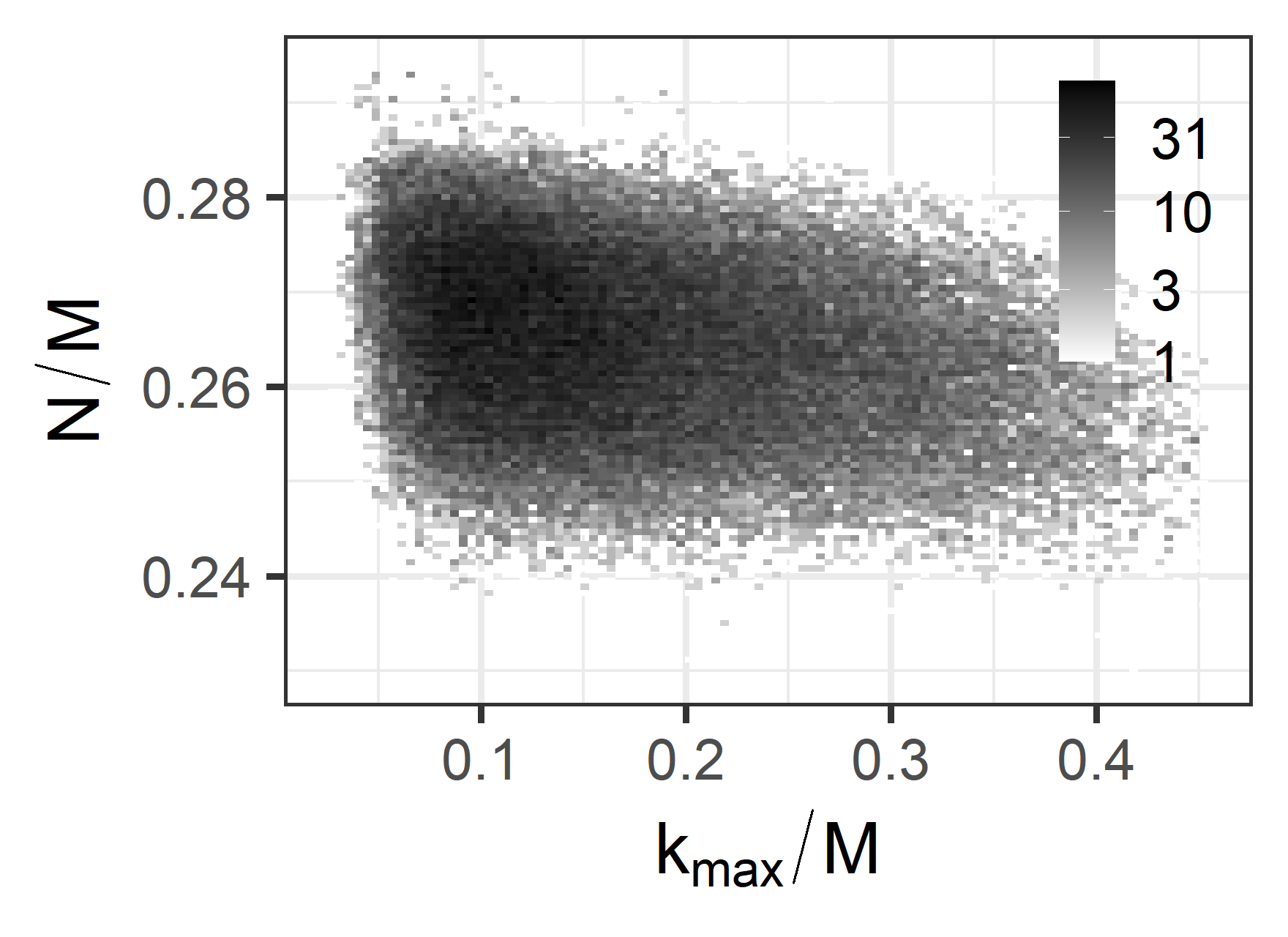}
        \end{minipage}
        \hspace{0.01\linewidth}
        \begin{minipage}{0.49\linewidth}
            \includegraphics[width=\linewidth]{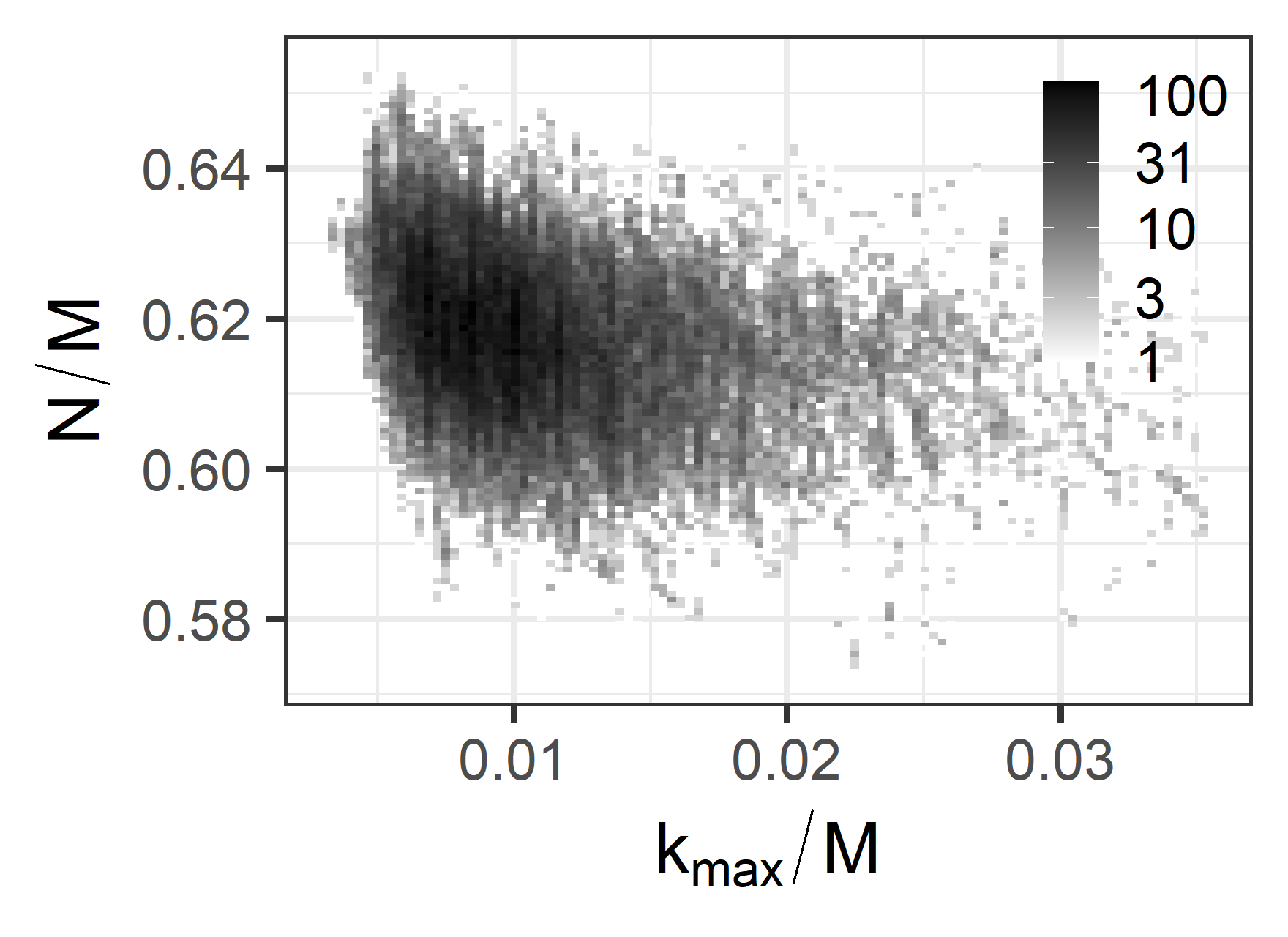}
        \end{minipage}
        \hspace{0.01\linewidth}
        \begin{minipage}{0.49\linewidth}
            \includegraphics[width=\linewidth]{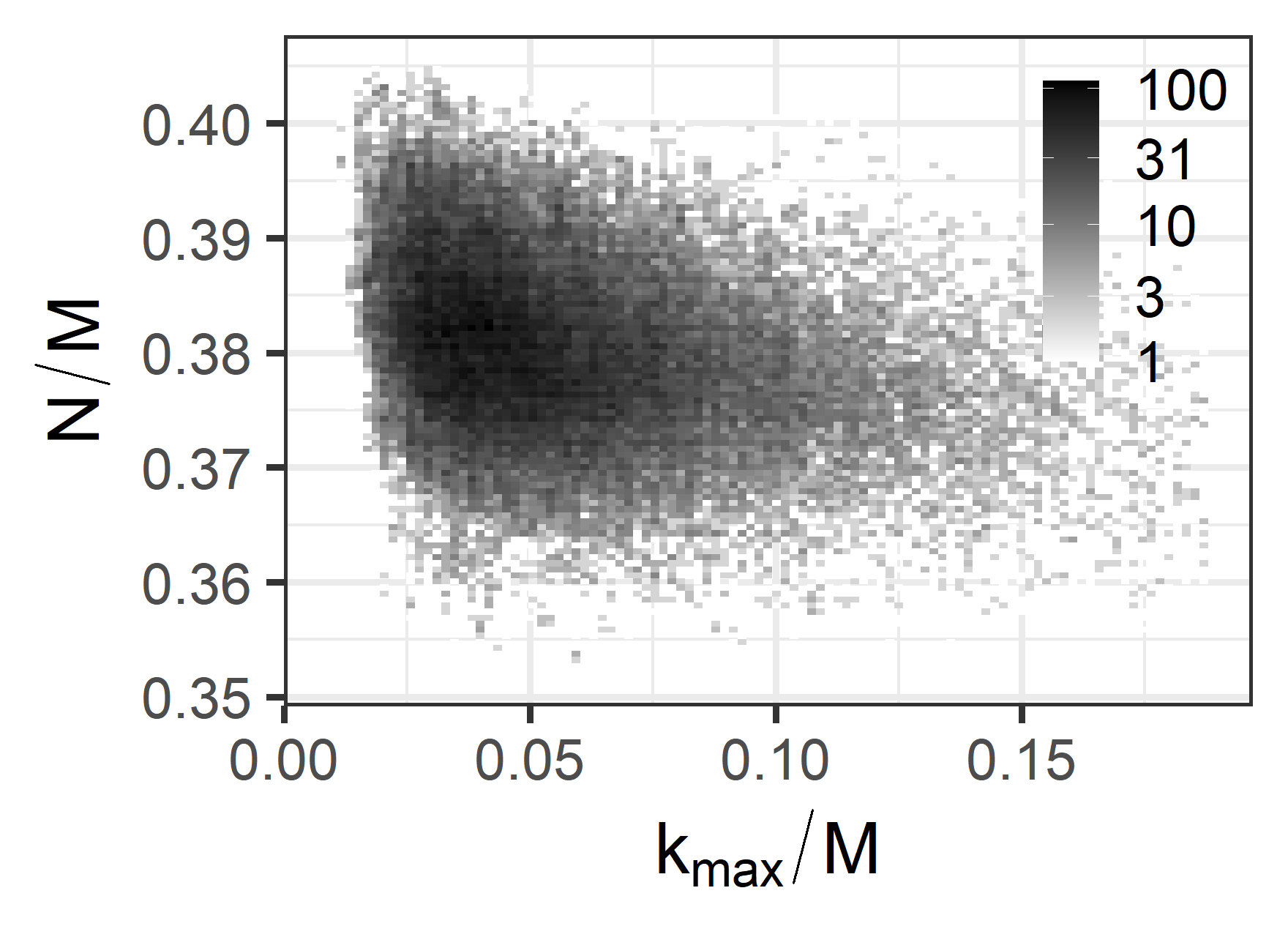}
        \end{minipage}
        \hspace{0.01\linewidth}
    \end{minipage}
    \caption[Coalescence and fragmentation, state space, ${M} = 10^4$, $\Prob{\Frag} = 0.2$, power-law fragmentation.]{
    Clockwise from top-left: stick-breaking coalescence, stick-breaking fragmentation, CRP fragmentation with $\theta = 1.2$, and CRP fragmentation with $\theta = 1.8$.
    }
    \label{fig:ststpl}
\end{figure}

Figure \ref{fig:ststpl} contains the results of stick-breaking coalescence (with base model fragmentation; top-left) and stick-breaking fragmentation (with base-model coalescence; top-right). Initial inspection of the former might lead one to believe that stick-breaking coalescence achieved no change from the base case other than a slightly damped cyclicity, $\calK = 0.094$ from $0.123$ and a slightly lowered maximum group size, with $95\%$ of KS-MLE and MLE estimates of $\alpha \in [2.534, 2.768]$ and $[2.725, 2.829]$ respectively.  We will return to this case in a moment.
On the other hand, stick-breaking fragmentation causes a larger perturbation to the system than does partition fragmentation, with $\calK = 0.412$ and $95\%$ of KS-MLE and MLE estimates of $\alpha \in [1.890, 2.166]$ and $[2.241, 2.402]$ respectively. Notice that whereas partition fragmentation reduced the MLE more, stick-breaking fragmentation reduces the KS-MLE more. This suggests that the tail induced by stick-breaking is far heavier than that produced by partitioning. 
In contrast, partitioning has a larger effect on the number of monomers in the system.

Finally, the combination of stick-breaking coalescence and fragmentation helps reveal some of the unseen effects of stick-breaking coalescence.
While not entirely capable of reining in the effects of stick-breaking fragmentation, stick-breaking coalescence has a large impact, bringing $\calK$ to $0.236$ and $95\%$ of KS-MLE and MLE estimates to $\alpha \in [2.208, 2.393]$ and $[2.298, 2.369]$. 
Stick-breaking coalescence permits a modest decrease in the speed of gelation while making the system less aggregated in general.
This in turn makes it harder to reverse the effects of a single stick-breaking fragmentation because there is a smaller reservoir of medium-large clusters (in exchange for a larger reservoir of small-medium clusters).
It also appears that this stick-breaking coalescence does not have a large effect on the $\alpha$ estimates by itself because the exchange of small-medium clusters for medium-large clusters does not greatly influence either the tail (KS-MLE) or the number of monomers (MLE).

Moving to a beta-binomial distribution, thus allowing the individual breaks to be more biased towards larger or smaller clusters, does not change the results greatly. Moving the distribution's parameters from (1, 1) (which is the specialization to the uniform distribution) to (2, 4), (3, 3), and (4, 2) results in $95\%$ KS-MLE $\alpha$ estimates of $[2.044, 2.221]$, $[1.941, 2.151]$ and $[1.853, 2.147]$ respectively.

Another interpretation of discrete stick-breaking gives us smooth control of fragmentation: the Chinese Restaurant Process (CRP).
The CRP is named for its metaphorical construction of a partition. For a partition of size $n$, we consider a queue of $n$ people who are permitted to seat themselves one at a time at circular tables. 
After the first individual is seated, the second then has a choice to sit with the first, or to initiate a new table, and so on for each customer.
Further, it is presumed that the probability that a new customer picks an occupied table is proportional to the number of people sitting there.
This mechanistic process has natural analogues to the manner in which a sub-group in insurgency warfare might fragment under external pressure.
The CRP has two parameters, which when positive can be interpreted as controlling how intrinsically attractive it is to start a new table (the `strength') and slightly penalising tables already occupied (the `discount').
Of these two, the latter, say $\phi \in [0, 1]$, is more important for our purposes, as it controls the power-law distribution over the number of customers seated at each table with exponent $\theta = 1 + \phi$ \cite{Pitman06_CombinatorialStochasticProcesses,Goldwater11_ChineseRestaurantProcess}.
(In principle, power laws above $\theta = 2$ could be accessed via \eg Price's network model \cite{Newman10_NetworkModels}, but such power laws did not appear to substantially alter our results.)

Replacing our stick-breaking fragmentation with CRP fragmentation for various $\theta$ proves particularly tractable for navigating various summary statistics. 
Cyclicity $\calK$ now appears to vary smoothly: compare 
stick-breaking fragmentation's $\calK = 0.412$ with 
$\theta = 1.2$ yielding $\calK = 0.436$, 
$\theta = 1.5$ $\calK = 0.246$, and 
$\theta = 1.8$ $\calK = 0.156$.
Similarly $\alpha$ estimates by both KS-MLE and MLE methods proceed smoothly:
our $95\%$ intervals for KS-MLE $\alpha$ go from 
$[1.890, 2.166]$, through 
$[2.075, 2.257]$ and 
$[2.299, 2.426]$, to
$[2.459, 2.620]$,
while MLE $\alpha$ proceeds as
$[2.241, 2.402]$,
$[2.274, 2.376]$,
$[2.434, 2.511]$, and
$[2.640, 2.734]$.
This is reflected in the bottom panels of Figure \ref{fig:ststpl}, which show results for $\theta = 1.2$ and $\theta = 1.8$. Just as in Figure \ref{fig:ststSplit}, the system appears to be returning to the original steady state as we go from stick-breaking fragmentation (top-right; largest $\kmax / M$ and smallest $N / M$) through CRP with $\theta = 1.2$ (bottom-right; intermediate values) to CRP with $\theta = 1.8$ (bottom-left; smallest $\kmax / M$ and largest $N / M$).

It is notable that one power-law distribution producing process has such a strong and well controlled effect when used as a part of a wider power-law distribution producing process. 
Presumably, the reason why this works so well is, building on the above, a power-law distribution with exponent $\theta$ in $[1, 2]$ is not so clustered as to have no effect (mostly shattered, $\theta > 2$) while providing a consistent control on shape.
This is in contrast to beta-binomial which clumps too much, while a significant amount of shattering takes too long to reverse (since monomers are slow and difficult to coalesce back together).

\subsection{Summary}\label{sec:StSt:Summary}
Is it safe to model a CF system as being in steady state?
In this section, we have considered how important unmodelled systematic perturbations can be to the steady state of a CF system that is believed to be in or close to its steady state.

Often, the CF system proved quite robust. 
Small amounts of accretion and erosion do not have disproportionate effects and dominate the system. Greater accretion and erosion can, however, shift the system away from its original power-law steady state.
For large amounts of accretion and erosion to cancel out each other's effects would require fine-tuning. 
Further, shattering does not need to be total in order to replenish the re-supply of monomers: a small amount of shattering effectively replicates the effects of total shattering.

At the extreme of simplicity, replacing shattering with halving altered the system beyond recognition, placing it firmly in a regime of simple forced cycles from which it cannot escape.
More subtly, there are distributions of fragments which leave enough medium-sized clusters that the system is delayed from reassembling without forcing it to reset (nearly) completely or not at all.
While this can be achieved with random partitioning or varieties of stick-breaking, this regime is most accessible by a power-law distribution generating process such as the Chinese Restaurant Process, a mechanistic process originally motivated by an analogy to people's behaviour. 
In such circumstances the cyclicity parameter $\calK$ can easily reach values significantly greater than zero, indicating some form of time-asymmetric cyclicity and, minimally, that the state is rather unsteady, so that a simple assumption of a steady state described by a power law is certainly not telling the full story. 

\begin{table} 
    \caption[Summary of robustness in steady state]{
    Summary of robustness results in steady-state coalescence and fragmentation. In all cases, ${M} = 10^4$, $\Prob{\Frag} = 0.20$, and four simulations were used.
    The ranges given for {KS-MLE} and {MLE} estimates of power-law $\alpha$ contain $95\%$ of empirical results. Perturbations used are explained in the text.
    We exclude large ($>5$) and small ($<1$) estimates of the {KS-MLE} $\alpha$ from our intervals.
    }
    \label{tab:cofr:robustnessSS}
    \centering
    \begin{tabular}{|l|ccc|} \hline
    Perturbation              & Cyclicity $\calK$   & {KS-MLE} ($\xmin > 1$) $\alpha$       & {MLE} $\alpha$ \\\hline 
    \hline
    Steady-state                       & $0.123$    & $\left[2.568, 2.754\right]$ & $\left[2.788, 2.932\right]$ \\\hline
    Accretion (3, Uniq.)               & $0.560$    & $\left[2.246, 2.685\right]$ & $\left[1.907, 2.135\right]$ \\\hline
    Erosion (3, Uniq.)                 & $0.009$    & $\left[2.849, 3.047\right]$ & $\left[3.122, 3.229\right]$ \\\hline
    Accr.\ and Eros.\ (3, Uniq.)       & $0.266$    & $\left[2.341, 2.769\right]$ & $\left[2.105, 2.256\right]$ \\\hline
    Accretion (3)                      & $0.564$    & $\left[2.248, 2.682\right]$ & $\left[1.906, 2.136\right]$ \\\hline
    Erosion (3)                        & $-0.011$   & $\left[2.118, 4.848\right]$ & $\left[3.528, 3.635\right]$ \\\hline
    Accr.\ and Eros.\ (3)              & $0.179$    & $\left[2.468, 2.806\right]$ & $\left[2.178, 2.315\right]$ \\\hline
    Coal.\ Stick-breaking              & $0.094$    & $\left[2.534, 2.768\right]$ & $\left[2.725, 2.829\right]$ \\\hline
    Frag.\ Stick-breaking              & $0.463$    & $\left[1.890, 2.166\right]$ & $\left[2.241, 2.402\right]$ \\\hline
    Coal.\ and Frag.\ Stick-breaking   & $0.239$    & $\left[2.208, 2.393\right]$ & $\left[2.298, 2.369\right]$ \\\hline
    {CRP} $\theta = 1.20$              & $0.454$    & $\left[2.075, 2.257\right]$ & $\left[2.274, 2.376\right]$ \\\hline
    {CRP} $\theta = 1.50$              & $0.247$    & $\left[2.299, 2.426\right]$ & $\left[2.434, 2.511\right]$ \\\hline
    {CRP} $\theta = 1.80$              & $0.157$    & $\left[2.459, 2.620\right]$ & $\left[2.640, 2.734\right]$ \\\hline
    \end{tabular}
\end{table}

\section{Effect of rule variations on base model with stochastic gel-shatter cycles} \label{sec:GlSh}

\begin{figure} 
    \begin{minipage}{\linewidth}
        \begin{minipage}{0.49\linewidth}
            \includegraphics[width=\linewidth]{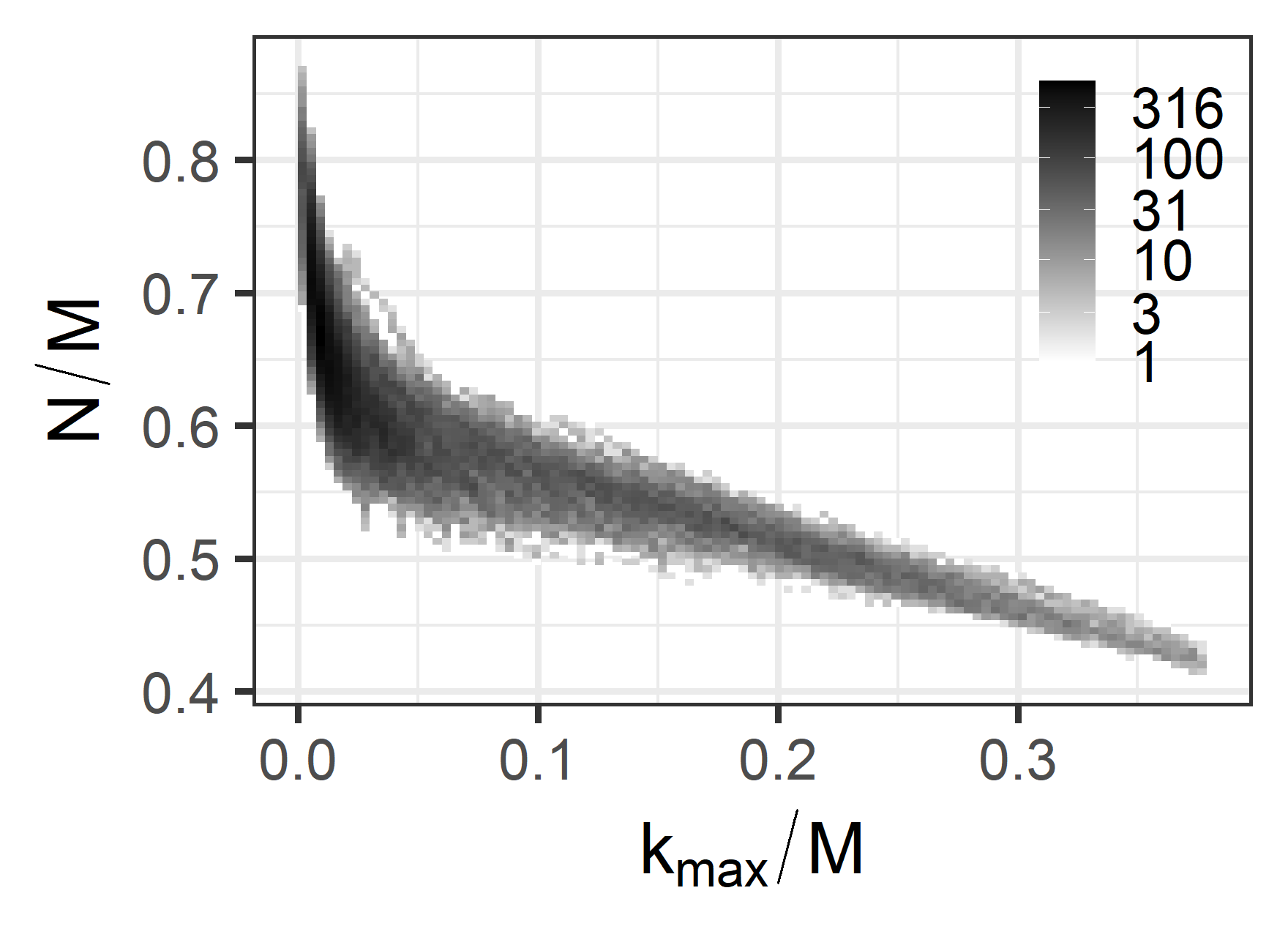}
        \end{minipage}
        \hspace{0.01\linewidth}
        \begin{minipage}{0.49\linewidth}
            \includegraphics[width=\linewidth]{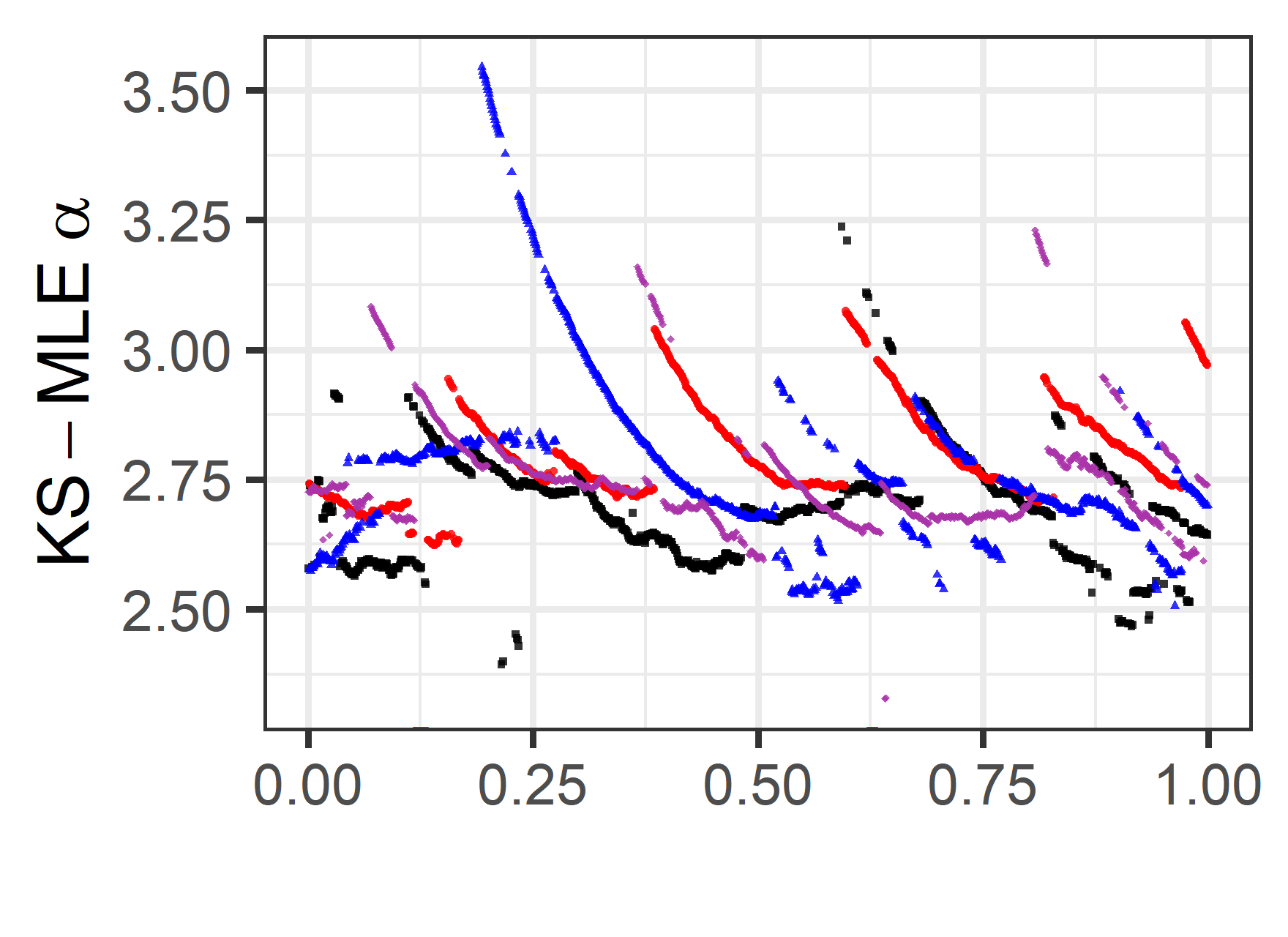}
        \end{minipage}
        \hspace{0.01\linewidth}
        \begin{minipage}{0.49\linewidth}
            \includegraphics[width=\linewidth]{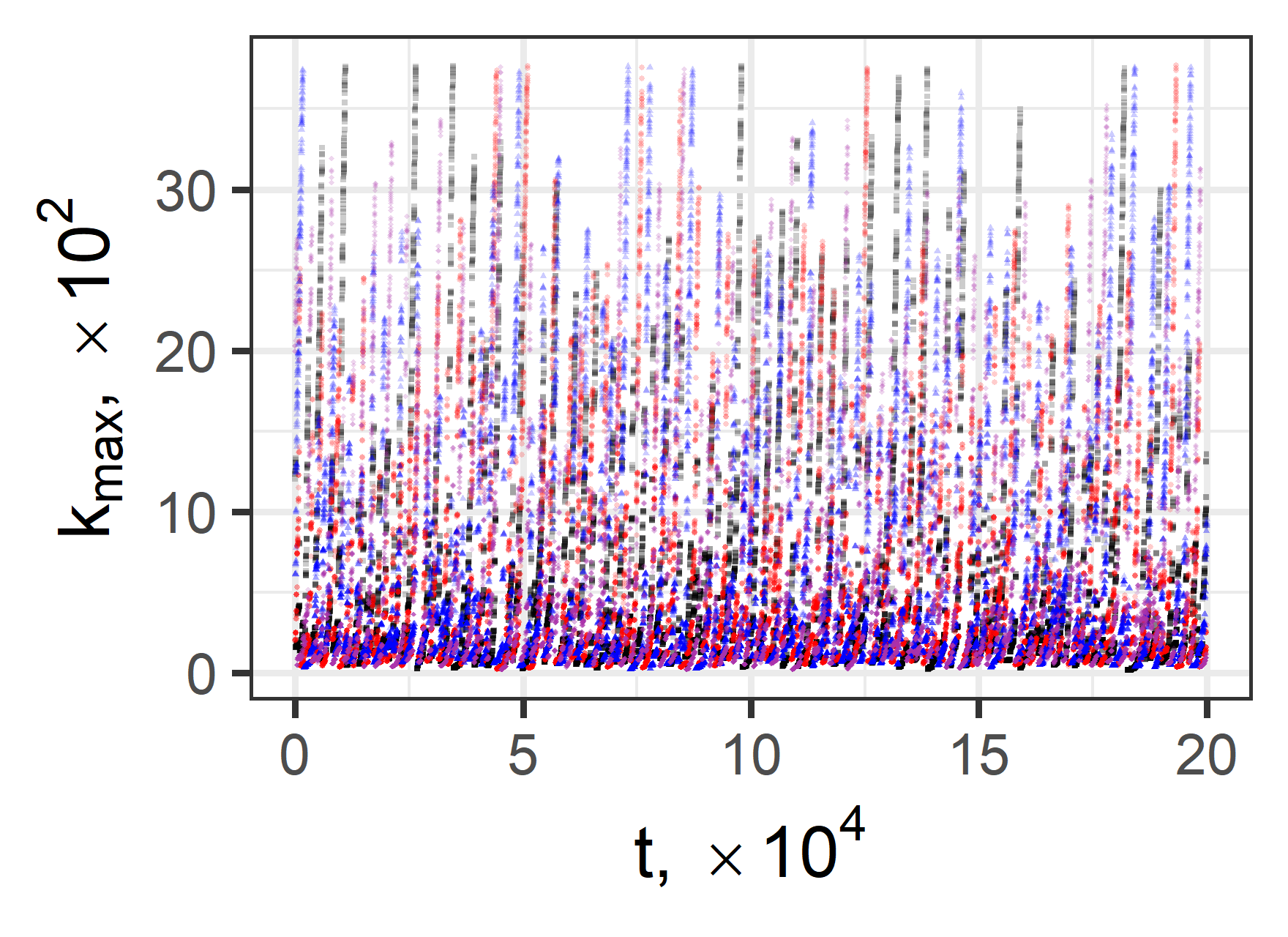}
        \end{minipage}
        \hspace{0.01\linewidth}
        \begin{minipage}{0.49\linewidth}
            \includegraphics[width=\linewidth]{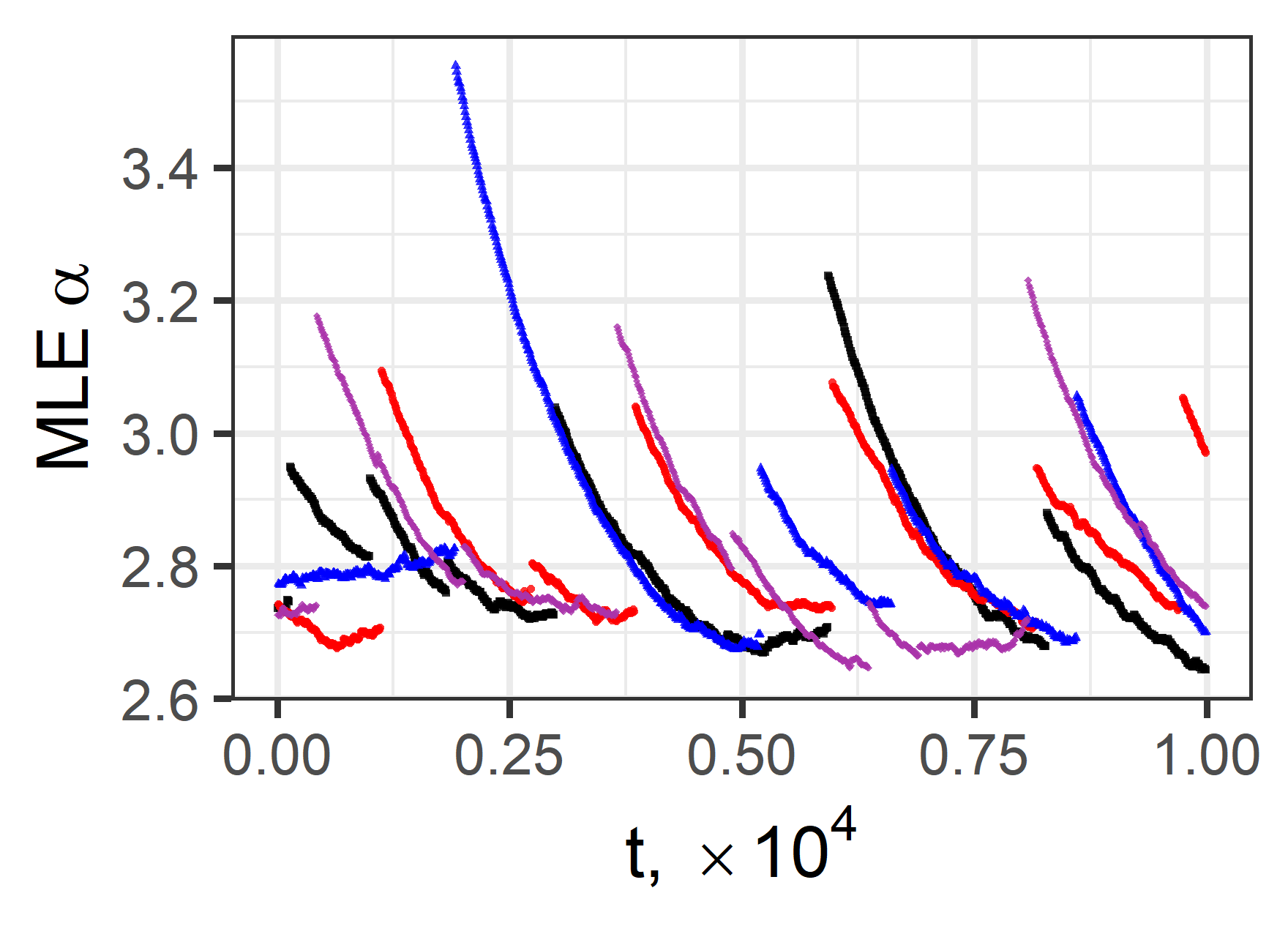}
        \end{minipage}
        \hspace{0.01\linewidth}
    \end{minipage}
    \caption[Coalescence and fragmentation, summary statistics, ${M} = 10^4$, $\Prob{\Frag} = 0.01$.]{
    Coalescence and fragmentation summary statistics, ${M} = 10^4$, $\Prob{\Frag} = 0.01$.
    Plot types are the same as in Figure \ref{fig:examplesteadystate}.
    The time series shown for estimates of the exponent are shorter than those for $\kmax$ or those in Figure \ref{fig:examplesteadystate} to better show details.
    }
    \label{fig:examplegelshatter}
\end{figure}

As we saw in the last section, it is not guaranteed that a steady state will emerge from a coalescence and fragmentation process.
A prominent alternative is the formation of {\em stochastic gel-shatter cycles}, in which the system is dominated by gelling coalescence and then stochastically resets due to shattering fragmentation.
We use a model gel-shatter system to show that 
this regime is more susceptible to perturbations of the sort described in Section \ref{sec:StSt}.
For the model system seen in Figure \ref{fig:examplegelshatter}, we set $M = 10^4$ and $\hat F = 0.01 = 1 - \hat K$ for $r \approx 101$, yielding $\calK = 0.490$.
This system starts in a very disaggregated state, grows, and then reaches a critical gelation point beyond which it rapidly aggregates before stochastically shattering and repeating.
As it does so, the fitted exponent $\alpha$ decreases smoothly before jumping upwards.
Its exponent summary statistics are the confidence intervals for KS-MLE $\alpha$ and MLE $\alpha$ from $95\%$ of samples. For this case, they are respectively $\left[2.474, 2.829\right]$ and $\left[2.656, 3.163\right]$.
We proceed in parallel to Section \ref{sec:StSt}: we first show that additional processes can significantly alter the $\alpha$ statistics before again observing that power-law distributed fragmentation greatly affects the gel-shatter cycles.
Summary statistics will be presented in Table \ref{tab:cofr:robustnessGS} at the end of this section.

\begin{figure} 
    \begin{minipage}{\linewidth}
        \begin{minipage}{0.49\linewidth}
            \includegraphics[width=\linewidth]{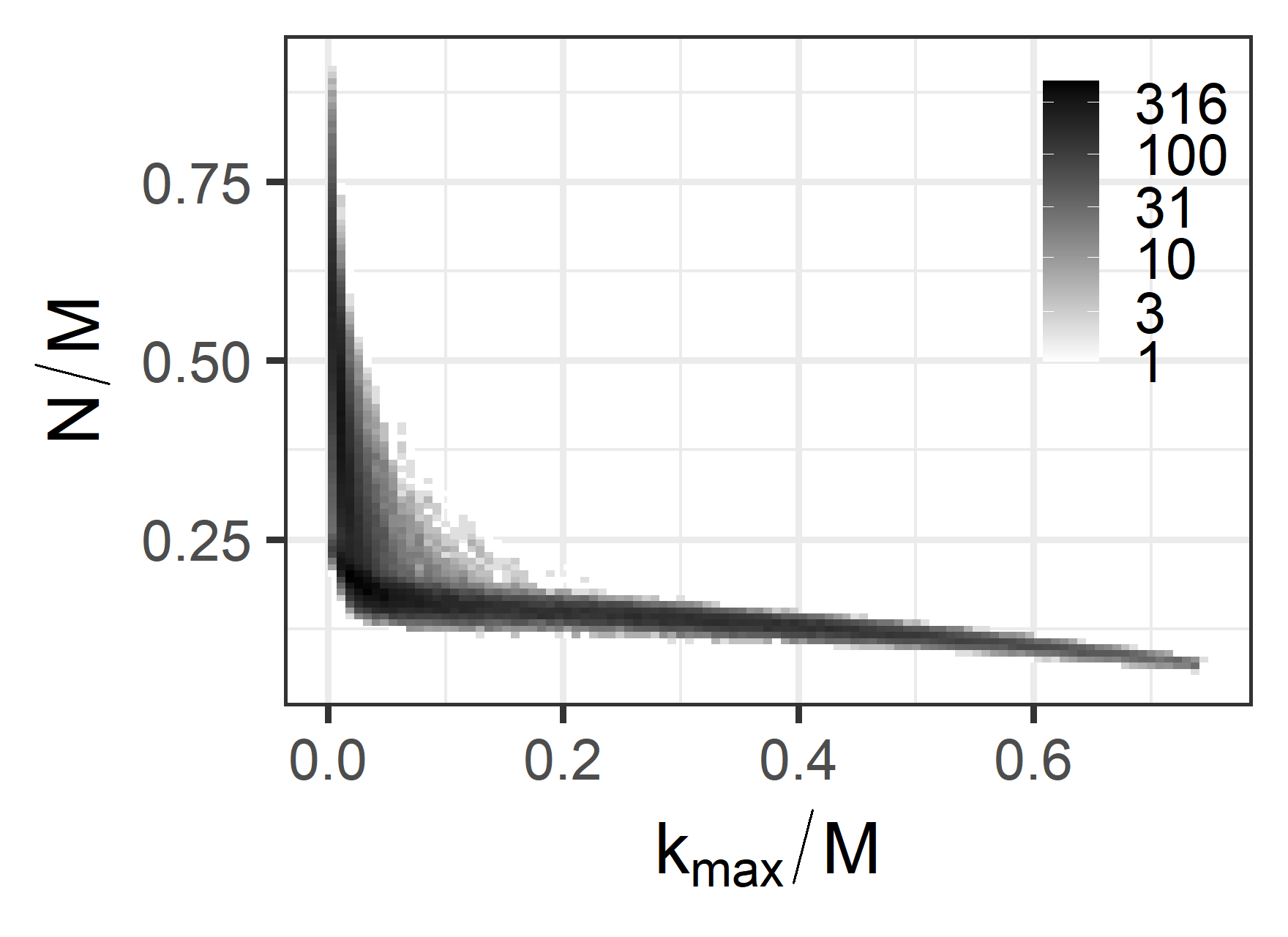}
        \end{minipage}
        \hspace{0.01\linewidth}
        \begin{minipage}{0.49\linewidth}
            \includegraphics[width=\linewidth]{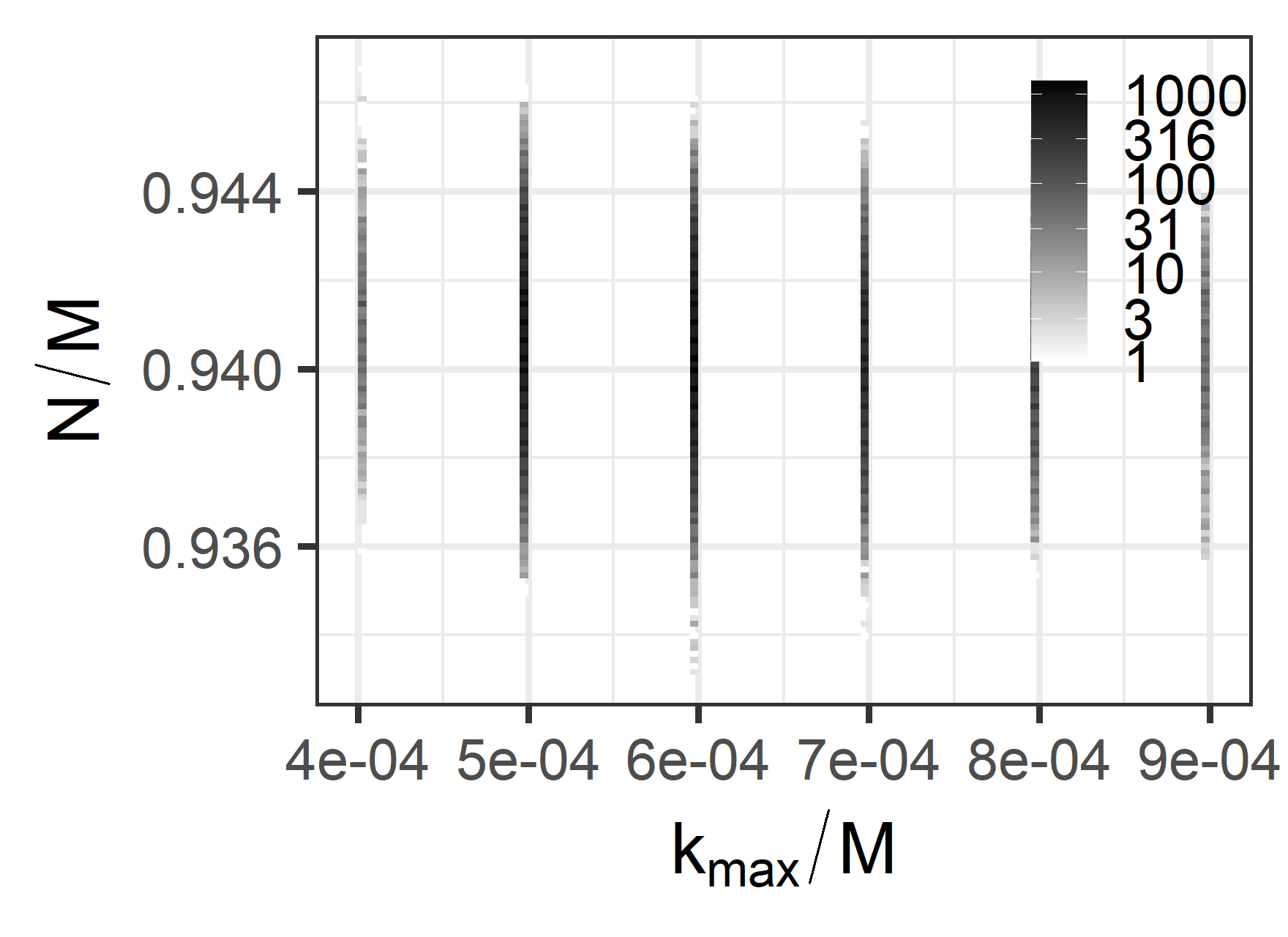}
        \end{minipage}
        \hspace{0.01\linewidth}
        \begin{minipage}{0.49\linewidth}
            \includegraphics[width=\linewidth]{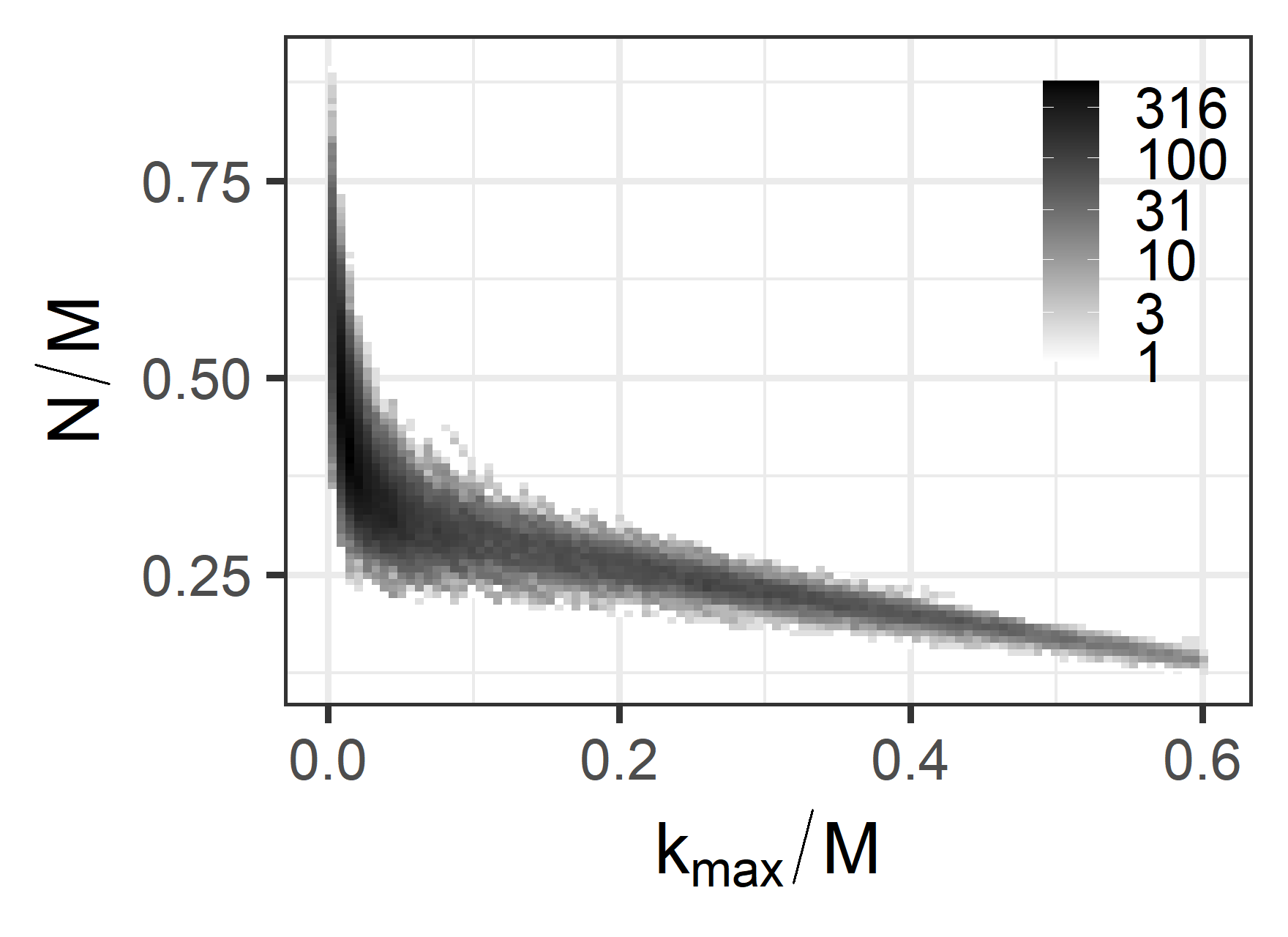}
        \end{minipage}
        \hspace{0.01\linewidth}
        \begin{minipage}{0.49\linewidth}
            \includegraphics[width=\linewidth]{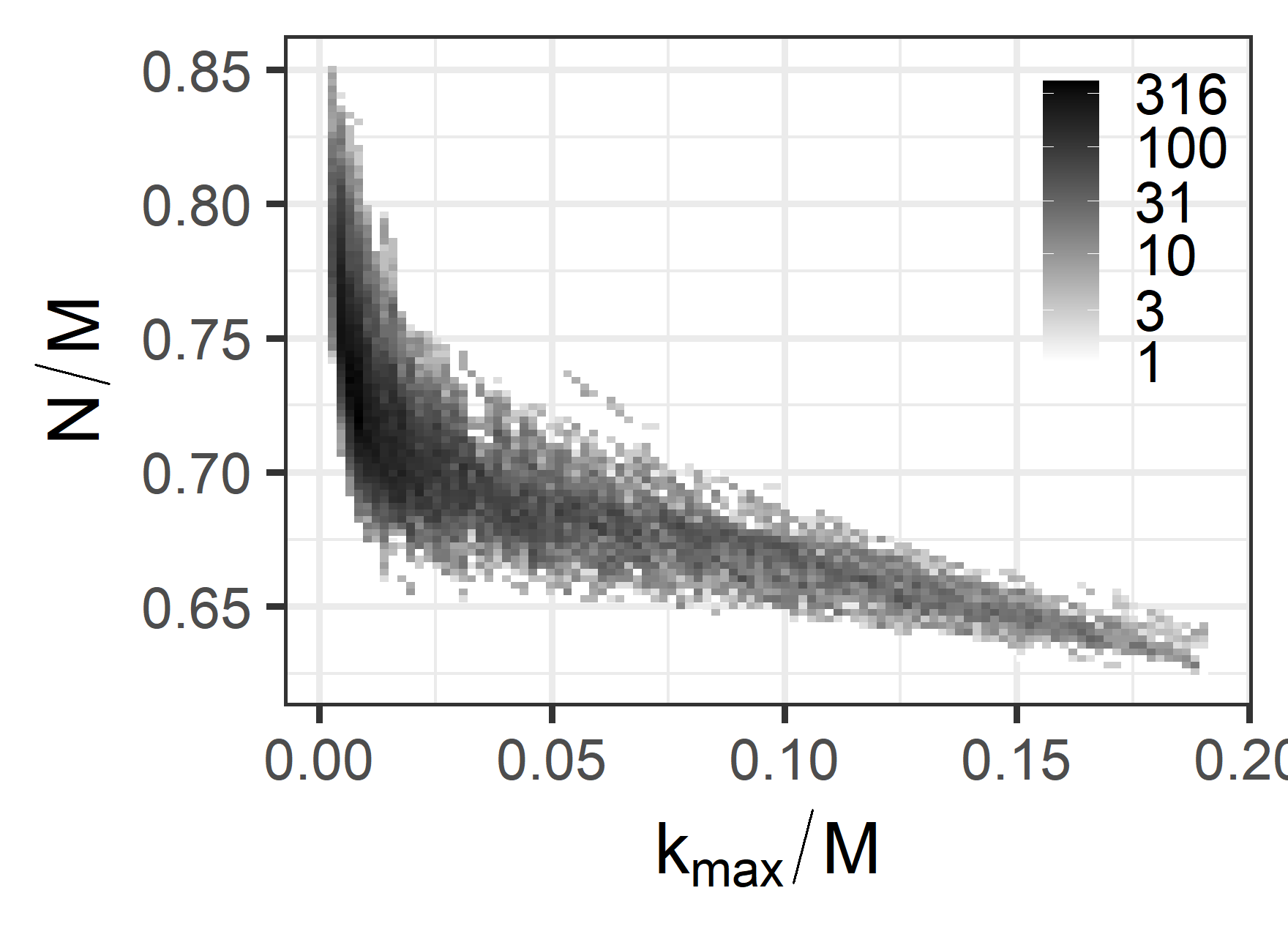}
        \end{minipage}
        \hspace{0.01\linewidth}
    \end{minipage}
    \caption[Coalescence and fragmentation, state space, ${M} = 10^4$, $\Prob{\Frag} = 0.01$, accretion and erosion.]{
    Clockwise from top-left: accretion, erosion amongst any clusters, erosion amongst unique clusters, and combined accretion and erosion.
    In each case, we set $n_{\mathrm{ac}} = 9$ and $n_{\mathrm{er}} = 9$ to best show the effects.
    }
    \label{fig:glshacat}
\end{figure}

We begin by adding Becker-D\"{o}ring mechanics to our base model. The effects of extreme accretion, $n_{\mathrm{ac}} = 9$, can be seen in Figure \ref{fig:glshacat} (top-left).
Accretion exaggerates the gel-shatter cycles, greatly increasing $\calK$ with $\calK_{n_{\mathrm{ac}} = 1} = 0.623$, $\calK_{n_{\mathrm{ac}} = 3} = 0.721$, and $\calK_{n_{\mathrm{ac}} = 9} = 0.797$ and widening the sampled KS-MLE $\alpha_{n_{\mathrm{ac}} = 3}$ to $[2.261, 3.043]$.
Naturally, accretion decreases $n_1$, more substantially influencing the MLE $\alpha$, which predicts a flatter distribution than KS-MLE $\alpha$ does.
The gel-shatter system is fairly robust to accretion, in the sense that $\alpha$ does not change dramatically and behaviour is similar to, albeit an exaggeration of, the typical behaviour for this regime.
Identifying that a system was undergoing accretion as well as coalescence and shattering would be a difficult task as a result, although there are substantial differences in the behaviour of $N$ and $\kmax$ that might be relevant for specific applications.

Erosion has similar effects on gel-shatter and steady-state systems.
When erosion occurs amongst any clusters, it rapidly forces the system to a steady state of only small clusters for ${n_{\mathrm{er}} = 9}$, while requiring erosion amongst separate unique clusters forces the system nearer to but not fully into a steady state, as shown in Figure \ref{fig:glshacat} (top-right and bottom-right).
This is reflected in the differences in summary statistics in Table \ref{tab:cofr:robustnessGS}:
$\calK = -0.079$ and KS-MLE $\alpha \in [2.904, 3.191]$ versus $\calK = 0.178$ and KS-MLE $\alpha \in [2.615, 2.991]$ respectively.
When erosion is amongst any clusters, this predicts that it is more likely for the system's largest cluster to shrink on any given time-step.
In effect, $\calK$ suggests that frequent erosion can act as a suitable alternative to fragmentation.
In either case, we must conclude that erosion can perturb the expected power-law exponent and even cause a gel-shatter system to behave as a (modified) steady-state system.
We conclude our discussion of accretion and erosion with the note that, as before, combining accretion and erosion mostly moderates the two constituents, resulting in somewhat higher $\calK$ and depleting monomers but generally similar behaviour as seen in Figure \ref{fig:glshacat} (bottom-left).

\begin{figure} 
    \begin{minipage}{\linewidth}
        \begin{minipage}{0.49\linewidth}
            \includegraphics[width=\linewidth]{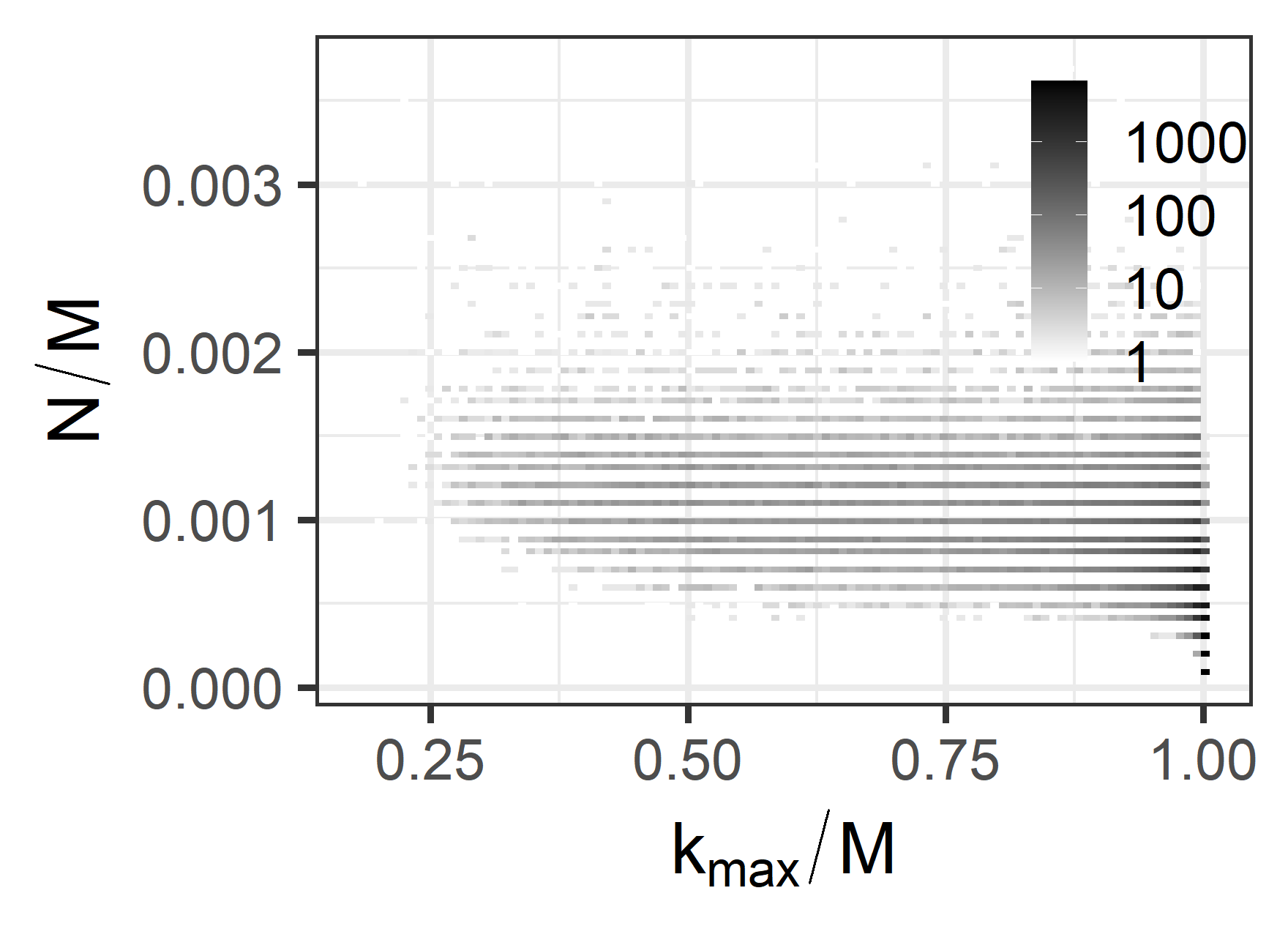}
        \end{minipage}
        \hspace{0.01\linewidth}
        \begin{minipage}{0.49\linewidth}
            \includegraphics[width=\linewidth]{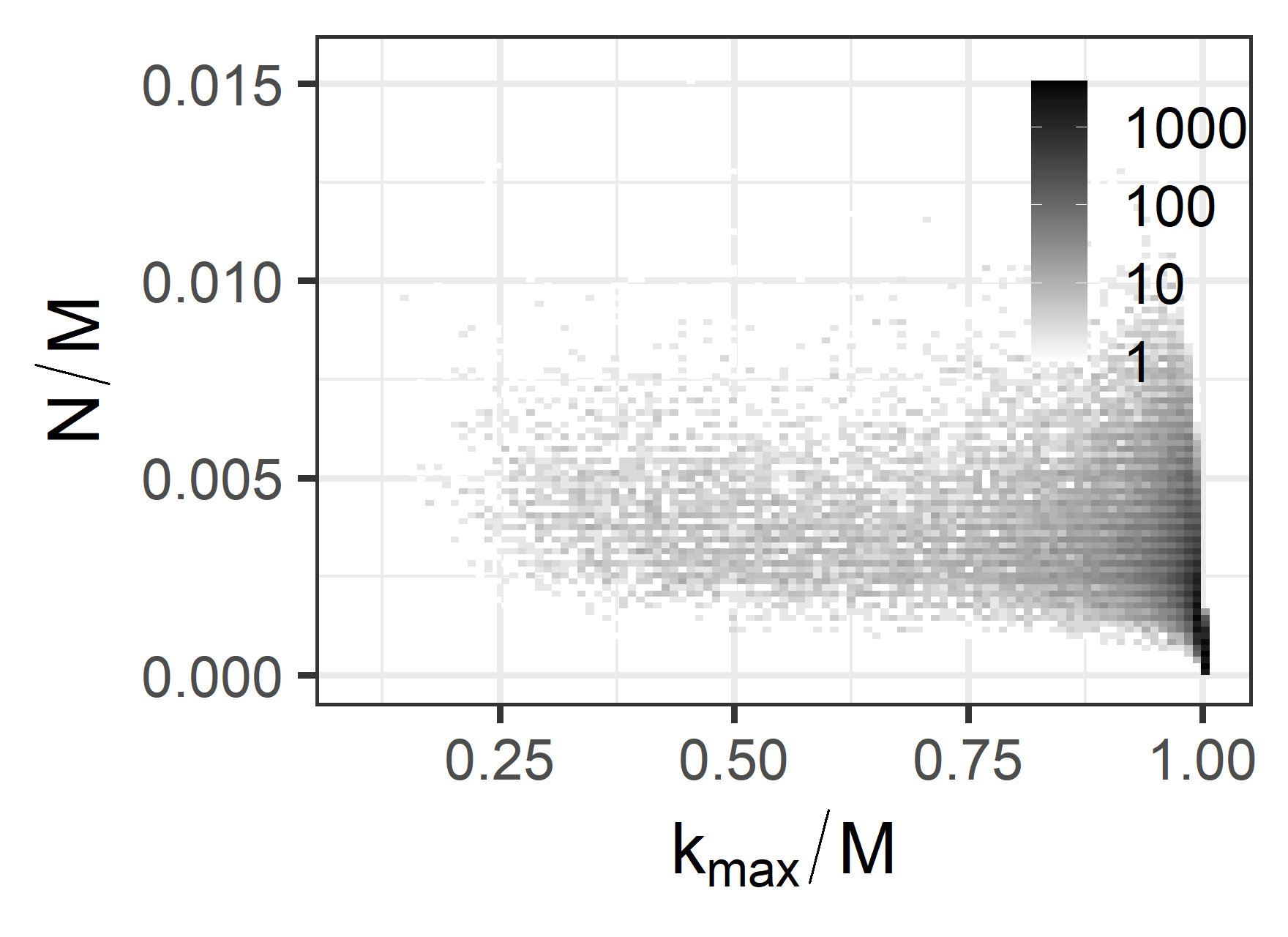}
        \end{minipage}
        \hspace{0.01\linewidth}
        \begin{minipage}{0.49\linewidth}
            \includegraphics[width=\linewidth]{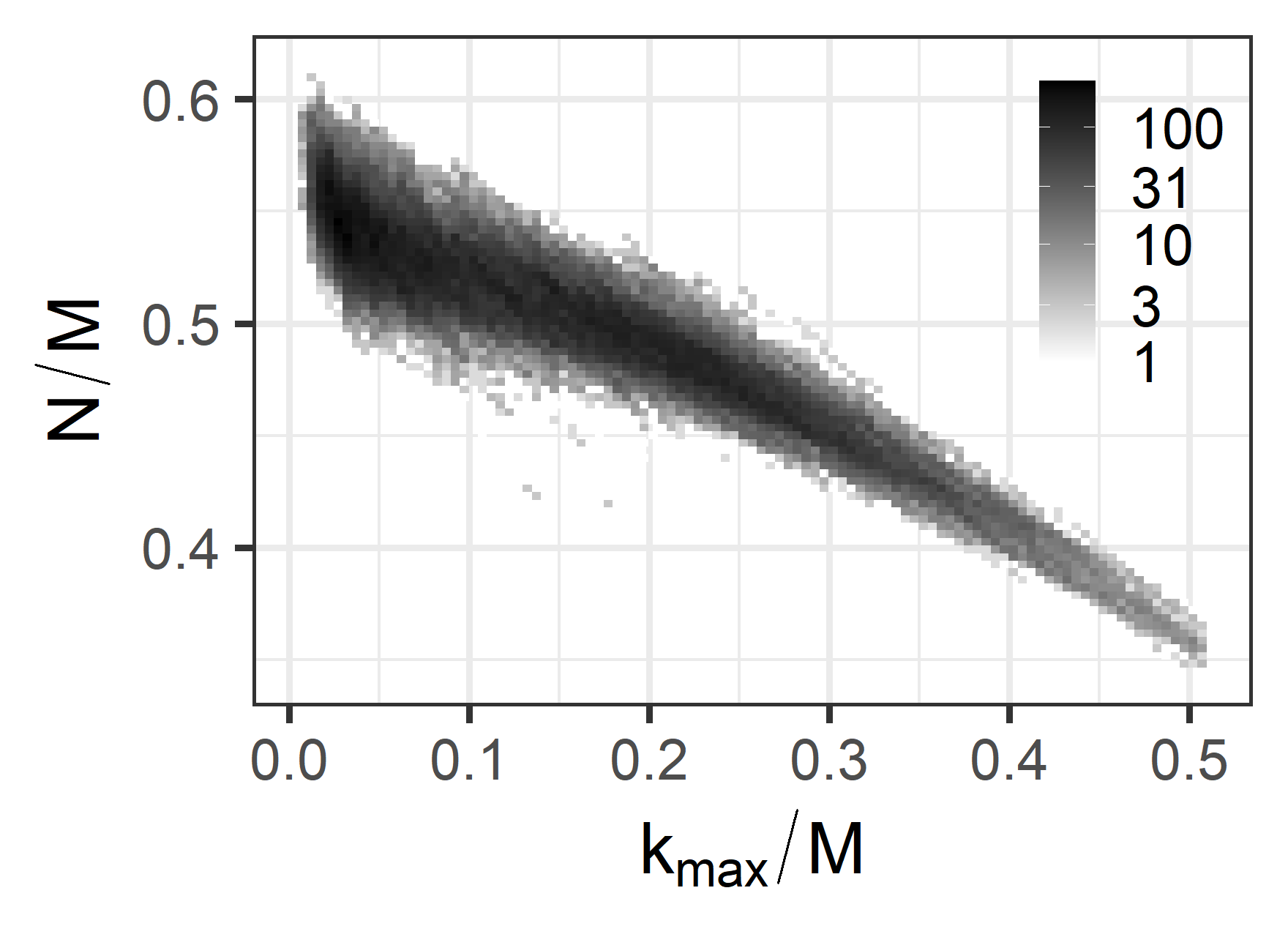}
        \end{minipage}
        \hspace{0.01\linewidth}
        \begin{minipage}{0.49\linewidth}
            \includegraphics[width=\linewidth]{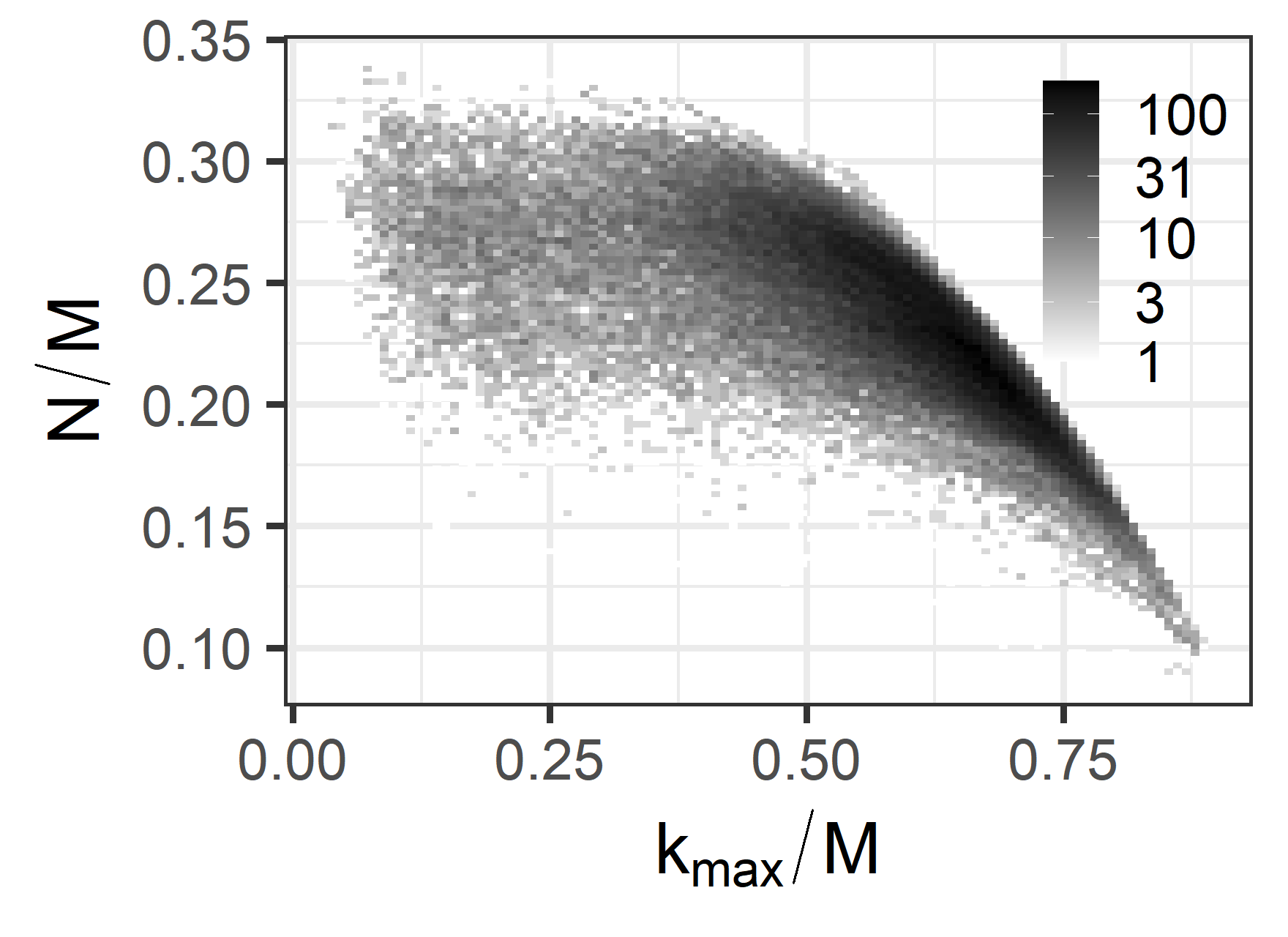}
        \end{minipage}
        \hspace{0.01\linewidth}
    \end{minipage}
    \caption[Coalescence and fragmentation, state space, ${M} = 10^4$, $\Prob{\Frag} = 0.01$, power-law fragmentation.]{
    Clockwise from top-left: stick-breaking fragmentation, CRP fragmentation with $\theta = 1.2$, CRP fragmentation with $\theta = 1.5$, and CRP fragmentation with $\theta = 1.8$.
    }
    \label{fig:glshpl}
\end{figure}

Clearly, simple Becker-D\"{o}ring dynamics can significantly affect gel-shatter dynamics.
It is then not surprising that using power-law variations should also significantly alter the dynamics observed.
We focus on power-law distributed CRP fragmentation.
At the extreme is stick-breaking fragmentation, seen in Figure \ref{fig:glshpl} (top-left).
In this case, the system remains almost completely aggregated despite experiencing fragmentation.
Indeed, $\calK = -0.026$ indicates that fragmentation is the dominant process in the system, reflected in the extremely low fitted exponents.
It is hard to argue that the system is power-law distributed, however, rendering the exponent summary statistics superfluous. 
Stick-breaking fragmentation effectively results in a system that is almost always aggregated into a single cluster, which rarely fragments slightly before quickly re-aggregating the fragments.
In this sense, stick-breaking fragmentation removes any gel-shatter cyclicity.

\begin{figure} 
    \begin{minipage}{\linewidth}
        \begin{minipage}{0.49\linewidth}
            \includegraphics[width=\linewidth]{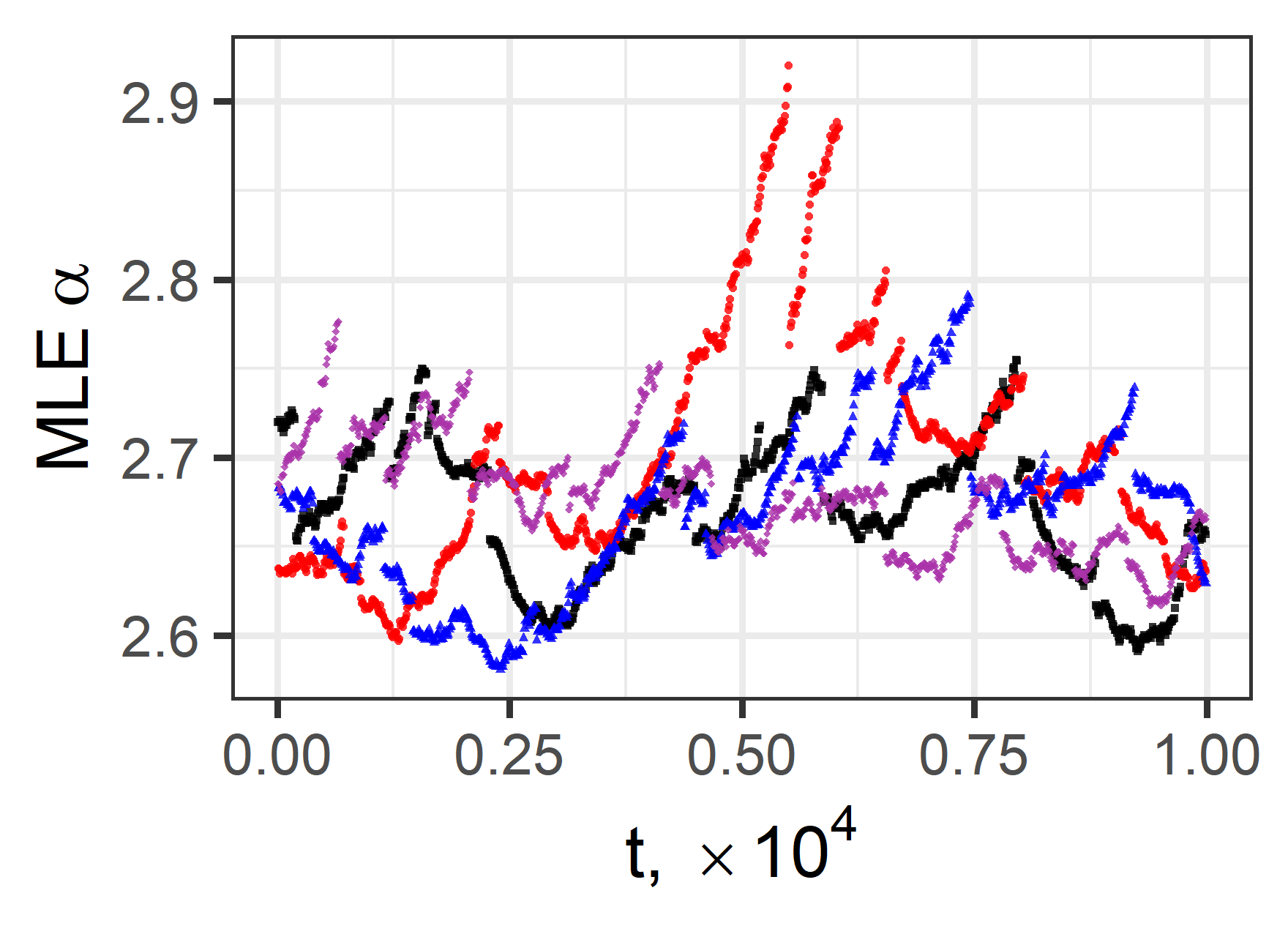}
        \end{minipage}%
        \hspace{0.01\linewidth}
        \begin{minipage}{0.49\linewidth}
            \includegraphics[width=\linewidth]{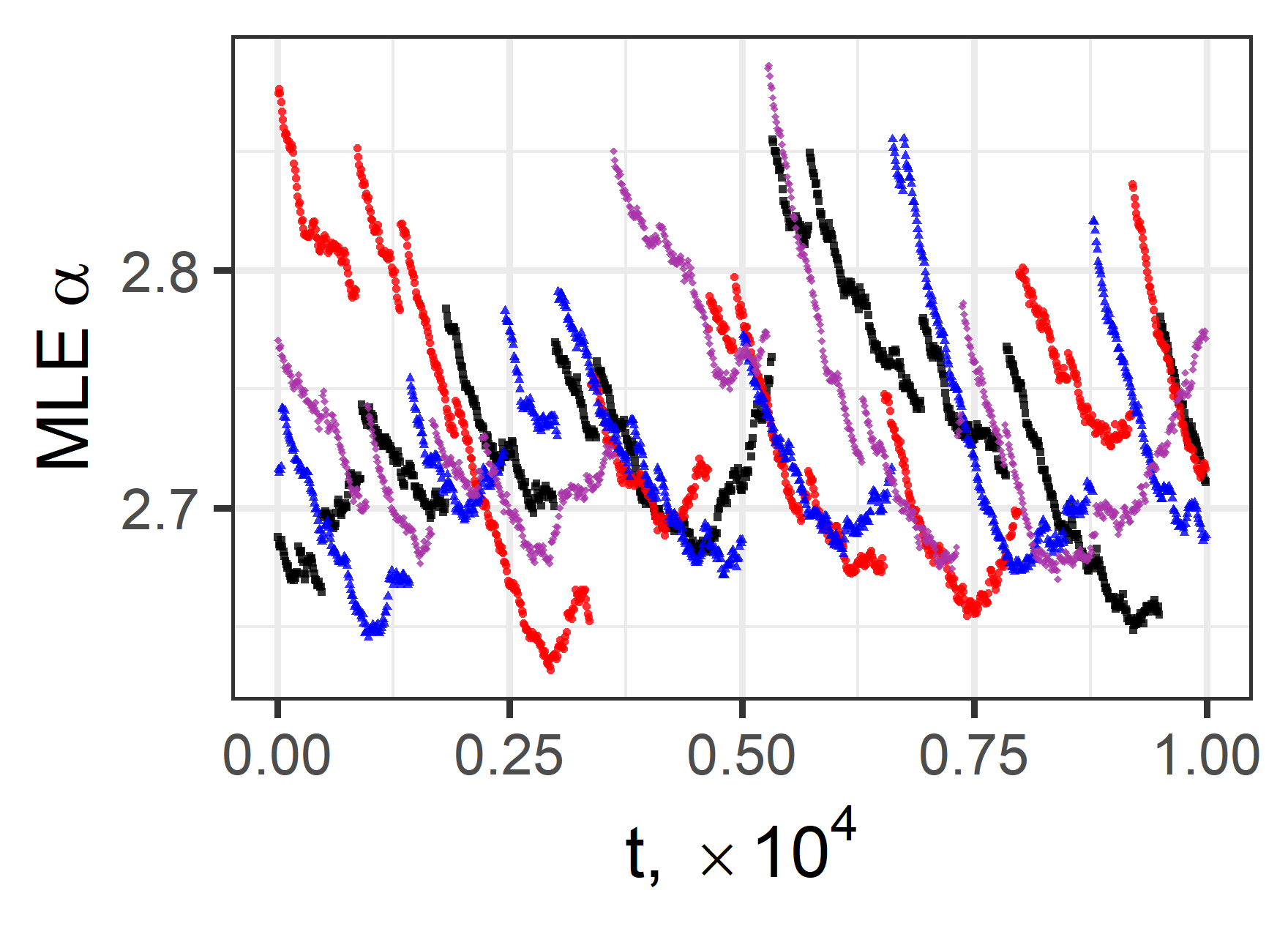}
        \end{minipage}%
    \end{minipage}
    \caption[Coalescence and fragmentation, transition in {MLE} at high $\theta$.]{
    Coalescence and fragmentation {MLE} exponent summary statistics combining gel-shatter cycles with power-law distributed fragmentation, ${M} = 10^4$, $\Prob{\Frag} = 0.01$, and $\theta = 1.80$, left, and $\theta = 1.90$, right.
    Four simulations were conducted (black, blue, red, and purple).
    Compare with Figure \ref{fig:examplegelshatter}. 
    }
    \label{fig:gelshatter:transition}
\end{figure}

Low fragmentation power-law exponents $\theta$ reproduce results akin to that of stick-breaking fragmentation.
As $\theta$ increases, the system becomes more disaggregated after fragmentation and requires longer periods of time to re-aggregate fully.
Large clusters are still common after fragmentation, aiding the reaggregation of the system by decreasing the number of steps that do not contribute directly to $\kmax$ and thus increasing $\calK$ beyond that of the standard system: $\calK_{\theta = 1.2} = 0.151$, $\calK_{\theta = 1.5} = 0.910$, and $\calK_{\theta = 1.8} = 0.811$.
The system also experiences a smooth change in the fitted power-law exponents as the system transitions back to the original gel-shatter system:
stick-breaking has KS-MLE $\alpha \in [0.686, 1.563]$,
$\theta = 1.2$ has KS-MLE $\alpha \in [0.979, 1.838]$,
$\theta = 1.5$ has KS-MLE $\alpha \in [1.875, 2.519]$, and
$\theta = 1.8$ has KS-MLE $\alpha \in [2.362, 2.609]$.
Discussions of summary statistics and state spaces do not show the full range of effects of power-law fragmentation, however. Comparing Figure \ref{fig:gelshatter:transition} with the distinctive cycles observed in the base gel-shatter model of Figure \ref{fig:examplegelshatter}, one still sees some cyclicity, but over a smaller range of $\alpha$, less distinctive, and with greater stochasticity both overall and in the rougher variation in $\alpha$ observed within each cycle. There is also a subtle transition between falling $\alpha$, Figures \ref{fig:examplegelshatter} and \ref{fig:gelshatter:transition} right, and rising $\alpha$, Figure \ref{fig:gelshatter:transition} left.

\begin{table} 
    \caption[Summary of robustness in gel-shatter cycles]{
    Summary of robustness results in gel-shatter cycle coalescence and fragmentation. In all cases, ${M} = 10^4$, $\Prob{\Frag} = 0.01$, and four simulations were used.
    The ranges given for {KS-MLE} and {MLE} estimates of power-law $\alpha$ contain $95\%$ of empirical results. Perturbations used are explained in the text.  $\calK > 0.2$ is characteristic of gel-shatter cycling.
    We exclude large ($>5$) and small ($<1$) estimates of the {KS-MLE} $\alpha$ from our intervals.
    }
    \label{tab:cofr:robustnessGS}
    \centering
    \begin{tabular}{|l|ccc|} \hline
    Perturbation                   & Cyclicity $\calK$    & {KS-MLE} ($\xmin > 1$) & {MLE} \\\hline \hline
    Gel-shatter cycles             & $0.490$   & $\left[2.474, 2.829\right]$ & $\left[2.656, 3.163\right]$ \\\hline
    Accretion (3, Uniq.)           & $0.729$   & $\left[2.248, 2.957\right]$ & $\left[1.559, 2.800\right]$ \\\hline
    Erosion (3, Uniq.)           & $0.178$   & $\left[2.615, 2.991\right]$ & $\left[2.970, 3.316\right]$ \\\hline
    Accr. and Eros. (3, Uniq.)     & $0.590$   & $\left[2.370, 2.884\right]$ & $\left[1.960, 2.710\right]$ \\\hline
    Accretion (3)                  & $0.721$   & $\left[2.261, 3.043\right]$ & $\left[1.603, 2.849\right]$ \\\hline
    Erosion (3)                  & $-0.079$  & $\left[2.904, 3.191\right]$ & $\left[3.276, 3.435\right]$ \\\hline
    Accr. and Eros. (3)            & $0.524$   & $\left[2.420, 2.883\right]$ & $\left[2.062, 2.673\right]$ \\\hline
    Frag. Stick-breaking           & $-0.026$  & $\left[0.686, 1.563\right]$ & $\left[0.902, 1.440\right]$ \\\hline
    {CRP} $\theta = 1.20$ & $0.151$   & $\left[0.979, 1.838\right]$ & $\left[1.027, 1.956\right]$ \\\hline
    {CRP} $\theta = 1.50$ & $0.910$   & $\left[1.875, 2.519\right]$ & $\left[2.357, 2.964\right]$ \\\hline
    {CRP} $\theta = 1.80$ & $0.811$   & $\left[2.362, 2.609\right]$ & $\left[2.588, 2.799\right]$ \\\hline
    \end{tabular}
\end{table}

\pagebreak
\section{Conclusion}\label{sec:Conc}

Descriptions of coalescence and fragmentation (CF) are ubiquitous across many scientific and social-scientific applications, yet there is limited understanding of the connections between the microscopic processes, theorized or actual, and the observed fragment distributions.
There are various specific models, often analytically tractable at the level of deterministic treatment of means, that result in (approximate) power-law distributions of group sizes. 
Empirical distributions, too, are often observed to be approximately power-law, so that power laws \cite{Clauset09_PowerLaw} have become the standard, go-to method for describing the outcomes of CF processes \cite[\emph{e.g.}][]{Birnstiel11_Astrophysics}.

But this is not enough on its own to justify a claim that the models are correctly capturing the microscopic processes, or that the power-law-described steady state is the best approximation to the macroscopic statistics implied by these processes. 
The danger is of a `streetlight effect', in which the model is used because it is available, tractable and sufficient to match the empirical statistics, rather than because it is correctly modelling the underlying process.

The purpose of this article was to take a standard model with an approximately power-law steady-state distribution and fully explore the finite-population stochastic simulations of the model with variations of its microscopic rules, observing the effect on the steady state and summary statistics.  
We examined the maximum likelihood estimators for the power-law exponent $\alpha$ for the whole distribution and (by minimizing the Kolmogorov-Smirnov distance between the actual and fitted cumulative distributions) for the most power-law-like part of the distribution, typically the tail.

Such a task is complicated by the possibility of cyclicity: is a steady state being imputed to a system which, on timescales important in the applied context, is not steady but rather cyclic? -- is there no detailed balance? 
As we saw in a precursor article \cite{Fagan21_gel-shatter}, the phenomenon of gelation (aggregation into a single large cluster), if accompanied by stochastic shattering of clusters, can create stochastic gel-shatter cycles. 
Not all stochastic cyclicity is necessarily of this form, but any cyclicity will still be a significant departure from a steady state. 
The best summary statistic to capture this was a measure $\calK$ of time-asymmetric cyclicity, the asymmetry between coalescing and fragmenting time steps.

\begin{figure} 
    \centering
    \includegraphics[width=\textwidth]{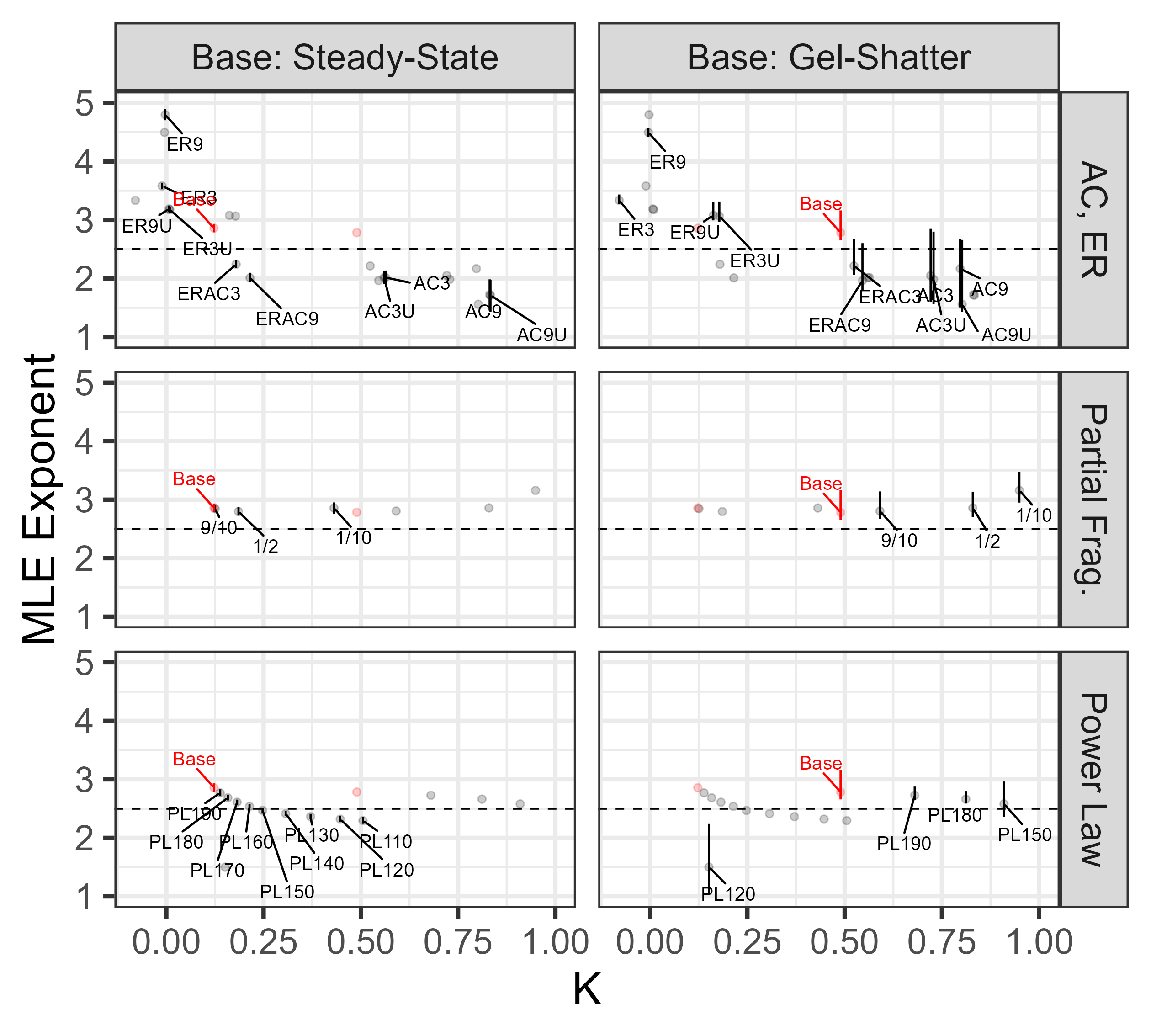}
    \caption{Visual summary of robustness results for both steady-state and gel-shatter cycles. Select results from Tables \ref{tab:cofr:robustnessSS} (left column) and \ref{tab:cofr:robustnessGS} (right column) are displayed as MLE fitted exponent against the cyclicity $\calK$. Fitted exponents are displayed as vertical $95\%$ ranges with points placed at the median estimates. The base case is highlighted in red, a horizontal line at the expected 2.5 exponent is plotted (but see Figures \ref{fig:steadystate:methodviolins} and \ref{fig:steadystate:methodviolinsSimulated}), and the points are repeated to highlight shifts between the regimes. Results for KS-MLE are similar, but show less curvature for accretion and erosion. The shorthand labels are: AC$n_{ac}$, accretion of $n_{ac}$; ER$n_{er}$, erosion of $n_{er}$; U, with the uniqueness constraint; $9/10$, $1/2$, $1/10$, shattering of proportion of cluster; PL$\theta$, fragmentation according to a CRP with parameter $\theta$ (units of hundredths here).}
    \label{fig:TablePlot}
\end{figure}

A full analysis of how any variation in the microscopic details gives rise to power law steady states and/or cyclicity would be a major research programme in itself.
Our numerical results, presenting unified data from Tables \ref{tab:cofr:robustnessSS} and \ref{tab:cofr:robustnessGS} in Figure \ref{fig:TablePlot}, enable us to infer some general trends.
We see that erosion tends to increase stochastic cyclicity $\calK$ and reduce alpha (giving a bigger tail), and accretion does the opposite, both quite significantly. 
Thus simply fitting a power law to a simulation or to empirical data requires some confidence that accretion and erosion processes are not significant in the modelled process.
Furthermore, partial shattering tends to leave the exponent little changed but enhances $\calK$. 
Power-law fragmentation can shift the exponent a little more but still enhances $\calK$.
Our results for such power-law fragmentation, where we have a continuous independent variable available to be adjusted, suggest that output should be a continuous function of the model variation, but it is beyond the scope of the present study to quantify this mapping precisely.
Overall, even if the power law seems to fit well, all sorts of non-shattering fragmentation can enhance the $\calK$ that we already know exists in the base model: simply put, a well-fitted power law does not indicate that this is necessarily all that is going on.

Gel-shatter cyclicity only occurs in certain narrow regimes in which coalescence can create a gel but size-biased shattering inevitably leads to its dissolution into monomers. 
Such regimes are, unsurprisingly, less robust. 
Again, small amounts of accretion and erosion do not have disproportionate effects, but larger amounts can reduce or remove gel-shatter cyclicity, and move the system close to a steady state. 
The various alternative fragmentation processes (random partition, stick-breaking and CRP) can easily prevent gel-shatter cyclicity: even when a single large cluster forms, it may merely fragment somewhat and re-form in a stochastic but time-symmetric, constant manner. 
It seems that some degree of shattering into monomers, rather than a more varied distribution of fragments, is necessary to observe true gel-shatter cycles.

Finally, let us return to our opening context, of war and political violence. 
On the basis of our results, we can say that Richardson's law for the distribution of wars \cite{Richardson60_SDQ} and modern results on deadly events in insurgencies \cite{Bohorquez09_Common} are both consistent with a broad class of CF models, broader than the base model and its generalizations treated deterministically in \cite{Ruszczycki2009_CF}.
Furthermore our findings suggest that the core dynamics are somewhat robust to the behaviour of individuals -- captured through the processes of accretion and erosion in Becker-D\"oring dynamics -- providing some explanation of the consistency of the power-law findings in the operations research literature.
Nevertheless one should also reasonably expect that an imputed steady state is not the full story, since endogenous (in addition to exogenous) factors can naturally produce not merely stochastic variation but some measure of time-asymmetric cyclicity, even if there is no clear gelation-like phenomenon.

\backmatter

\bmhead{Acknowledgments}

We should like to thank Stephen Connor, Gustav Delius and Dmitri (Mitya) Pushkin for discussions, and George Constable and Kevin Glazebrook for comments. 

This project was undertaken on the Viking Cluster, which is a high performance compute facility provided by the University of York. We are grateful for computational support from the University of York High Performance Computing service, Viking and the Research Computing team.

\section*{Declarations}

{\bf Funding:} Brennen Fagan would like to thank the Department of Mathematics at the University of York for a PhD studentship, and the Leverhulme Centre for Anthropocene Biodiversity at the University of York for a postdoctoral position, Grant Number RC-2018-021. For the purpose of open access a Creative Commons Attribution (CC BY) licence is applied to any Author Accepted Manuscript version arising from this submission.

\noindent{\bf Conflicts of interest:} None.

\noindent{\bf Authors' contributions:} BTF conducted the simulations. NJM and AJW conceived and supervised the project. All authors participated in writing and editing the article.

\noindent{\bf Availability of code and data:} Code and data to generate the plots are available at \url{https://github.com/Brennen-Fagan/Robustness-Coal-Frag/}.







\bibliography{references}

\end{document}